\documentclass[a4paper,11pt,oneside,fleqn]{article}

\usepackage[mathscr]{eucal}
\usepackage{amsmath,amssymb,amsthm}
\usepackage{graphicx,subfig}
\usepackage{caption}
\usepackage{color}

\usepackage{cite}
\usepackage{hyperref}
\hypersetup{colorlinks, linkcolor=blue, citecolor=blue, urlcolor=blue}

\allowdisplaybreaks

\makeatletter
\long\def\@makecaption#1#2{%
  \vskip\abovecaptionskip\footnotesize
  \sbox\@tempboxa{#1. #2}%
  \ifdim \wd\@tempboxa >\hsize
    #1. #2\par
  \else
    \global \@minipagefalse
    \hb@xt@\hsize{\hfil\box\@tempboxa\hfil}%
  \fi
  \vskip\belowcaptionskip}
\makeatother

\newcommand{\tsup}[1]{\textsuperscript{#1}}
\title{Wave propagation modelling in various microearthquake environments using a spectral-element method}

\author{Hom Nath Gharti$^{1,2}$\thanks{Email: hgharti@princeton.edu}, Volker Oye$^3$, Michael Roth$^3$, and Daniela K\"uhn$^3$\\
${}^1$ Department of Geosciences, Princeton University, Princeton, NJ 08544, USA\\
${}^2$ formerly at NORSAR, Gunnar Randers vei 15, N-2007 Kjeller, Norway\\
${}^3$ NORSAR, Gunnar Randers vei 15, N-2007 Kjeller, Norway}
\begin{document}

\maketitle

\begin{abstract}
Simulation of wave propagation in a microearthquake environment is often challenging due to small-scale structural and material heterogeneities. We simulate wave propagation in three different real microearthquake environments using a spectral-element method. In the first example, we compute the full wavefield in 2D and 3D models of an underground ore mine, namely the Pyh\"asalmi mine in Finland. In the second example, we simulate wave propagation in a homogeneous velocity model including the actual topography of an unstable rock slope at \r Aknes in western Norway. Finally, we compute the full wavefield for a weakly anisotropic cylindrical sample at laboratory scale, which was used for an acoustic emission experiment under triaxial loading. We investigate the characteristic features of wave propagation in those models and compare synthetic waveforms with observed waveforms wherever possible. We illustrate the challenges associated with the spectral-element simulation in those models.
\end{abstract}

\textbf{Keywords}: Wave propagation, microearthquakes, mining, rock slope, acoustic emission

\newpage

\section{Introduction}

Microearthquakes are small events having a magnitude less than 3.0 (e.g., \cite{lee1981,bohnhoff2010}). These events usually occur in large number, and are caused by the induced fracture, such as mining, excavation and hydrofracture (e.g., \cite{shapiro2011}) or natural fracture, such as earthquake faulting, volcanoes and landslide (e.g., \cite{bulut2009}). Collective interpretation of these events often provides a valuable information. Therefore the microearthquake monitoring is used in several applications, for example, location of potentially unstable zones in mines (e.g., \cite{mendecki1997,young1992}), monitoring $\text{CO}_2$ sequestration (e.g., \cite{maxwell2004}), and to study the internal structure of geothermal (e.g., \cite{phillips1997}) and hydrocarbon reservoirs (e.g., \cite{rutledge1998,maxwell2010}). Small earthquakes have a high-frequency content (e.g., \cite{aki1967,hanks1977}), and therefore small-scale structural or material heterogeneities may strongly influence the wavefield. Due to this reason, observed waveforms of microearthquakes are often complicated (e.g., \cite{oye2003}). The simulation of wave propagation helps to understand complexities of the waveforms and provides valuable information onIn many cases, the presence of strong velocity contrasts causes problems for the numerical stability. the wavefield interaction. Besides, synthetic data are necessary to assess the applicability and reliability of data processing algorithms. Similarly, synthetic Green's functions are necessary for moment-tensor inversion. However, wave propagation modelling is a challenging task for microearthquakes. An accurate geometry mapping is complicated due to the small-scale natural or human-made structural complexities. In many cases, the presence of strong velocity contrasts causes problems for the numerical stability.

The finite-difference method (FDM) is widely used for wave propagation modelling (e.g., \cite{virieux1986,dablain1986,bayliss1986,levander1988}). Direct discretization of the governing equation on structured grid (i.e., regular grid) makes the FDM simple and efficient, and easily adaptable to parallel processing (e.g., \cite{larsen1995,saenger2004}). However, it is difficult to accurately model complex boundaries and interfaces using the structured grid. In the presence of surface topography, the FDM is less accurate due to the approximation of boundary conditions (e.g., \cite{jih1988,ohminato1997}). The spectral-element method (SEM) is a higher order finite-element method (FEM) that uses a nodal quadrature, namely the Gauss-Lobatto-Legendre quadrature for numerical integration over an element. While the nodal quadrature results in a diagonal mass matrix enabling an efficient time-marching scheme, the higher-order elements give a high degree of spatial accuracy. Since the SEM solves a weak form of the wave equation, free-surface boundary conditions are automatically satisfied. It is also possible to model complex boundaries and interfaces accurately, using the unstructured mesh (e.g., \cite{seriani1998,komatitsch1998}). The nodal quadrature was initially limited to certain kind of elements, e.g., quadrilaterals in 2D and hexahedra in 3D. Hexahedral meshing is a challenging task and an area of active research (e.g., \cite{mercerat2006,pebay2008,shepherd2008}). Only a few hexahedral meshing tools are currently available, e.g., CUBIT \cite{cubit2011}, Gmsh \cite{geuzaine2009}, and TrueGrid \cite{truegrid2006}. The hexahedral meshing is usually not fully automated, and careful mesh design is necessary. There have been a few attempts to implement the SEM using other types of elements such as triangles in 2D and tetrahedra in 3D using so-called Fekete points (e.g., \cite{hesthaven1999,taylor2000b,komatitsch2001,mercerat2006}). Since the nodal quadrature includes end points of the numerical integration interval, the order of numerical integration may not always be sufficiently high~\cite{durufle2009}. However, the advantages of the high degree of spatial accuracy and the efficient time marching outweigh the disadvantage of the low-order integration (e.g., \cite{nielasen1997,seriani2008,debasabe2010}). The SEM was originally developed to solve fluid dynamics problems (e.g., \cite{patera1984,canuto1988,cohen2002,deville2002}). Since then, it has been widely used to simulate seismic wave propagation in various scales (e.g., \cite{faccioli1997,komatitsch1999,komatitsch2002a,tromp2008,oye2010,peter2011}). 

In this paper, we present the results of full wavefield simulations in an underground ore mine, an unstable rock slope, and a weakly anisotropic cylindrical sample used for an acoustic emission laboratory experiment. We discuss the characteristic features of the wavefields and compare the waveforms with observed data wherever possible. We illustrate the challenges associated with the application of the spectral-element method in these problems. We use the SPECFEM2D and SPECFEM3D packages~\cite{specfem2011} originally developed by Komatitsch and Tromp~\cite{komatitsch1999} for our simulations.

\section{Pyh\"asalmi ore mine}

\begin{figure}[htbp]
\centering
\subfloat[]{\includegraphics[scale=0.34]{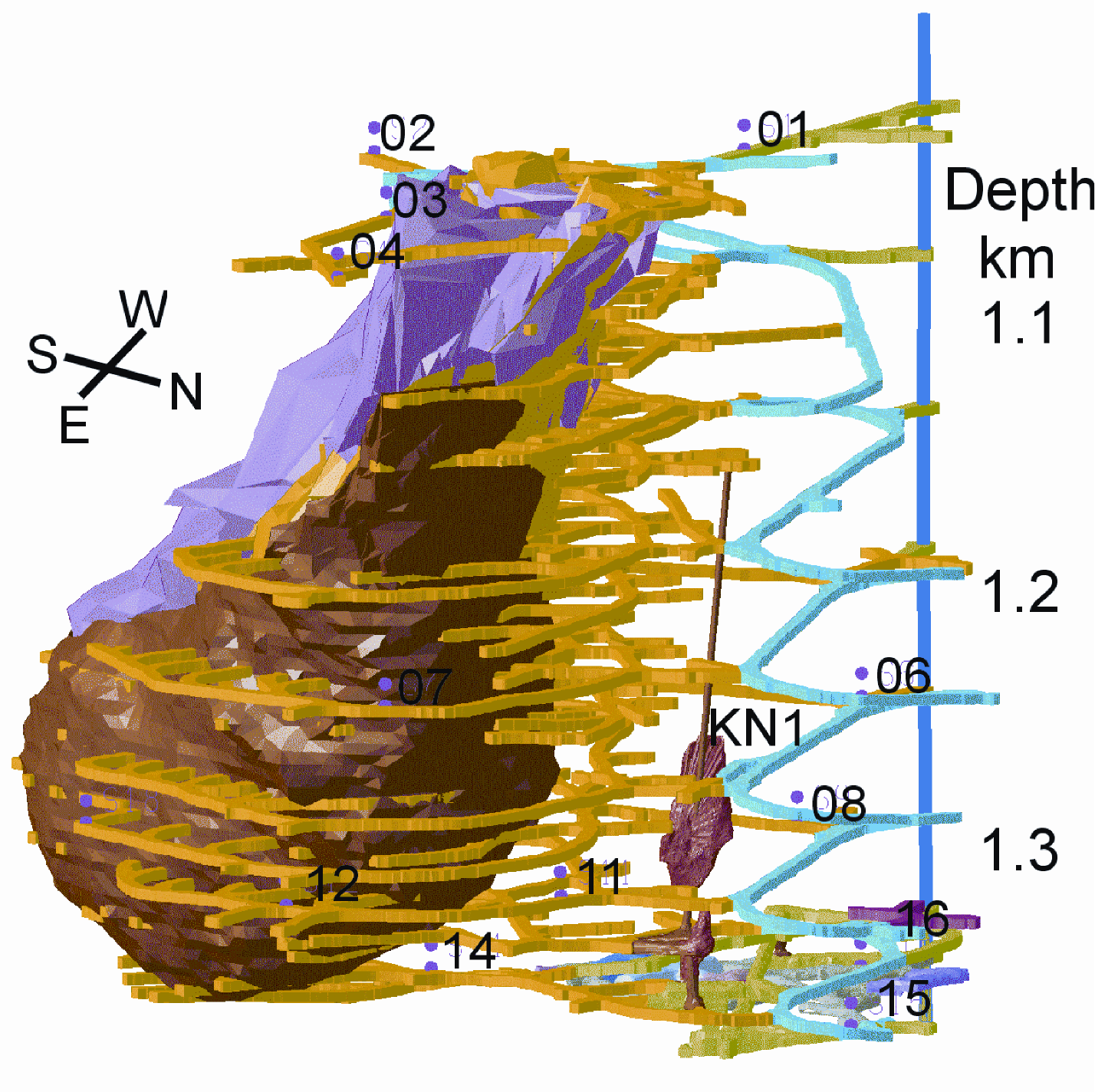}}\qquad
\subfloat[]{\includegraphics[scale=0.35]{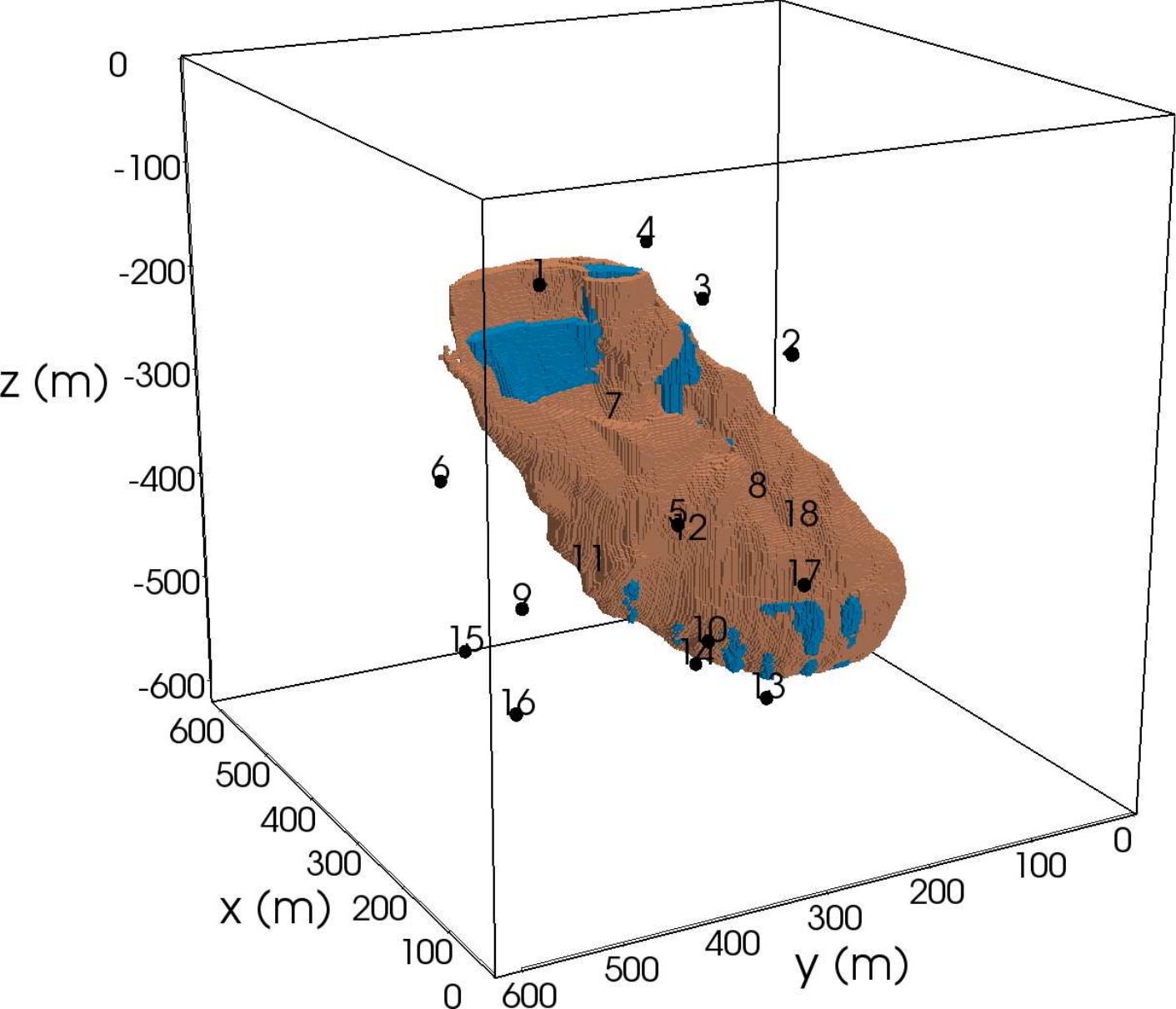}}
\caption{a) Pyh\"asalmi ore mine with surrounding
infrastructure: the copper/zink ore body is shown in brown/pink,
access tunnels are shown in yellow, the elevator shaft is shown in
dark blue, and seismic stations are numbered. The passage for the quarried
ore is marked by KN1. b) 3D velocity model (see Table \ref{tab:mine_prop}) of the mine
used to generate the synthetic data: stopes (i.e., mined-out cavities)
are shown in blue and the ore body is in brown. Remainder is the host rock. Geophone locations are shown in black and are numbered. The $x$, $y$, and $z$ axes represent East, North, and vertical directions, respectively.}
\label{fig:py}
\end{figure}

The Pyh\"asalmi mine is an underground ore mine located in central Finland. This mine consists of a volcanogenic massive sulphide (VMS) deposit, and produces mainly copper, zinc, and pyrite. The copper-zinc ore body in the mine extends down to a depth of $\sim$~1.4~km (Figure~\ref{fig:py}a). The in-mine seismic network consists of 18 geophones including 6 three-component instruments (geophones 1, 5, 9, 13, 17, and 18)~\cite{puustjarvi1999}. The geophones have a sampling rate of up to 3~kHz. For this mine, we have a detailed 3D velocity model and observed microseismic data at our disposal (for more details on the mine and the microseismic event characteristics, see \cite{oye2005}). The 3D velocity model of the mine is shown in Figure~\ref{fig:py}b and described in Table~\ref{tab:mine_prop}. Even though the model is a simplification of the original structure (Figure~\ref{fig:py}a), it still poses a challenging task for the modelling of wave propagation regarding geometrical discretization, numerical stability, and accuracy. Gharti et. al.~\cite{gharti2008} used this model to compute full waveforms and first-arrival times using a 3D visco-elastic finite-difference code~\cite{larsen1995} and a 3D finite-difference eikonal solver~\cite{podvin1991}, respectively, and used these data to locate the microearthquakes in the mine~\cite{gharti2010}.

\begin{table}
\centering
\begin{tabular}{c c c c}
\hline
     & $V_p$ & $V_s$ & $\rho$ \\
     & (m/s) & (m/s) & (kg/m\tsup{3})\\
\hline
Air & 300 & 0 & 1.25 \\
Rock & 6000 & 3460 & 2000 \\
Ore & 6300 & 3700 & 4400 \\
\hline
\end{tabular}
\caption{Material properties of the Pyh\"asalmi mine model.}
\label{tab:mine_prop}
\end{table}

\subsection{2D model}

\begin{figure}[htbp]
\centering
\includegraphics[scale=0.35]{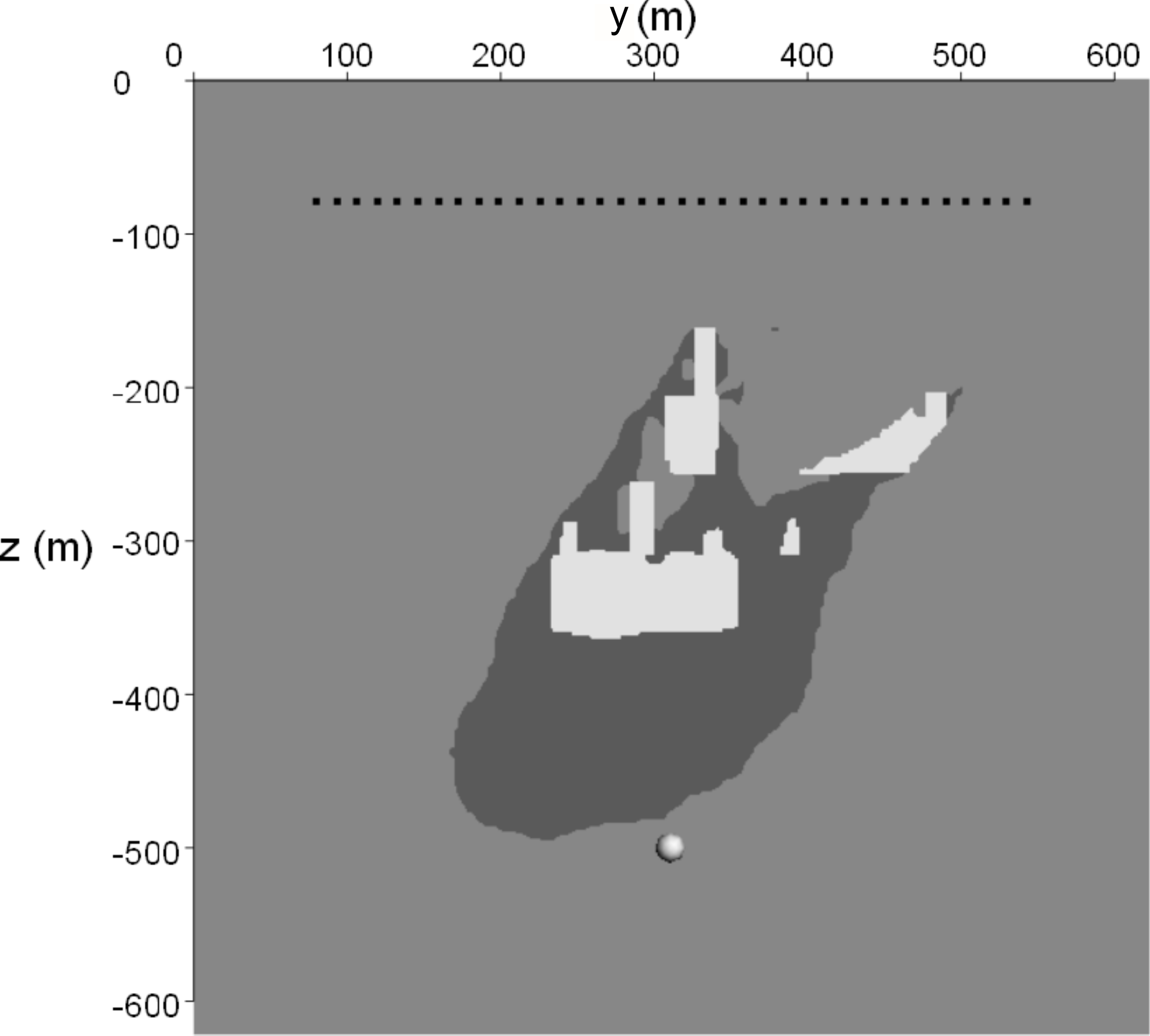}
\caption{2D model of the Pyh\"asalmi mine (North-South cross-section of the 3D model, i.e., $yz$-plane at $x$ = 311~m in Figure~\ref{fig:py}b). The model consists of host rock (light gray), ore body (dark gray), mined-out stopes (white), and source (sphere). The receiver line of 36 geophones (black) is used for the computation purpose.}
\label{fig:py2d_model}
\end{figure}

\begin{figure}[htbp]
\centering
\subfloat[]{\includegraphics[scale=2]{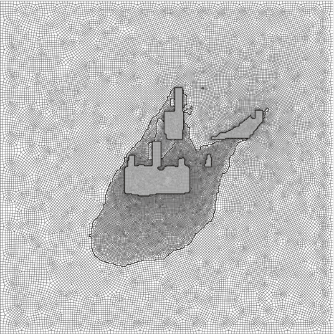}}\quad
\subfloat[]{\includegraphics[scale=2]{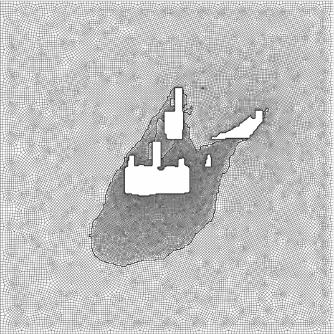}}
\caption{Spectral element mesh for the 2D model of the Pyh\"asalmi mine. a) Including air. b) Excluding air.}
\label{fig:py2d_mesh}
\end{figure}

In this example, we consider a 2D model of the Pyh\"asalmi mine taking a North-South (i.e., $yz$-plane at $x$ = 311~m in Figure~\ref{fig:py}b) cross-section, and perform two simulations. In the first simulation, we mesh the entire model including the air in the stopes (i.e., mined-out cavities). Since the S-wave velocity is zero in air, we cannot satisfy the dispersion condition in this region. The P-wave velocity for the air is also very low; therefore we need a very fine mesh within this region. In the second simulation, we mesh the model excluding the mined-out cavities. The primary purposes of these simulations are to assess the stability of the SEM in the presence of a strong velocity contrast, and to investigate whether there are any discrepancies in the waveforms when including or excluding the cavities during meshing of the model.

For the computation purpose, we place a receiver line with 36 equispaced geophones near the top surface (at $z$ = -80~m) as shown in Figure~\ref{fig:py2d_model}. We take a source represented by a Ricker wavelet having a central frequency of 250~Hz, which is located at $y=310$~m and $z=-500$~m. This source is characterised by a strike-slip mechanism on the $yz$-plane and a seismic moment ($M_0$) of $-10^{10}$~Nm. We select a sampling interval of 1~$\mu$s for the seismogram recordings. A quadrilateral mesh is required for the SEM simulation in 2D models. For the quadrilateral meshing, we use the mesh generation tool CUBIT~\cite{cubit2011}. For the case with air included, we use an average element size of 2~m resulting in a total of 53,071 spectral elements (Figure~\ref{fig:py2d_mesh}a). To preserve the same Courant number for numerical stability outside the stopes, we have to maintain the same degree of mesh fineness outside the stopes in both cases. Therefore, we simply remove the mesh on the stopes to obtain the mesh for the case with air excluded, resulting in a total of 40,594 spectral elements (Figure~\ref{fig:py2d_mesh}b). Excluding the stopes for the meshing reduces the number of elements by about 30\%, for this particular example, which is significant regarding the computational cost. Practically, a coarser mesh may be used when excluding the air.

\begin{figure}[htbp]
\centering
\subfloat{\includegraphics[scale=0.32]{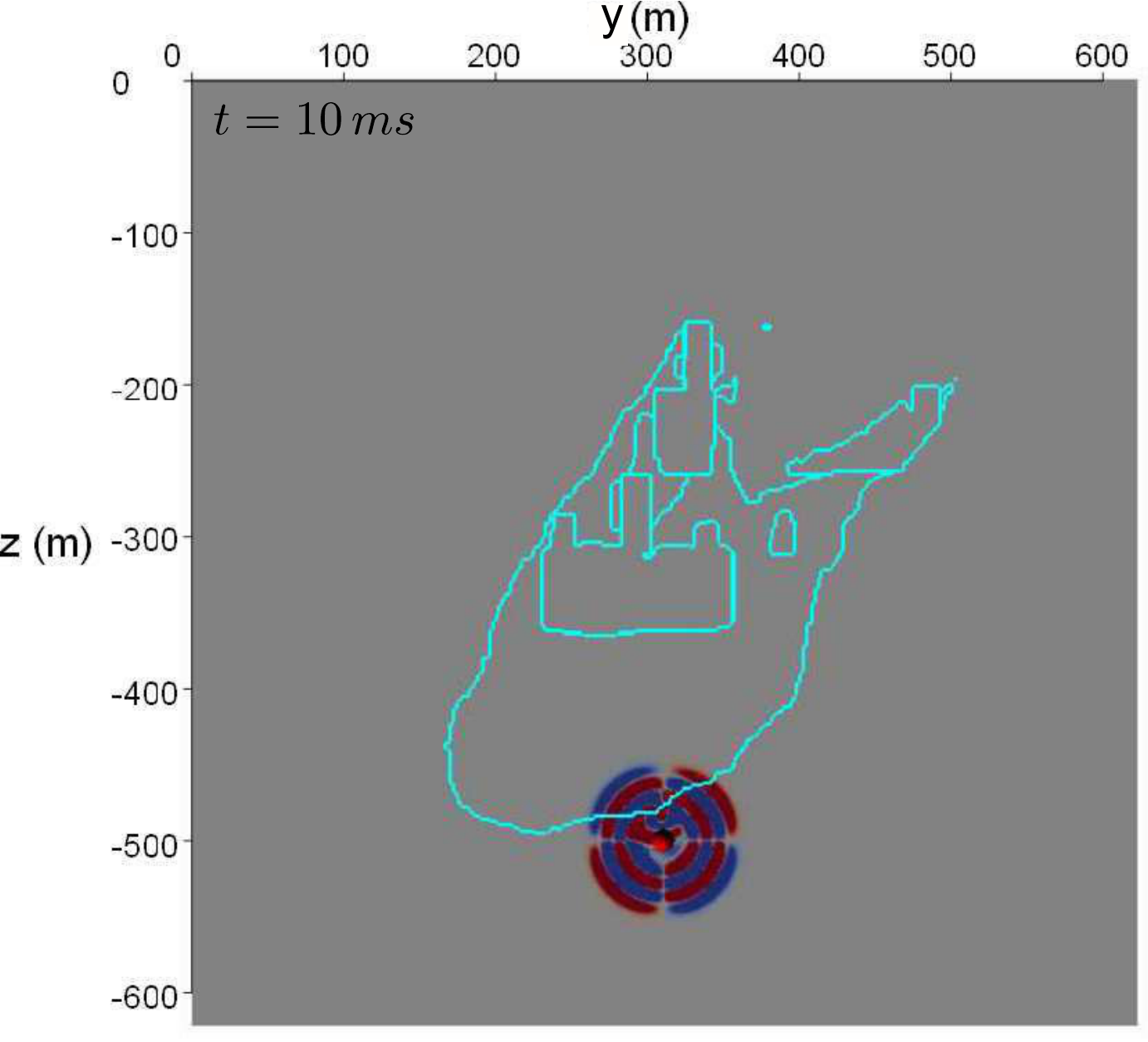}}
\subfloat{\includegraphics[scale=0.32]{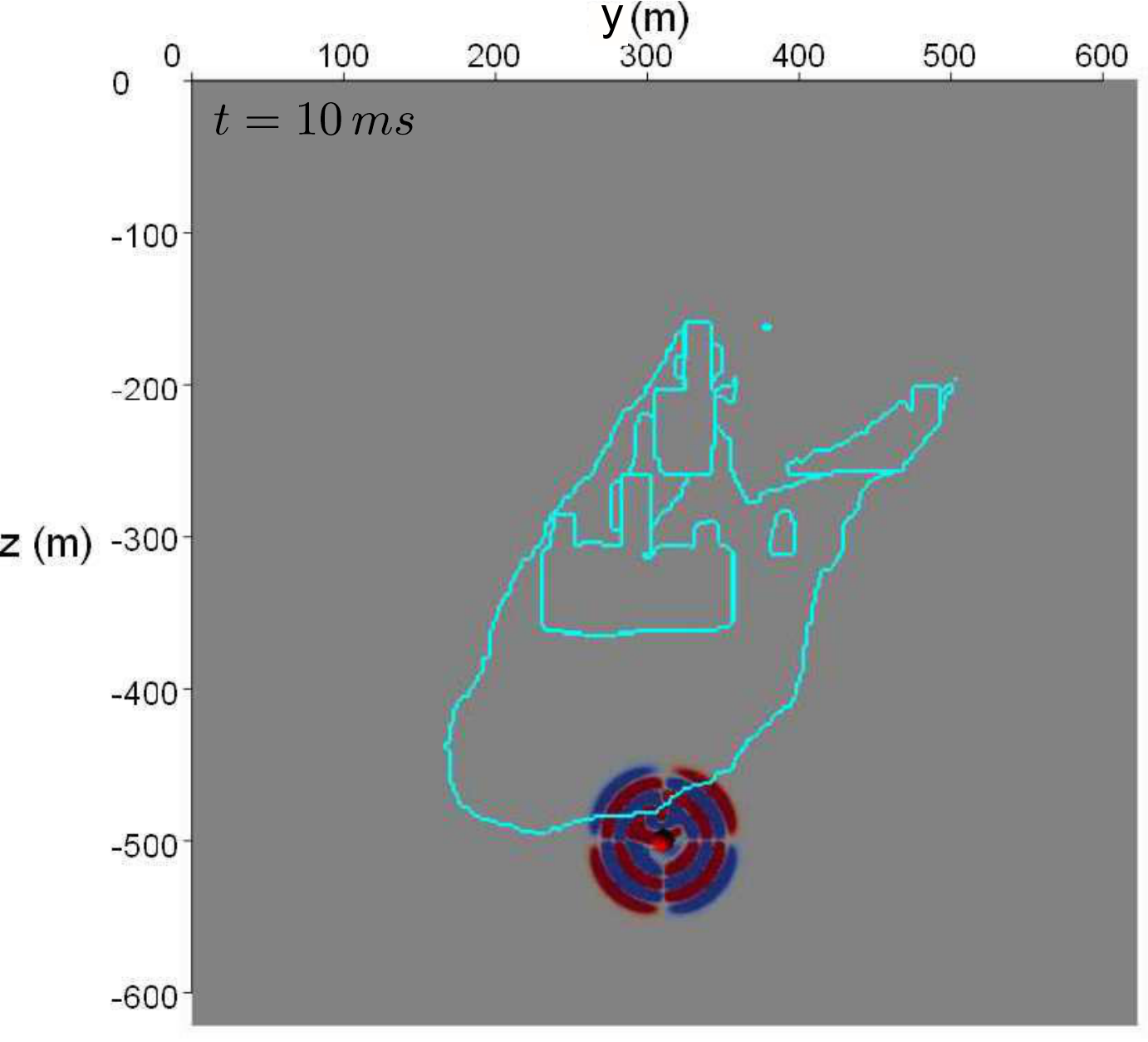}}\\
\subfloat{\includegraphics[scale=0.32]{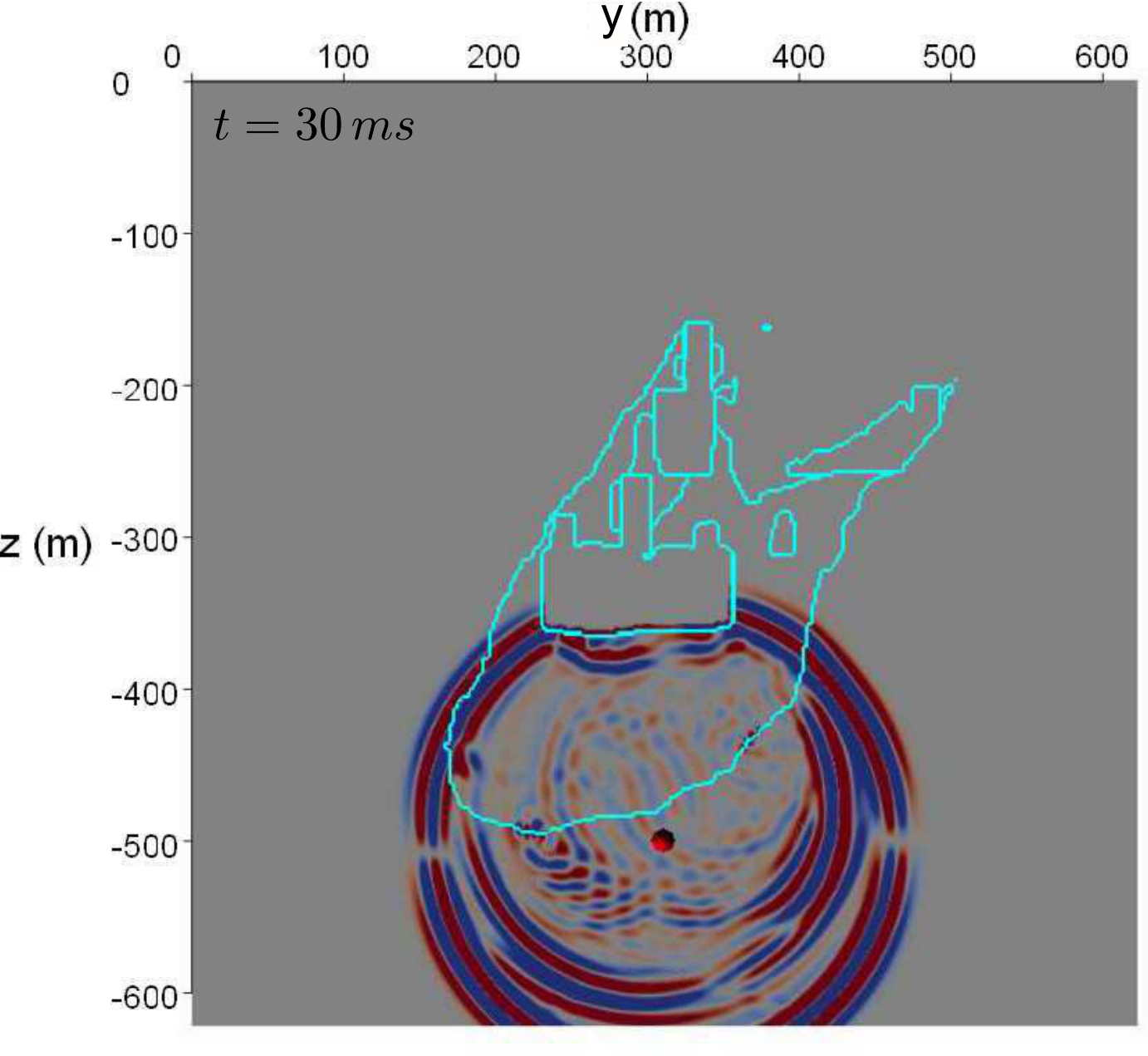}}
\subfloat{\includegraphics[scale=0.32]{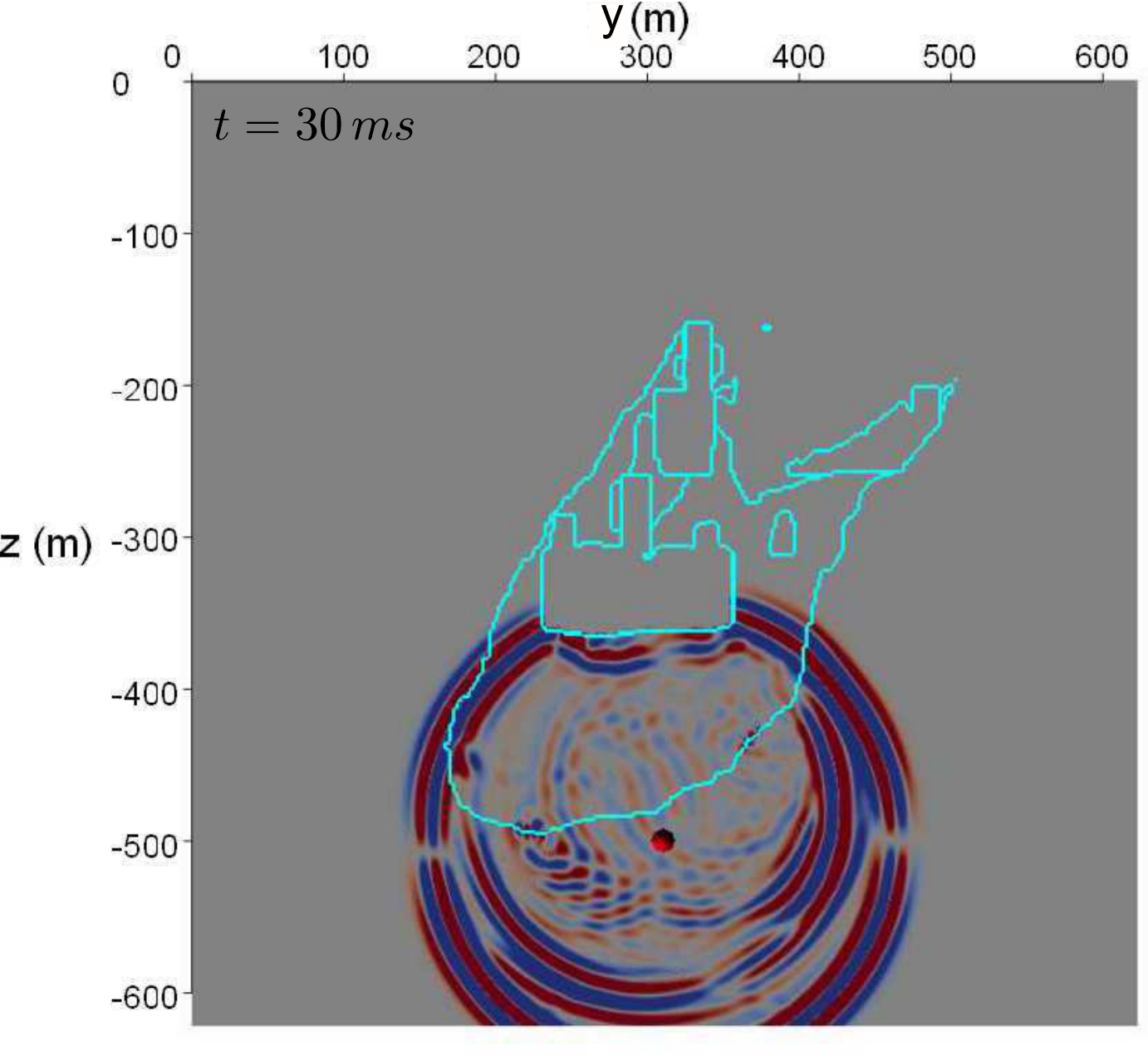}}\\
\subfloat{\includegraphics[scale=0.32]{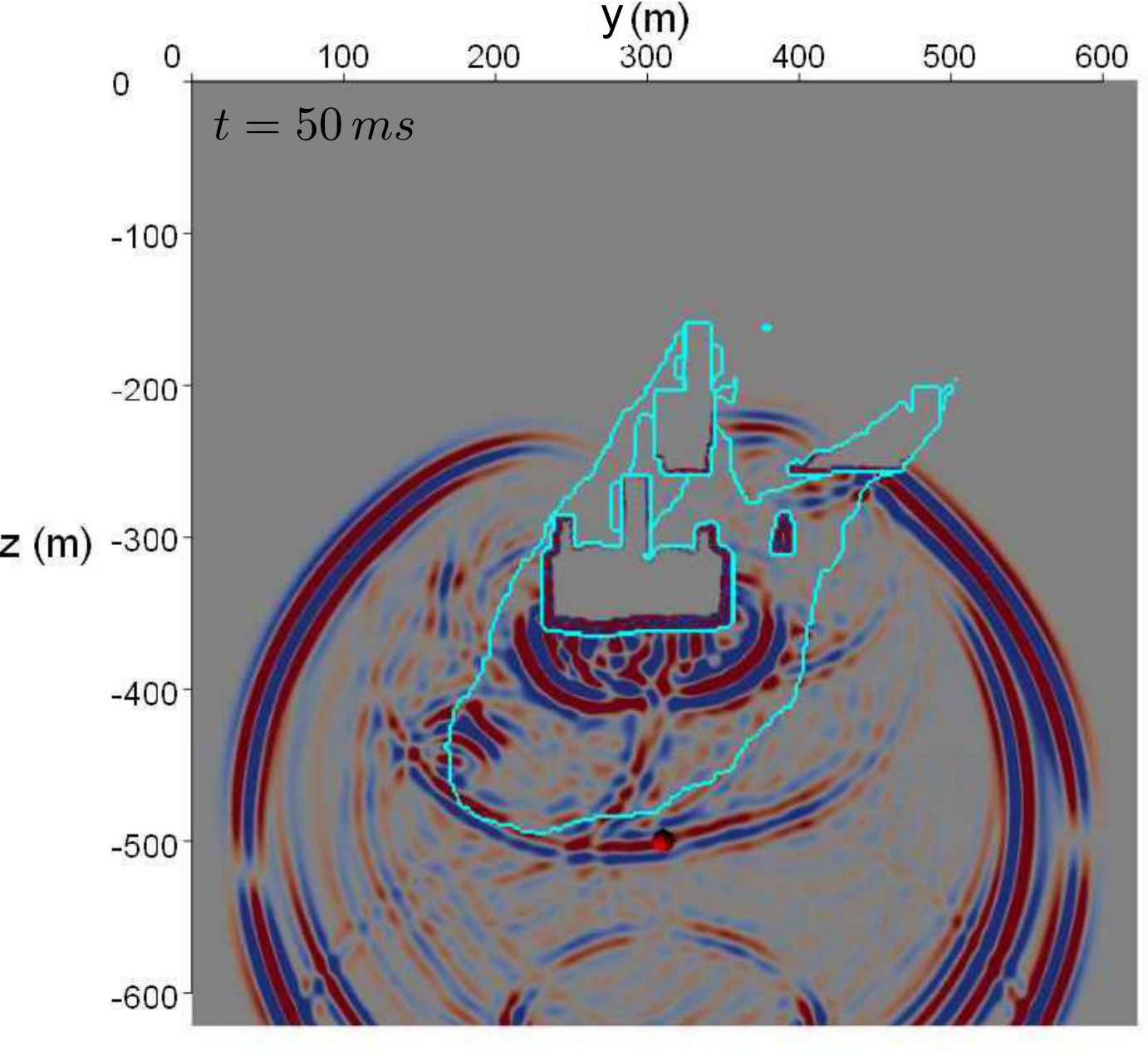}}
\subfloat{\includegraphics[scale=0.32]{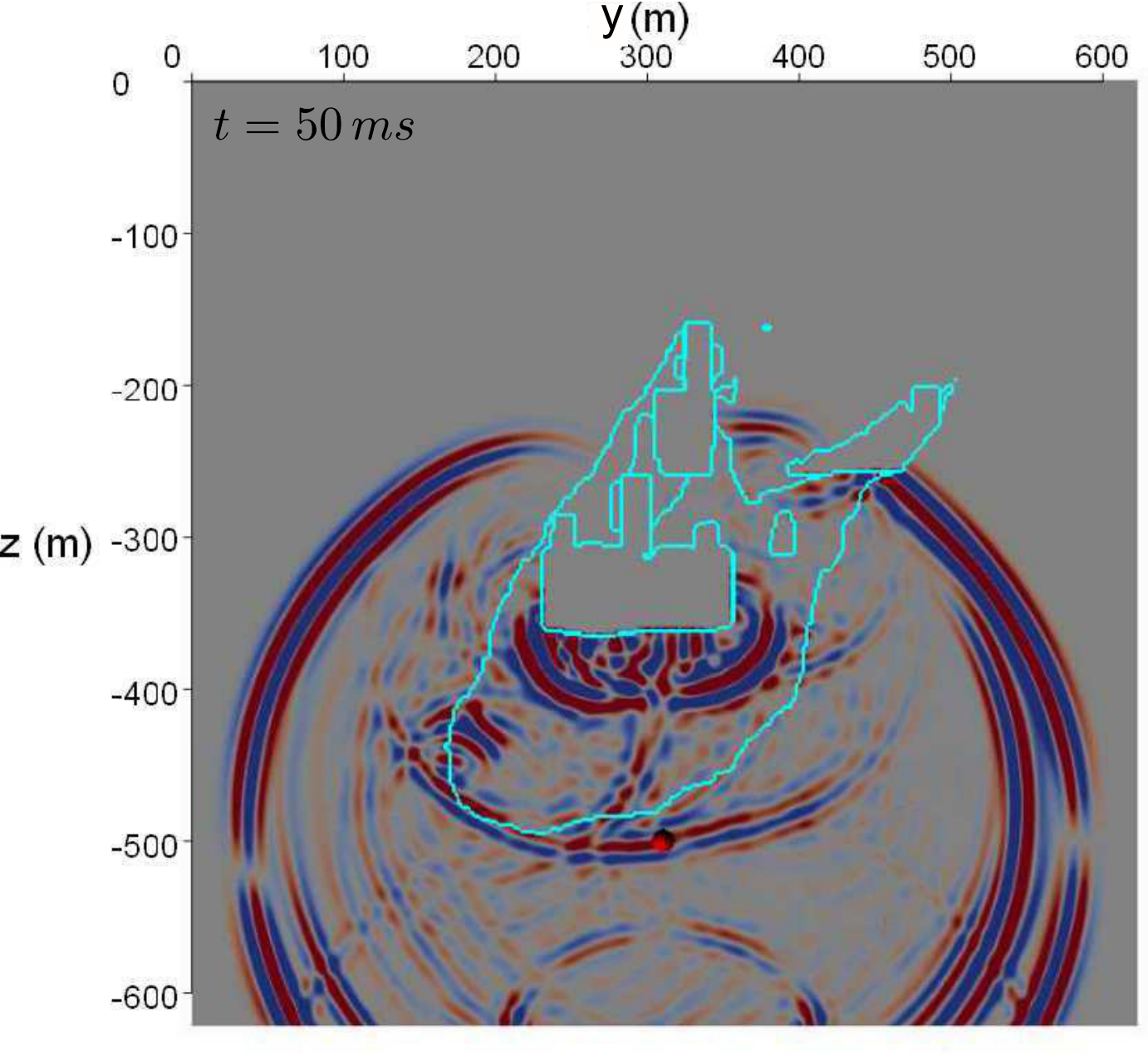}}\\
\subfloat{\includegraphics[scale=0.32]{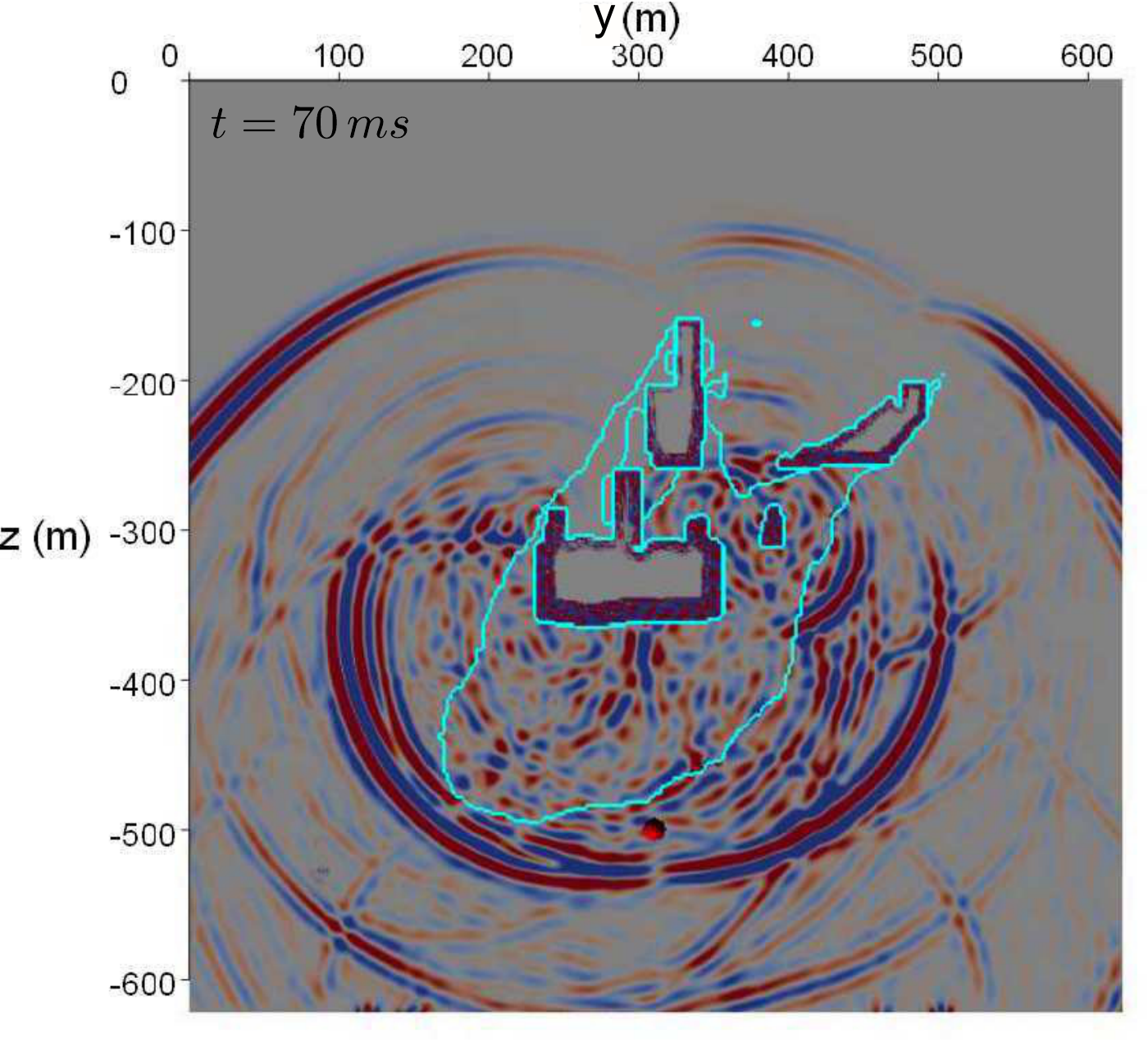}}
\subfloat{\includegraphics[scale=0.32]{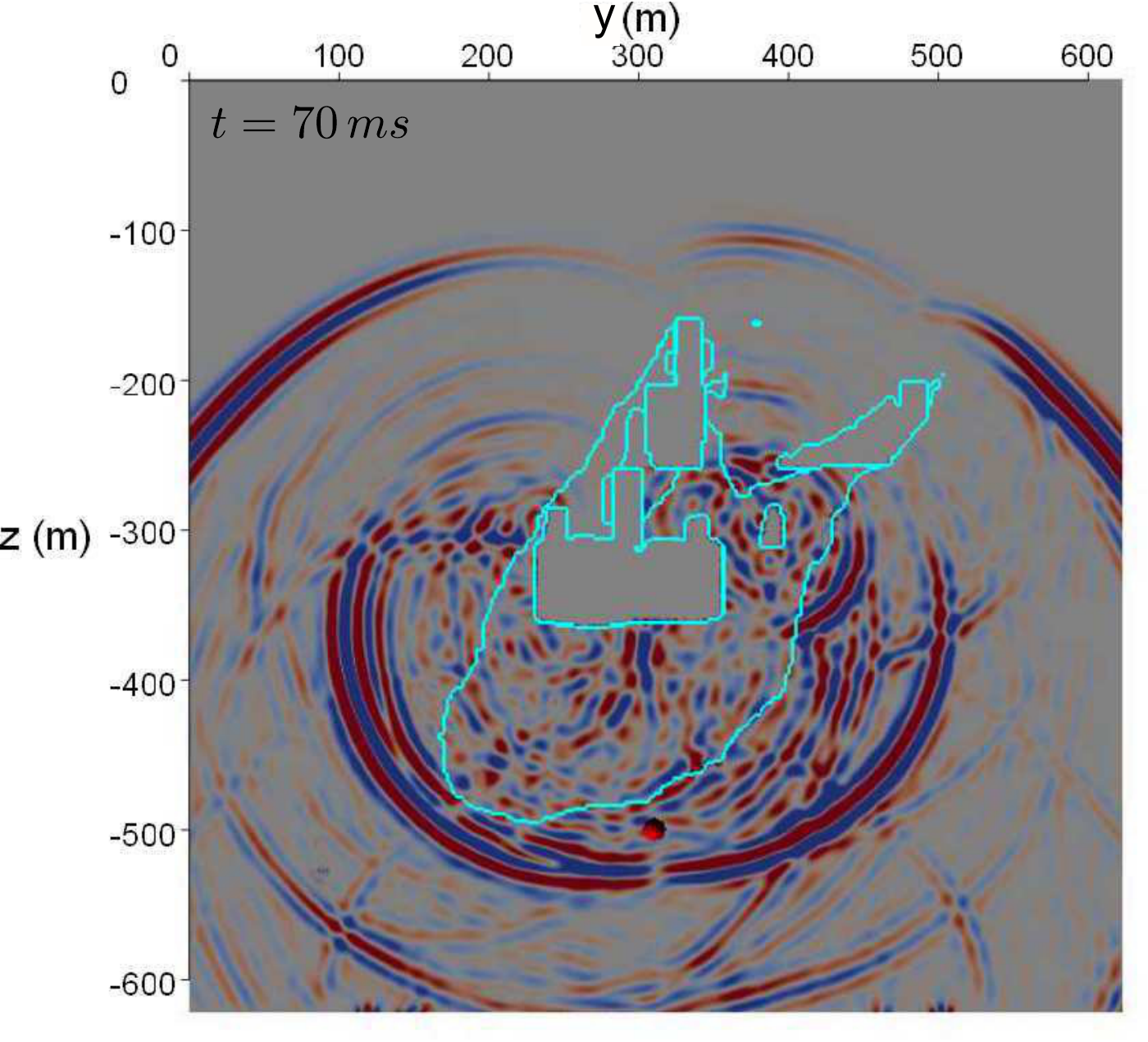}}
\caption{P-wave potential for the model with air included (left column) and with air excluded (right column). The lines in cyan represent the outlines of the ore body and the stopes. High values of the potential are shown in red and low values are shown in blue.}
\label{fig:py2d_pwave_air_noair}
\end{figure}

\begin{figure}[htbp]
\centering
\subfloat{\includegraphics[scale=0.32]{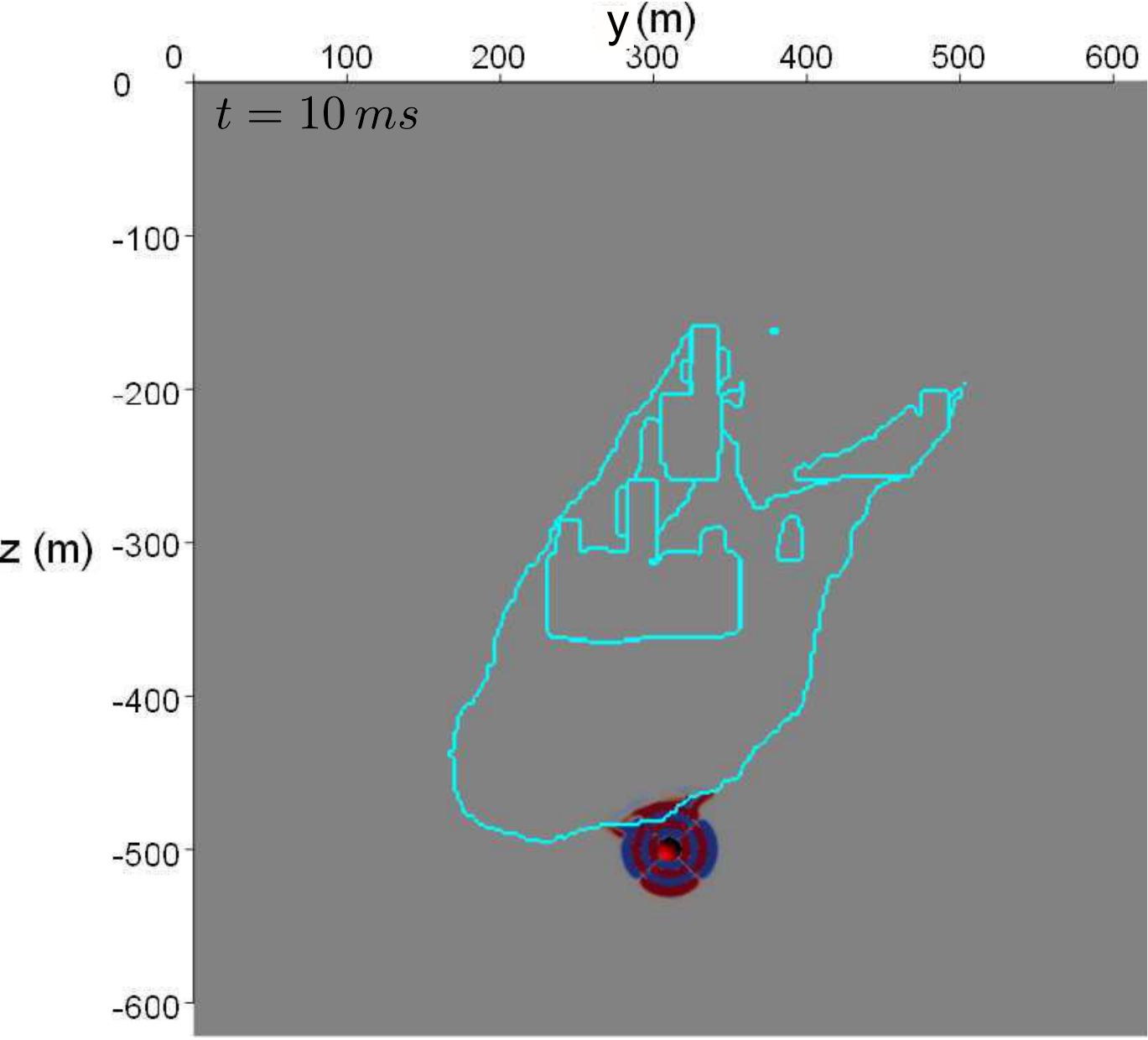}}
\subfloat{\includegraphics[scale=0.32]{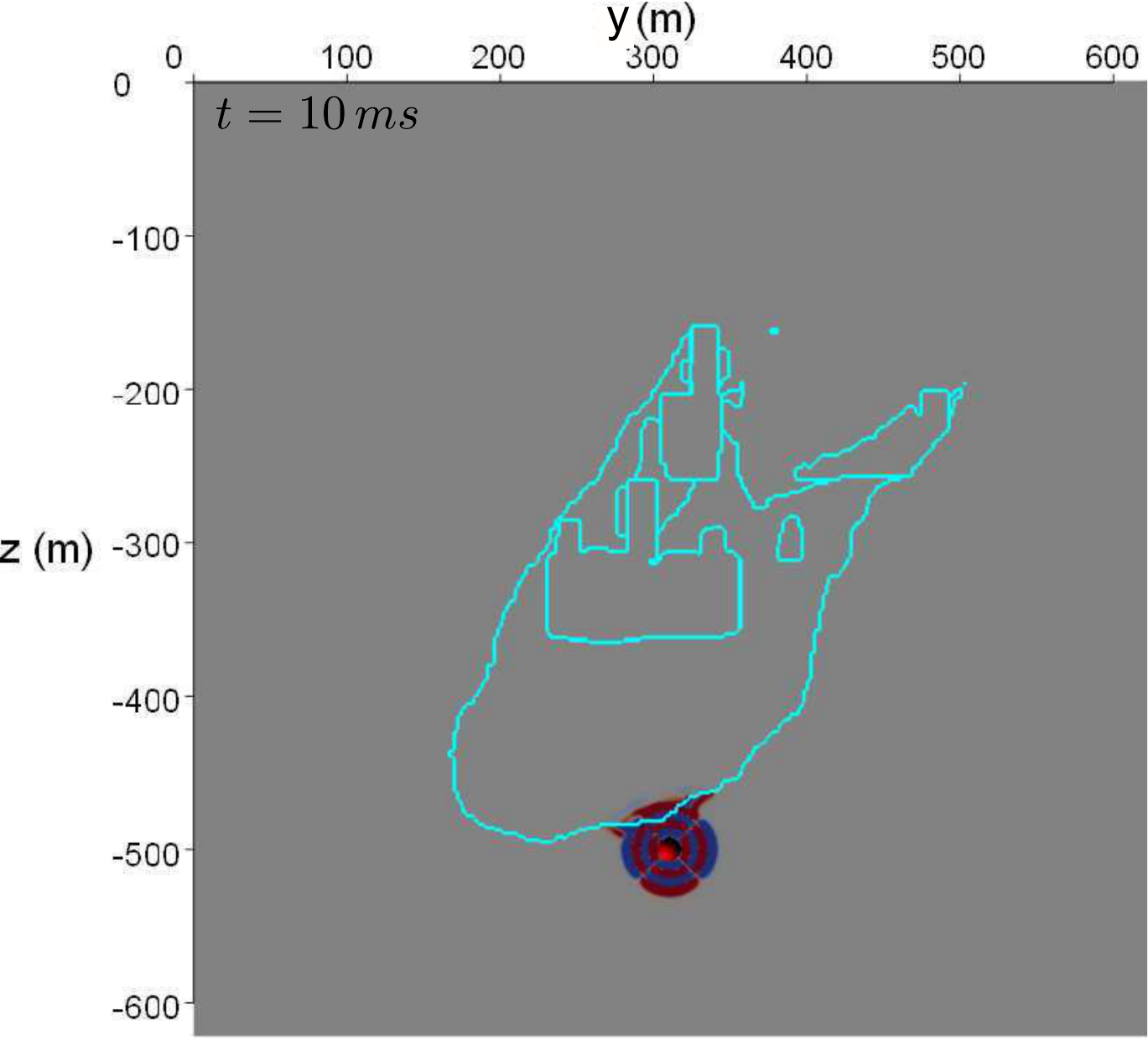}}\\
\subfloat{\includegraphics[scale=0.32]{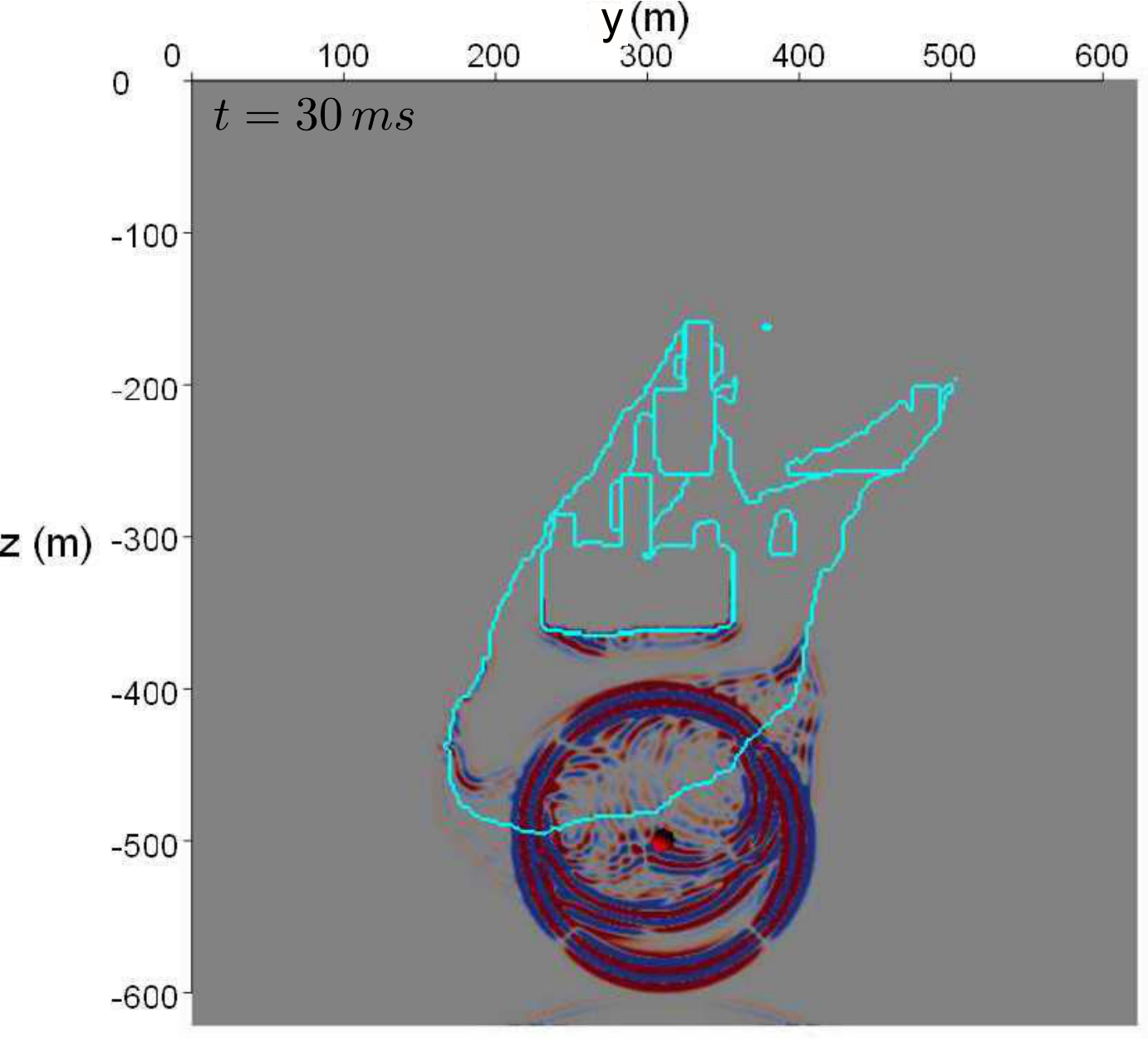}}
\subfloat{\includegraphics[scale=0.32]{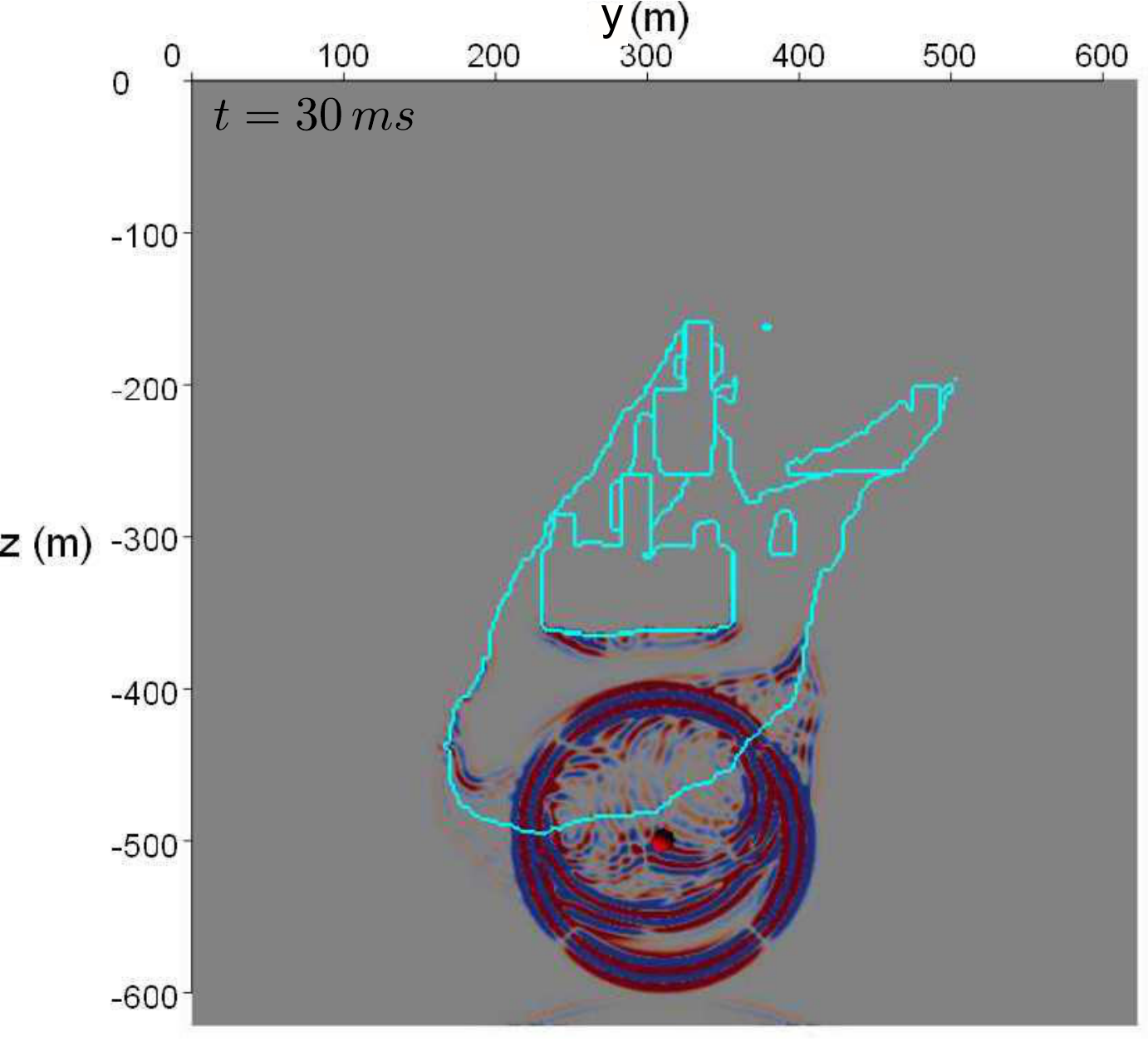}}\\
\subfloat{\includegraphics[scale=0.32]{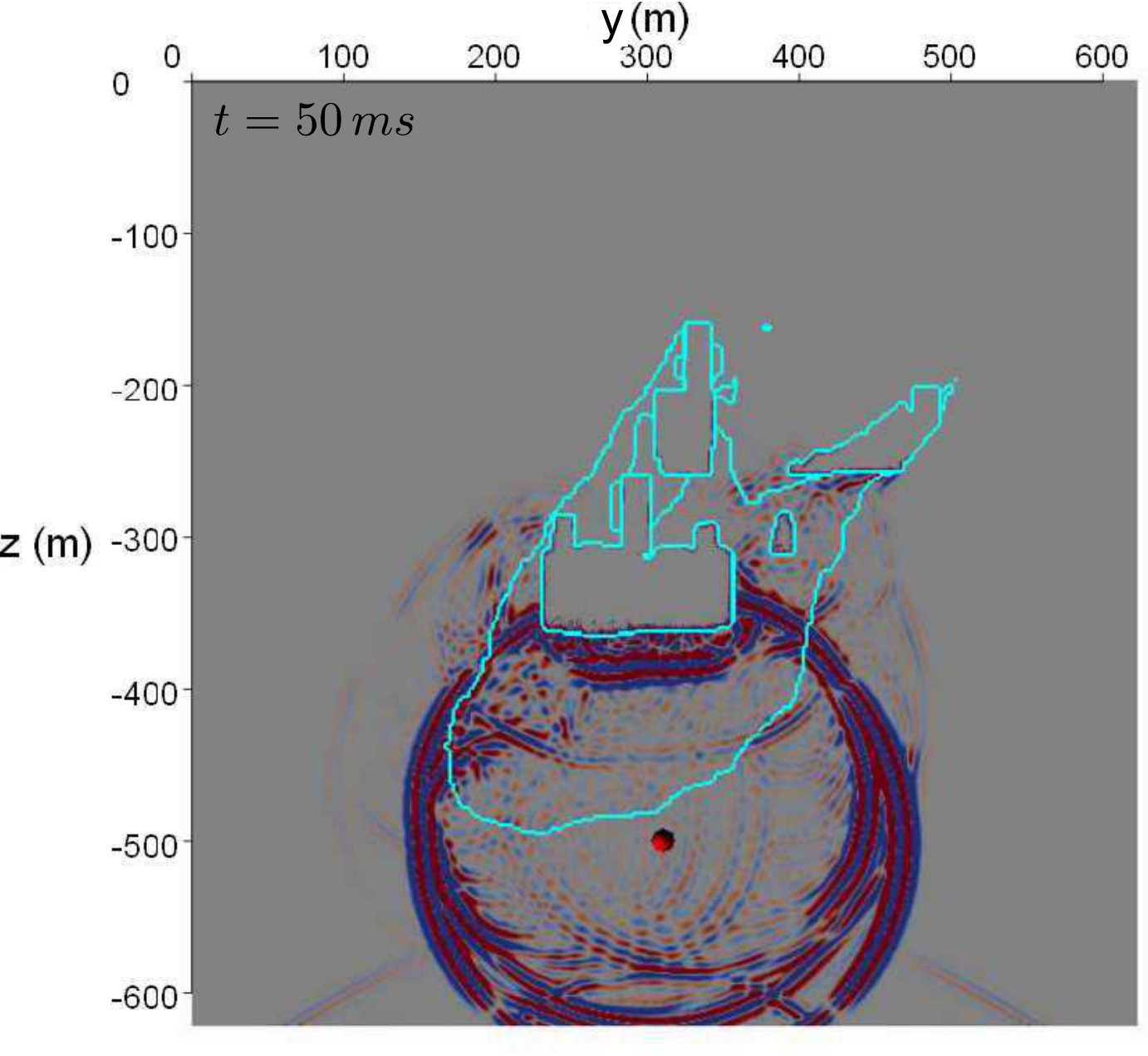}}
\subfloat{\includegraphics[scale=0.32]{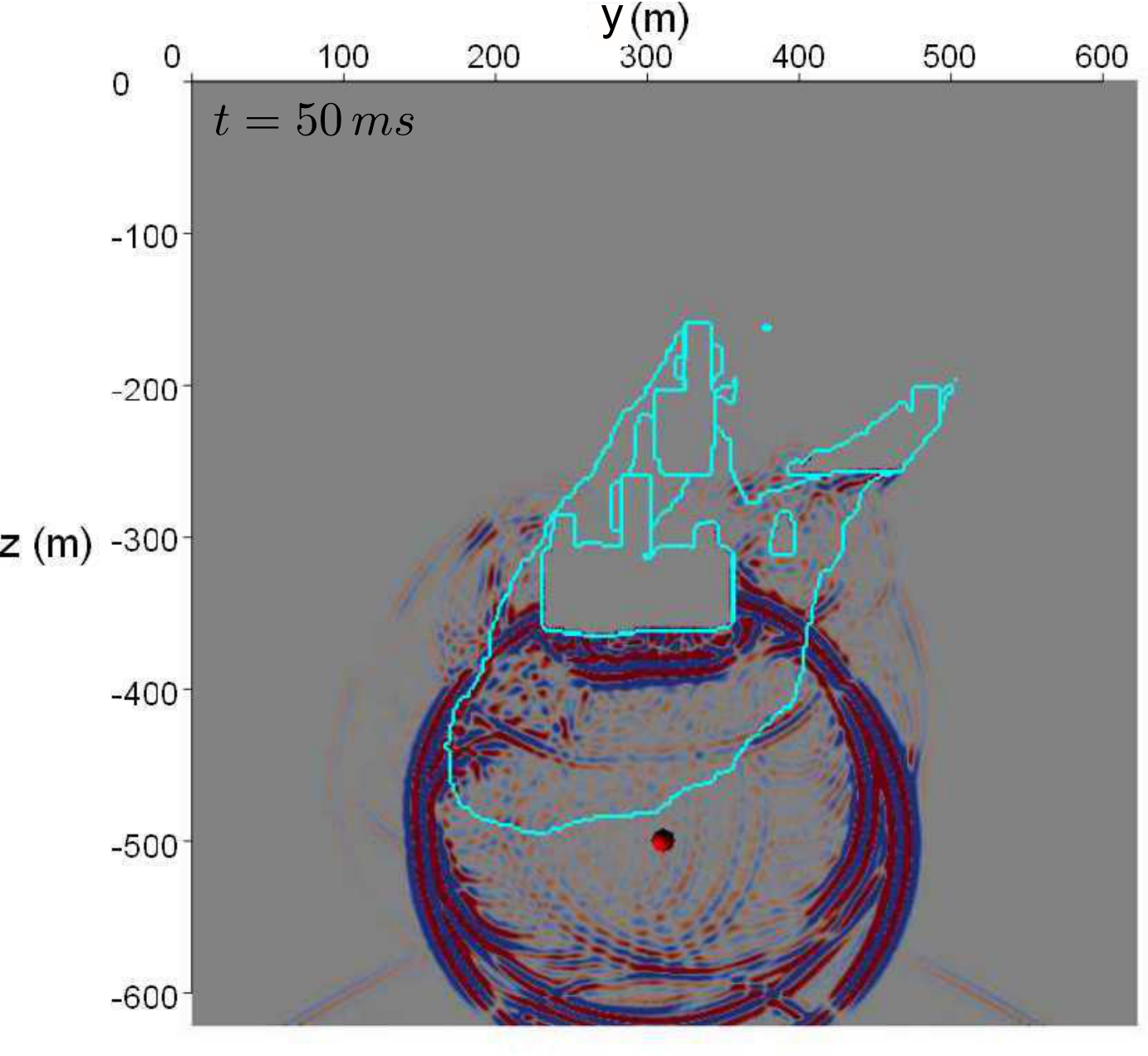}}\\
\subfloat{\includegraphics[scale=0.32]{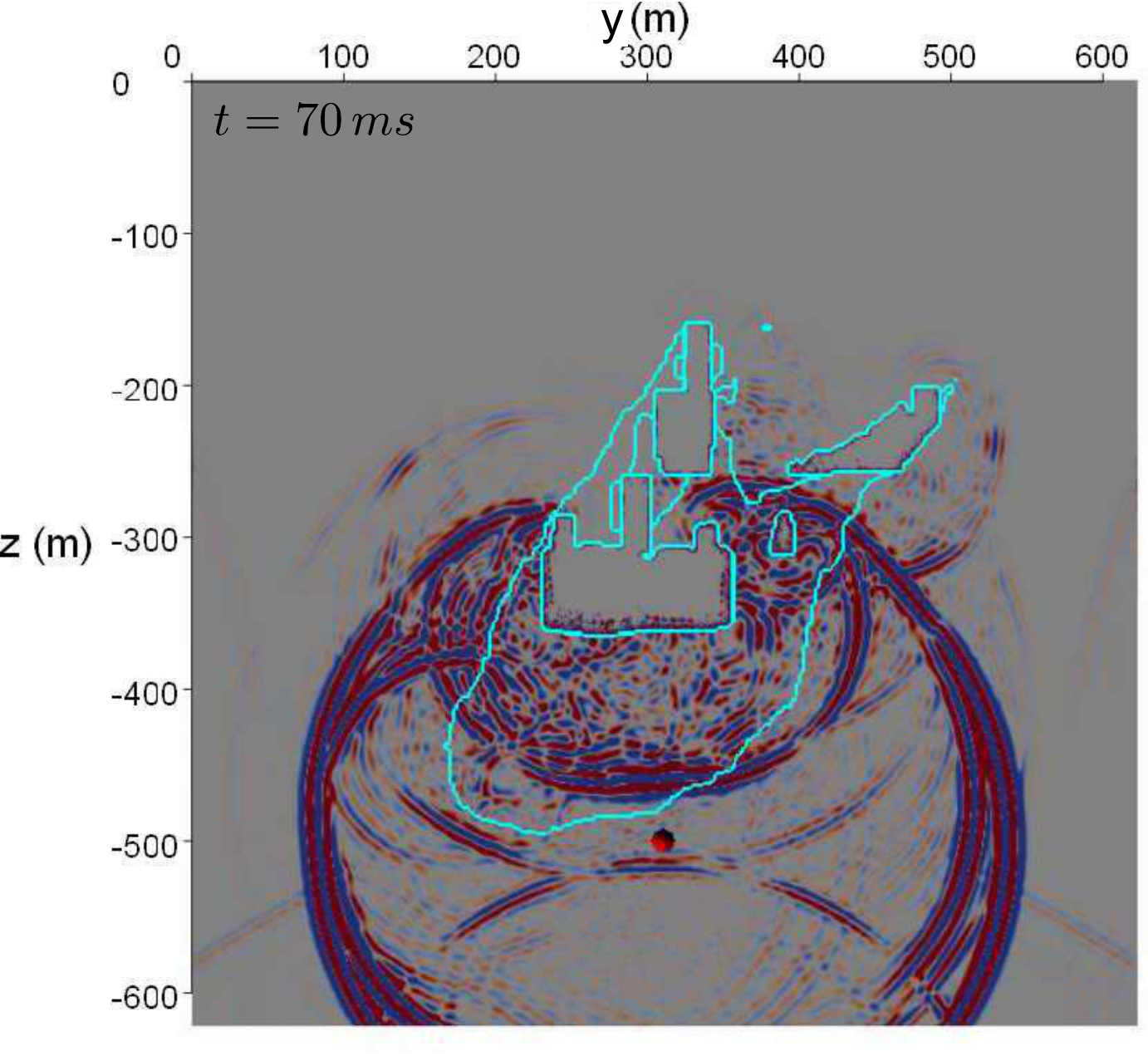}}
\subfloat{\includegraphics[scale=0.32]{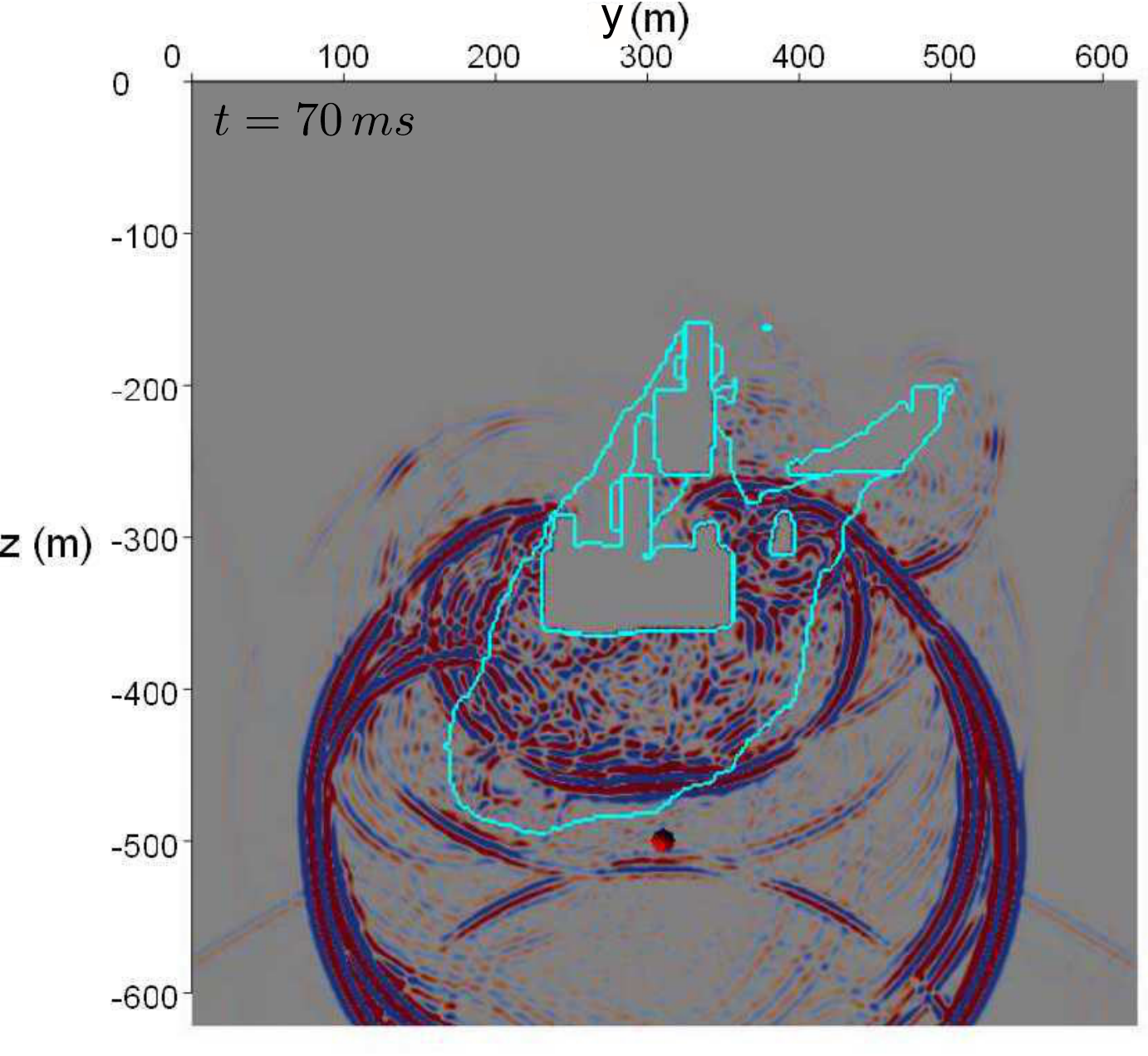}}
\caption{S-wave potential for the model with air included (left column) and with air excluded (right column). The lines in cyan represent the outlines of the ore body and the stopes. High values of the potential are shown in red and low values are shown in blue.}
\label{fig:py2d_swave_air_noair}
\end{figure}

\begin{figure}[htbp]
\centering
\subfloat[]{\includegraphics[scale=0.35]{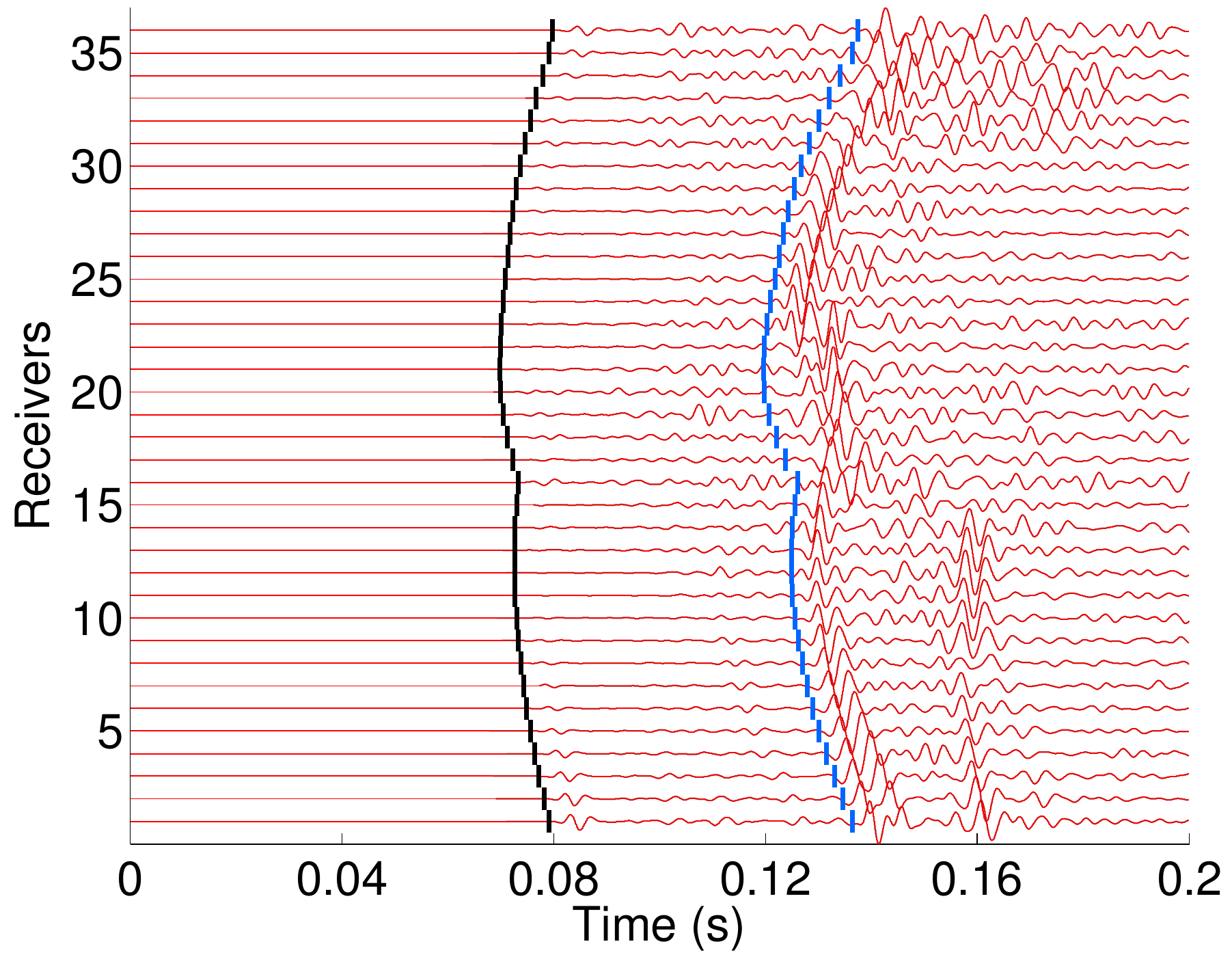}}
\subfloat[]{\includegraphics[scale=0.35]{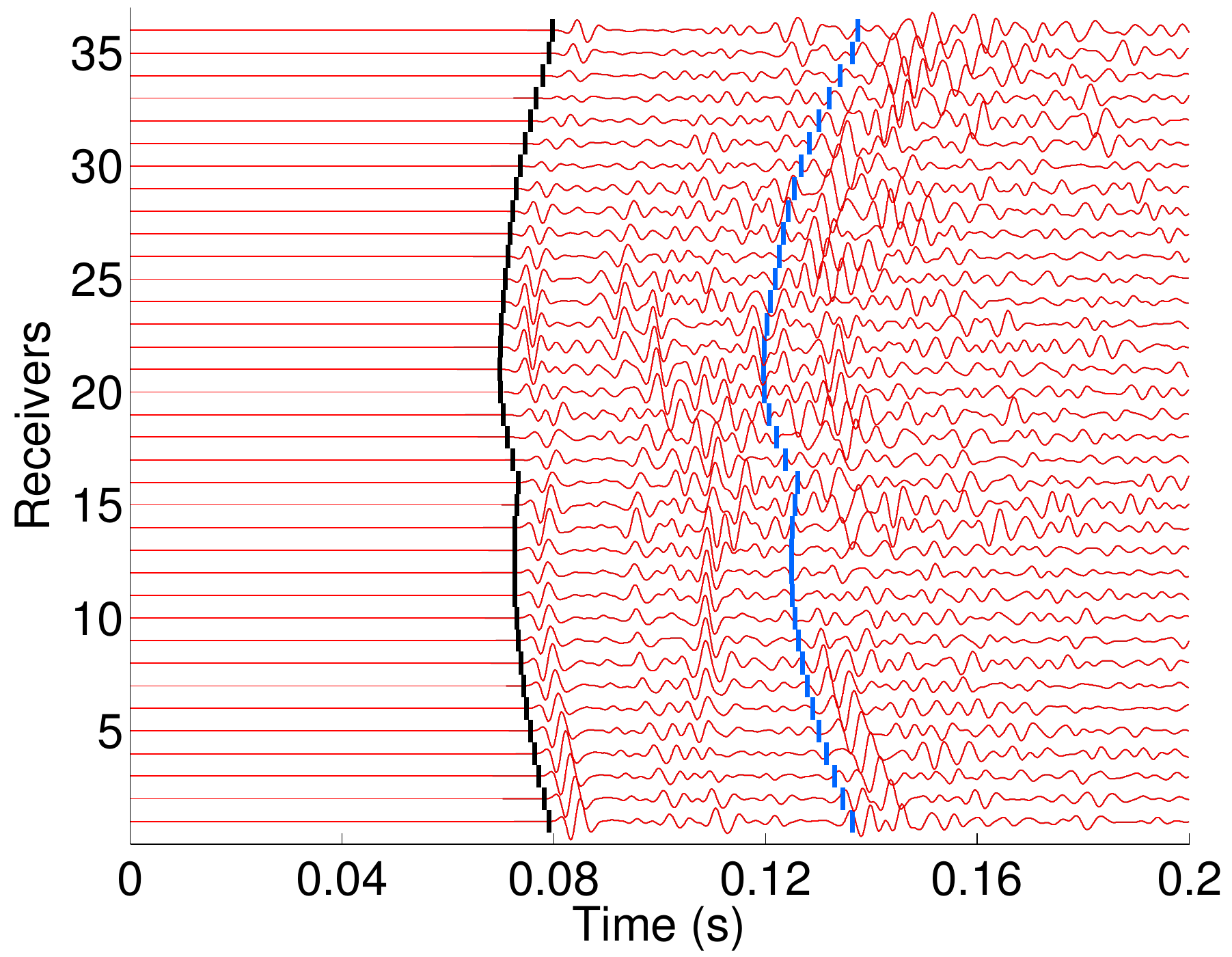}}
\caption{Synthetic waveforms computed for the 2D model of the Pyh\"asalmi mine without air (black) and with air (red). (a) Horizontal components. (b) Vertical components. Superimposed are the P- (black) and S-wave (blue) first-arrival times computed with a finite-difference eikonal solver. Seismograms are normalized to trace maximum.}
\label{fig:py2d_air_noair}
\end{figure}

Both simulations give stable results. For visualization, we calculate P-wave and S-wave potentials. The P-wave potential is computed as the divergence of the displacement field, and the S-wave potential is derived from the sum of the components of the curl of the displacement field~\cite{larsen1995}. Figure~\ref{fig:py2d_pwave_air_noair} and \ref{fig:py2d_swave_air_noair} show snapshots of P-wave and S-wave potentials, respectively. Wavefields are very similar in both cases of air-included and air-excluded. In case of air-included, only P waves travel at a very slow speed within the voids. We observe the diffracted waves around the stopes. The converted waves, i.e., P to S and S to P are also visible. Due to multiple reflections and conversions at the strong-contrast interfaces, the wavefield is very complicated. It is severely distorted by the voids but tends to heal further away from the voids. For both cases (i.e., with and without air), the computed waveforms are very similar (Figure~\ref{fig:py2d_air_noair}), and discrepancies are negligible. The P and S first arrival times calculated by an eikonal solver \cite{podvin1991} are in good agreement with the respective arrivals in computed waveforms. The strong signals observed between first P and S arrivals as well as after the S arrivals are the secondary waves generated by multiple reflections and conversions at the interfaces with strong-velocity contrast. Because of source radiation pattern, these secondary waves are stronger on vertical components than on horizontal components. 

These results show that the spectral-element method is stable even in the presence of high-velocity contrasts. For this particular mine model, we may safely exclude the air during meshing to compute the full waveforms unless we are interested in the acoustic waves in the stopes. In some cases, excluding the air drastically reduces the computational cost and avoids possible numerical instability due to the high-velocity contrast.

\subsection{3D model}

Now, we simulate wave propagation in a 3D model of the mine. We use an observed event for this simulation. The source location was estimated using a microseismic monitoring software (MIMO see~\cite{oye2003}), and the full moment-tensor inversion was performed using first motion polarities and P-wave amplitudes~\cite{manthei2005}. The source is located at $x=368$~m, $y=371$~m, and $z=-392$~m. It is characterised by a Ricker wavelet with a central frequency of 200~Hz and moment-tensor components $M_{xx}=-1.0158\times 10^8$~Nm, $M_{yy}=0.0858\times 10^8$~Nm, $M_{zz}=0.3540\times 10^8$~Nm, $M_{xy}=0.4394\times 10^8$~Nm, $M_{yz}=0.1025\times 10^8$~Nm, and $M_{zx}=0.0731\times 10^8$~Nm. Although the uncertainty in the moment tensor was large due to the complicated waveforms and very heterogeneous velocity model, we use this moment tensor for our simulation. The estimated moment tensor consists of both isotropic and deviatoric components implying a complex source mechanism. The sampling interval of the seismogram recordings is set to 1~$\mu$s.  

\begin{figure}[htbp]
\centering
\includegraphics[scale=0.5]{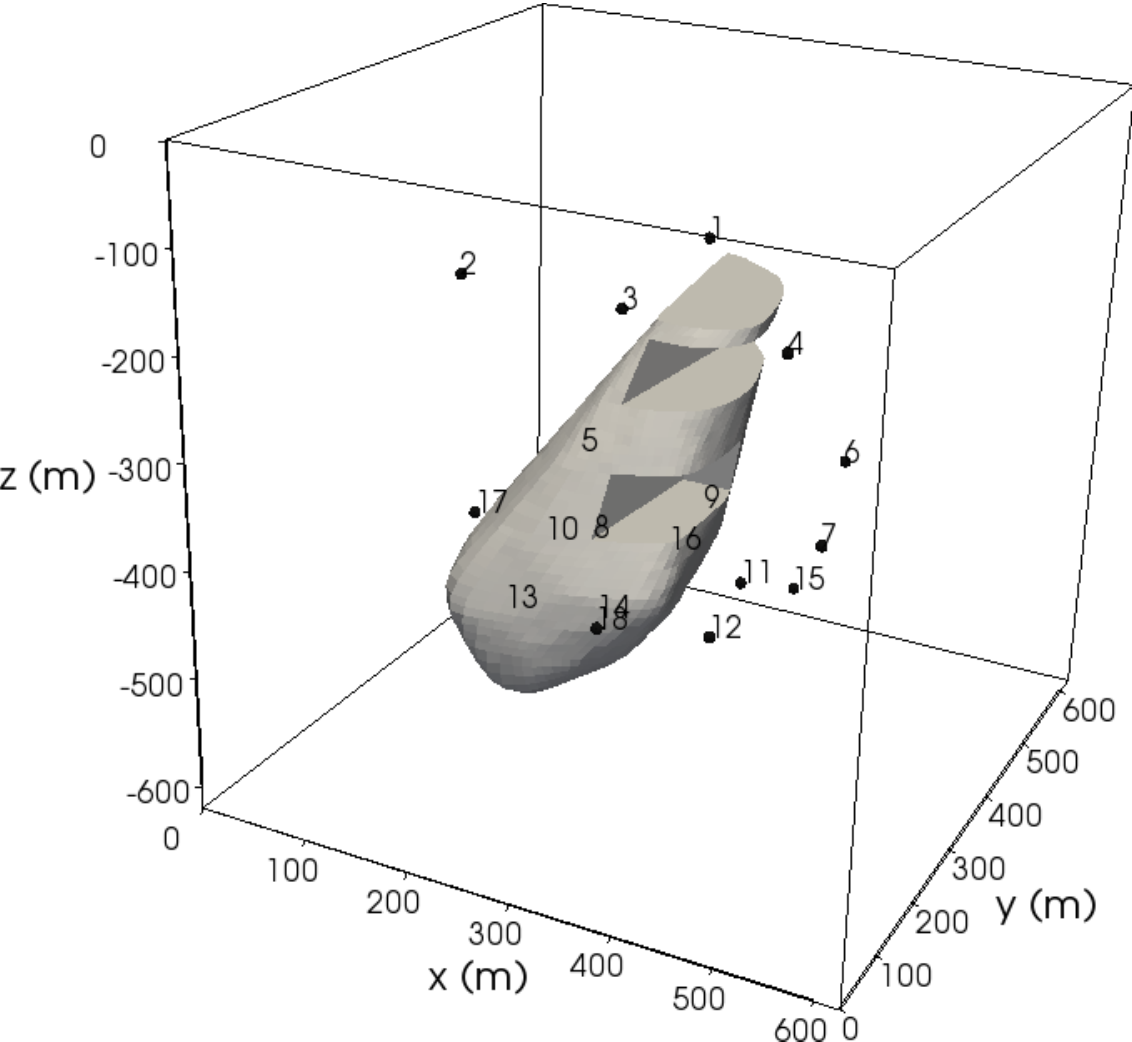}
\caption{3D model of the Pyh\"asalmi mine including ore body (solid), two major stopes, and 18 geophones (numbered).}
\label{fig:py_model}
\end{figure}

\begin{figure}[htbp]
\centering
\subfloat[]{\includegraphics[scale=0.4]{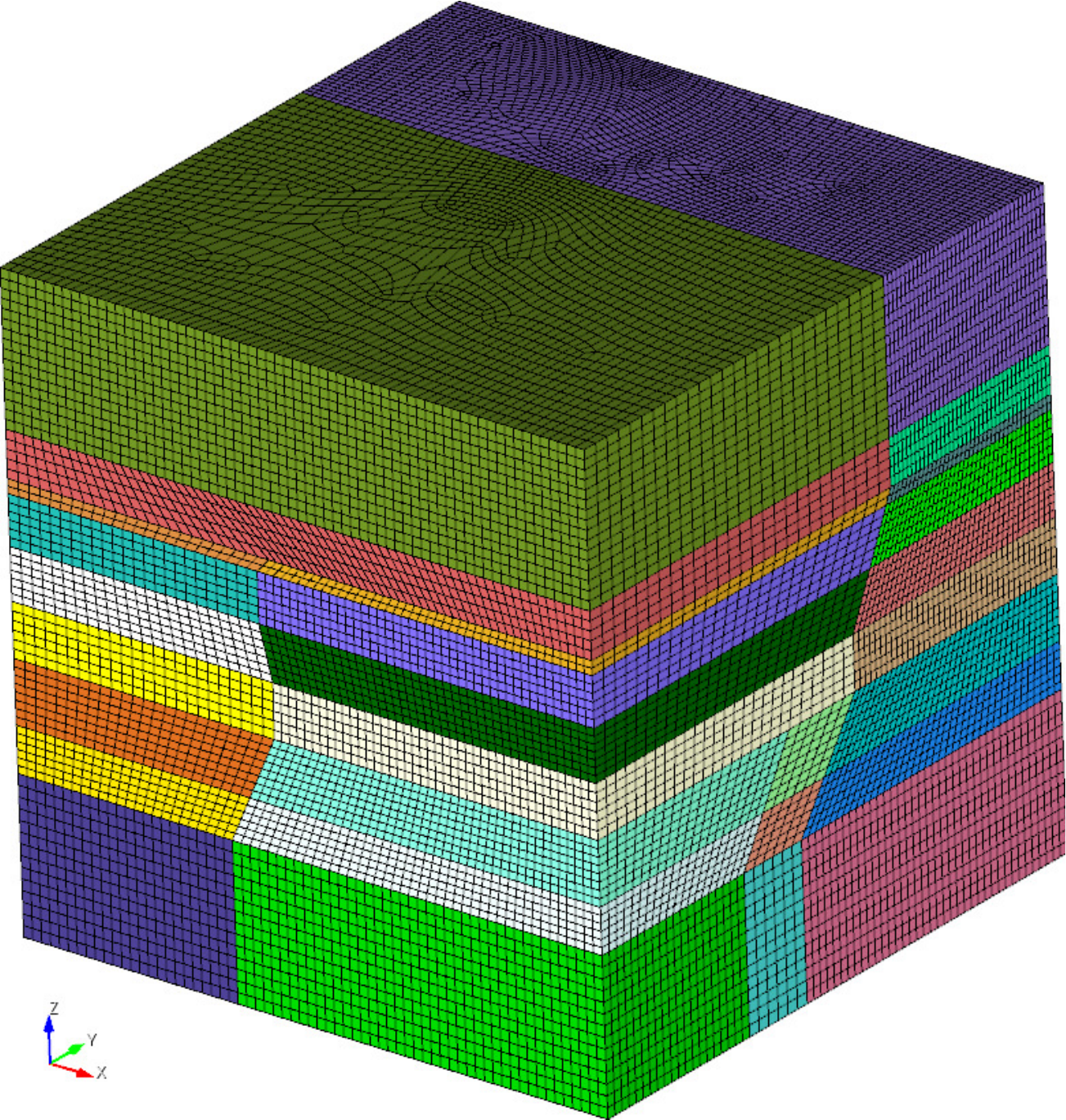}}\qquad
\subfloat[]{\includegraphics[scale=0.4]{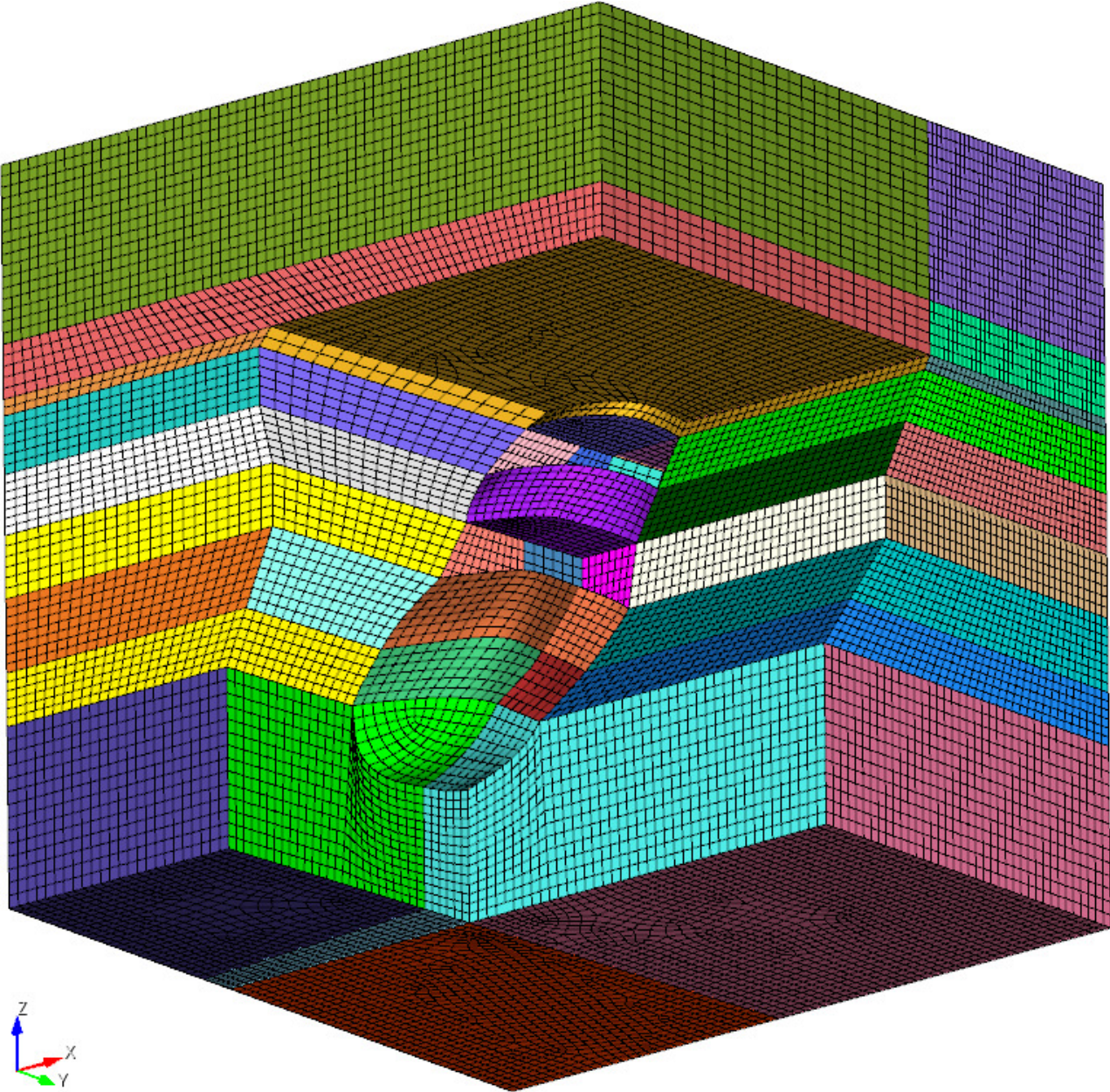}}
\caption{a) Spectral-element mesh for the 3D model of the Pyh\"asalmi mine. b) Interior section of the mesh visualizing the ore body and stopes. Colors represent different volumes created for meshing.}
\label{fig:py_mesh}
\end{figure}

\begin{figure}[htbp]
\centering
\includegraphics[scale=0.5]{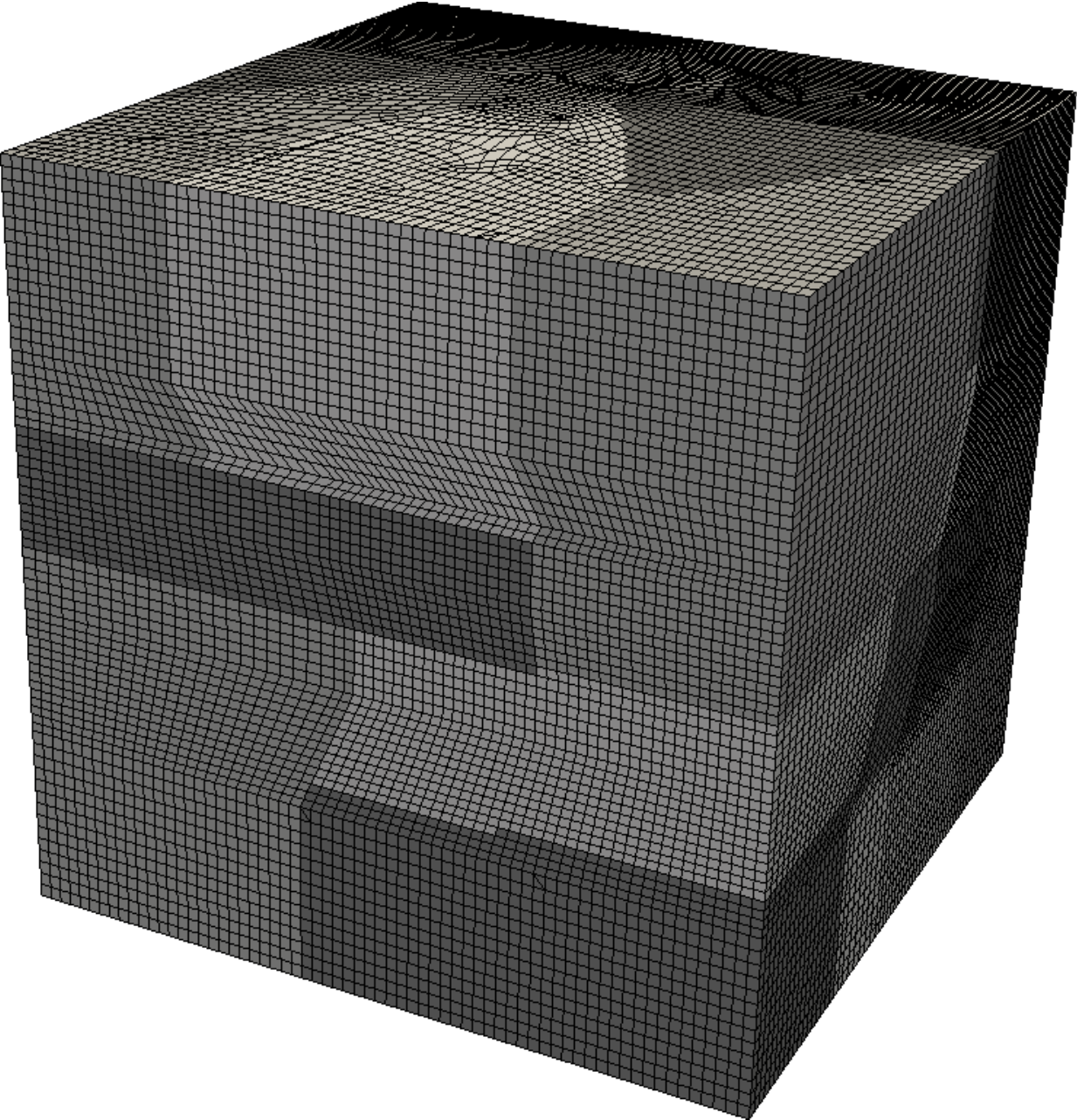}
\caption{Spectral-element mesh for the 3D model of the Pyhaesalmi mine partitioned into 24 domains for parallel processing.}
\label{fig:py_mesh_part}
\end{figure}

Although there are several tools for 2D quadrilateral meshing, only a few tools are available for 3D hexahedral meshing, and the functionalities of such tools are limited. Small cavities and inclusions make the hexahedral meshing very difficult. It is also unfeasible to honour these small-scale features during meshing due to the high computational cost caused by a large number of elements. Therefore, we simplify the original 3D model including only two major stopes (Figure~\ref{fig:py_model}). We exclude the stopes from the meshing. Automatic hexahedral meshing is currently not possible with the CUBIT for such a complex 3D model like the Pyh\"asalmi mine. The complex model has to be decomposed into several volumes which can be meshed with the functionalities available within the CUBIT. We decompose the 3D model into 78 volumes (Figure~\ref{fig:py_mesh}a). We use average element sizes of 9.5~m for rock and 10~m for ore body resulting in a total of 107,712 spectral elements and a total of 7,161,572 spectral nodes (Figures~\ref{fig:py_mesh}a-b). We partition the mesh into 24 domains for parallel processing (Figure~\ref{fig:py_mesh_part}). To balance the load among the processors, we use an open-source graph partitioning tool SCOTCH~\cite{chevalier2008} for the mesh partition.

\begin{figure}[htbp]
\centering
\subfloat{\includegraphics[scale=0.32]{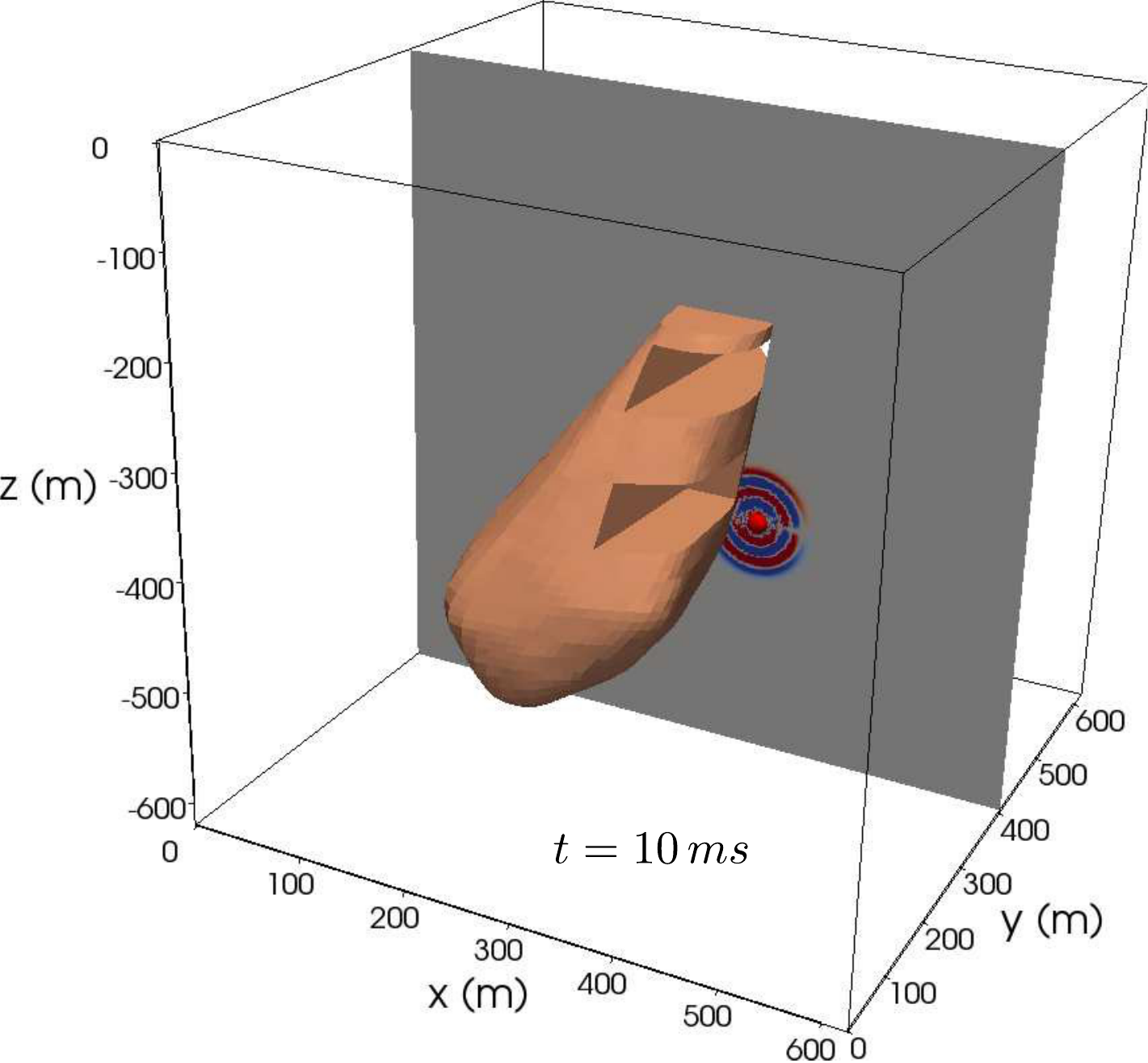}}
\subfloat{\includegraphics[scale=0.32]{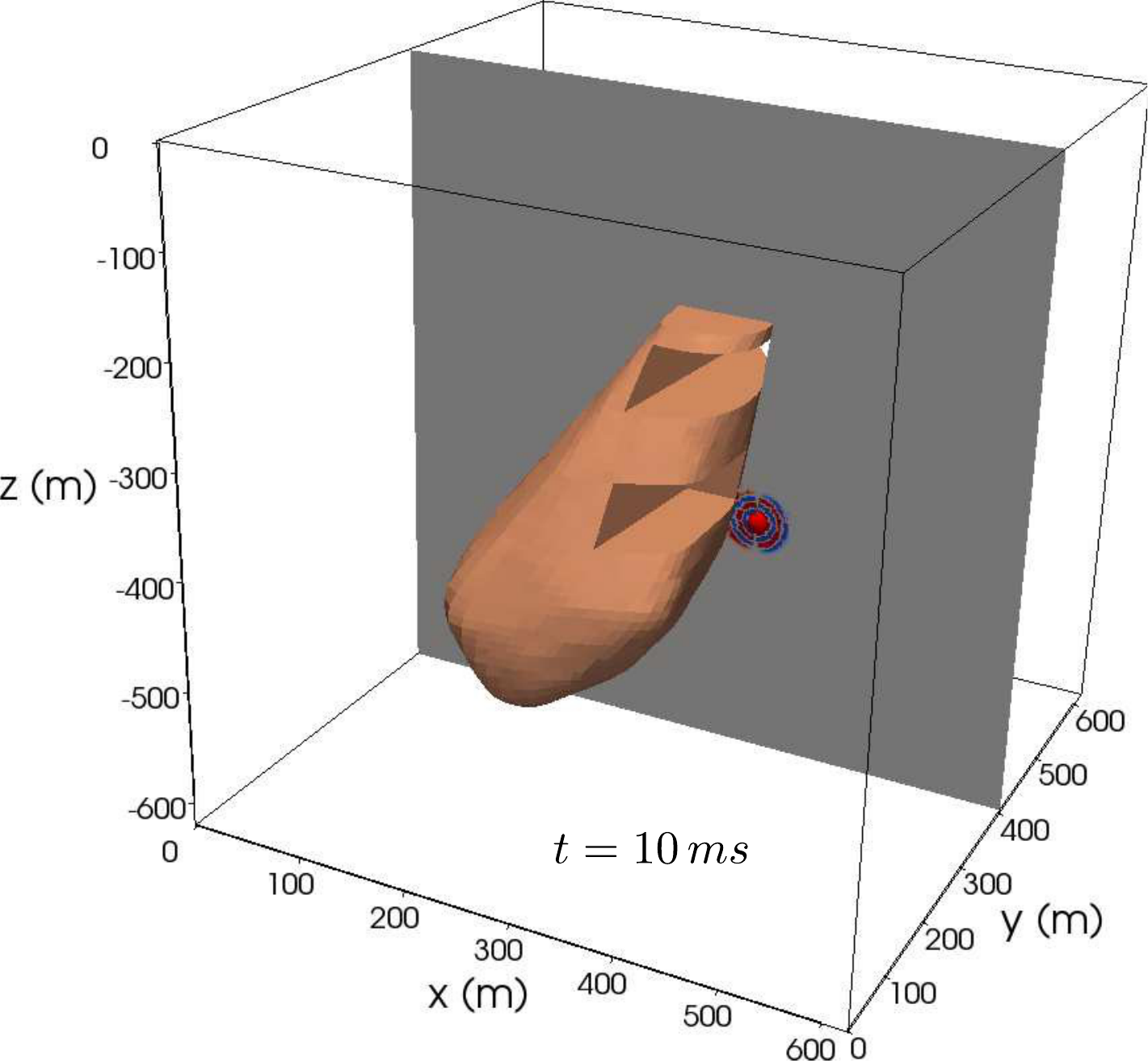}}\\
\subfloat{\includegraphics[scale=0.32]{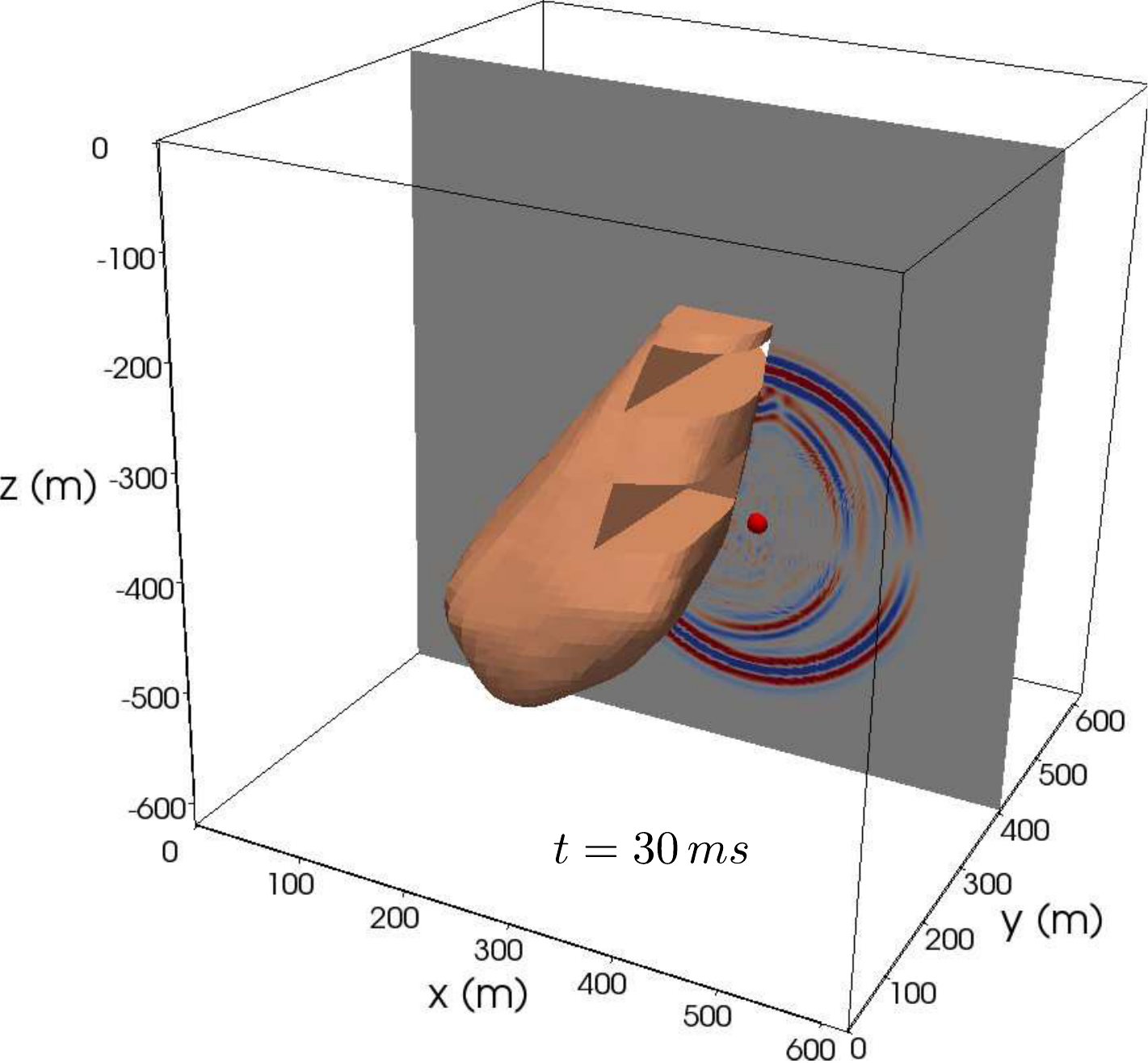}}
\subfloat{\includegraphics[scale=0.32]{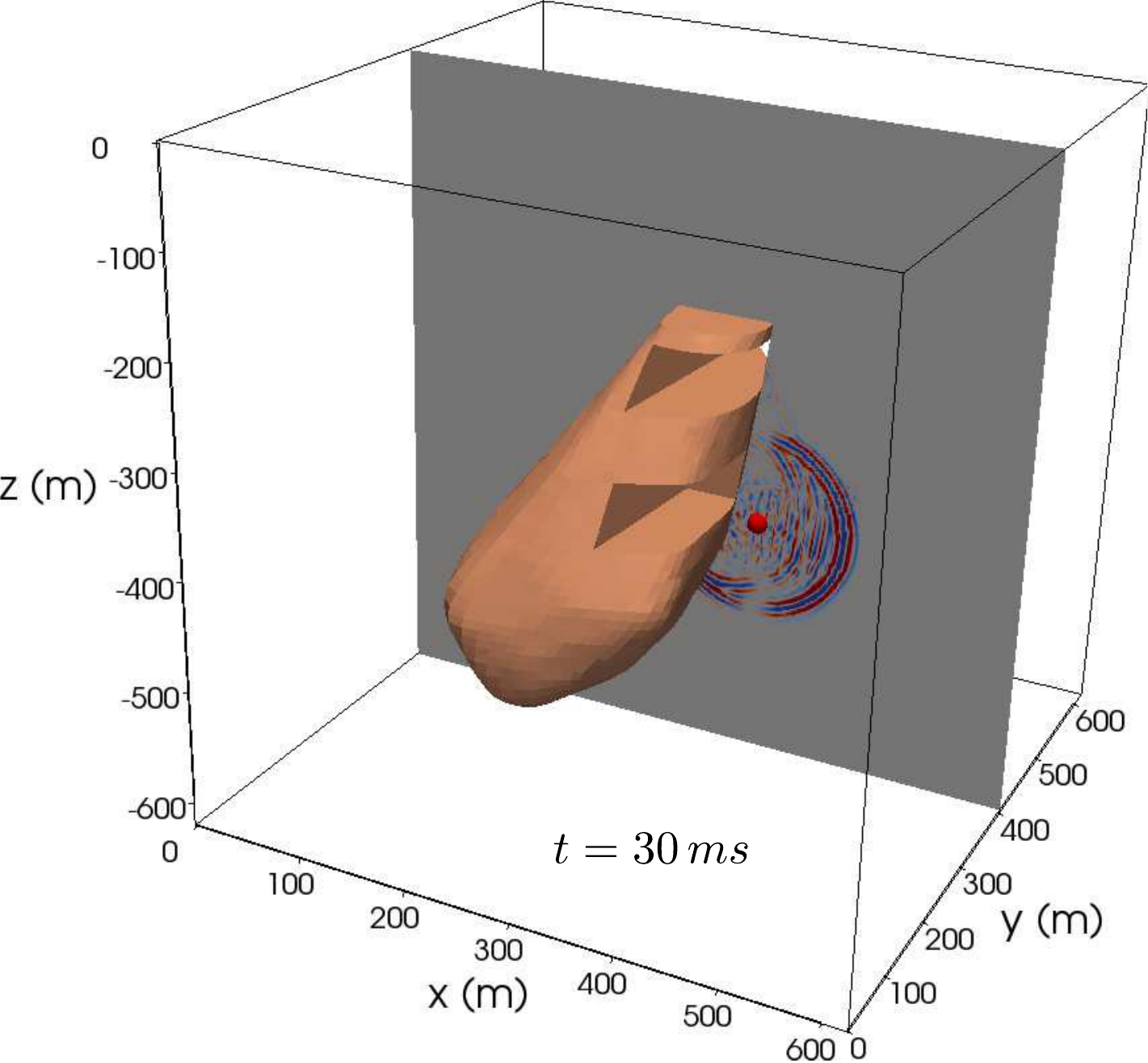}}\\
\subfloat{\includegraphics[scale=0.32]{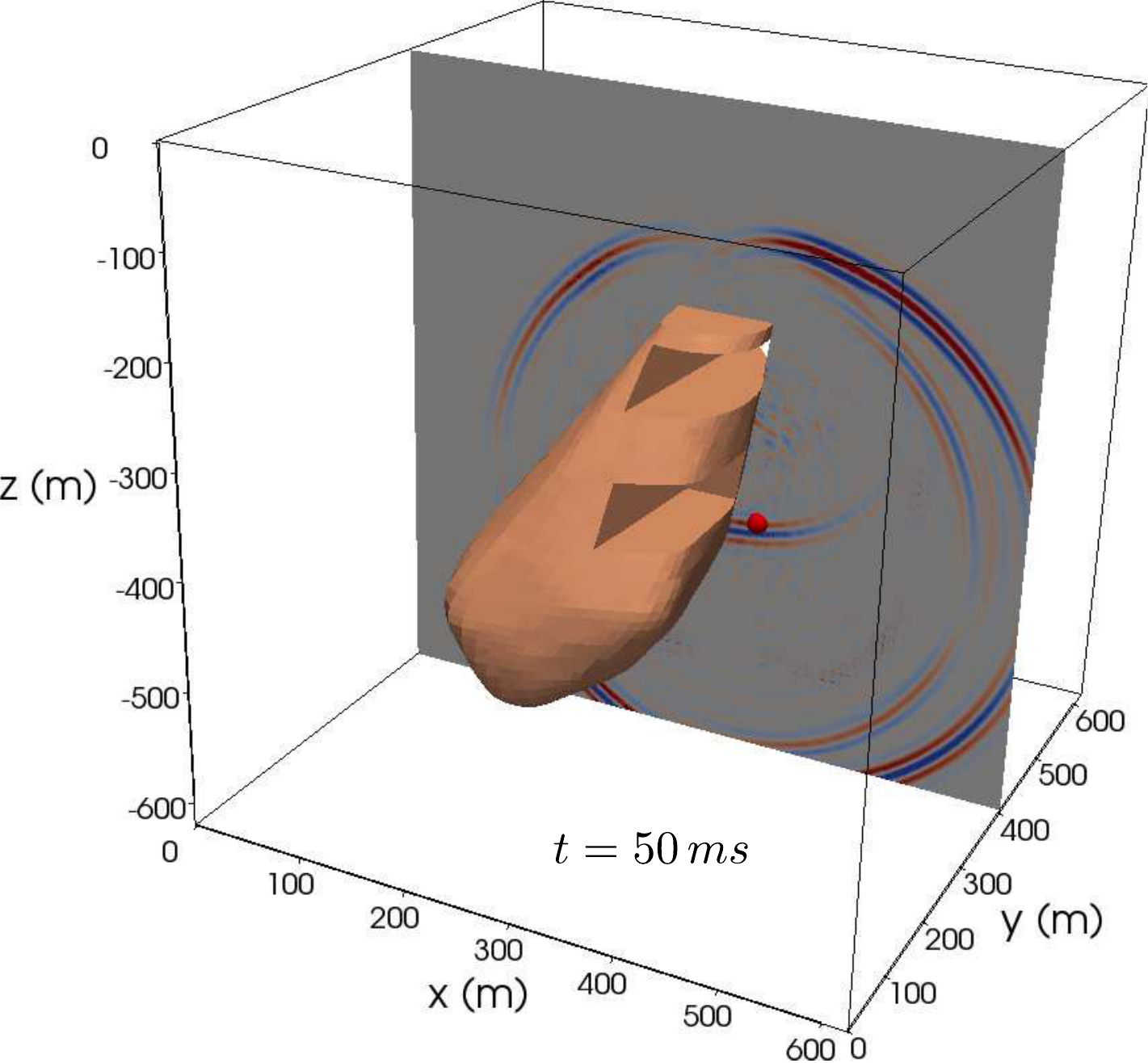}}
\subfloat{\includegraphics[scale=0.32]{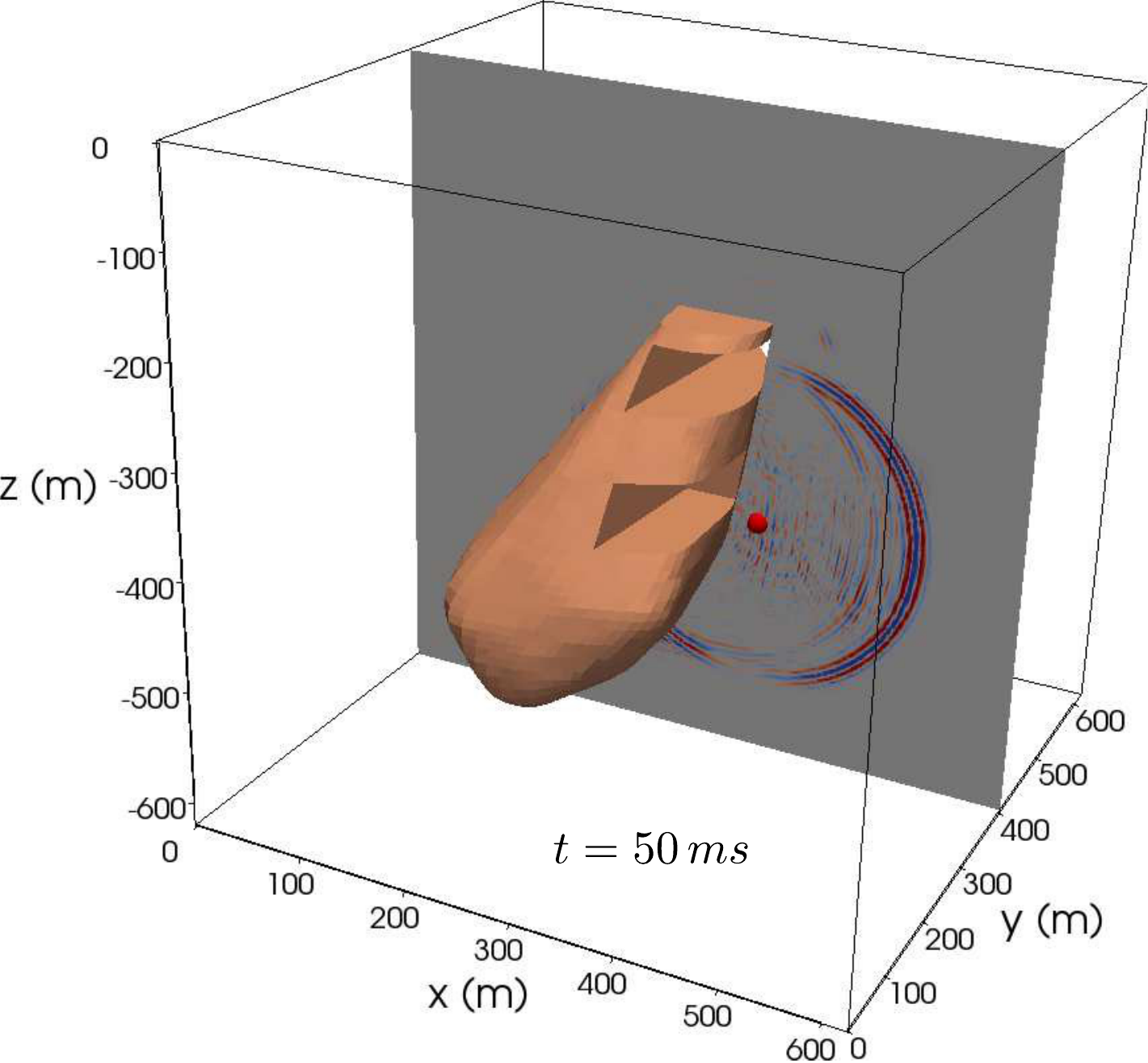}}\\
\subfloat{\includegraphics[scale=0.32]{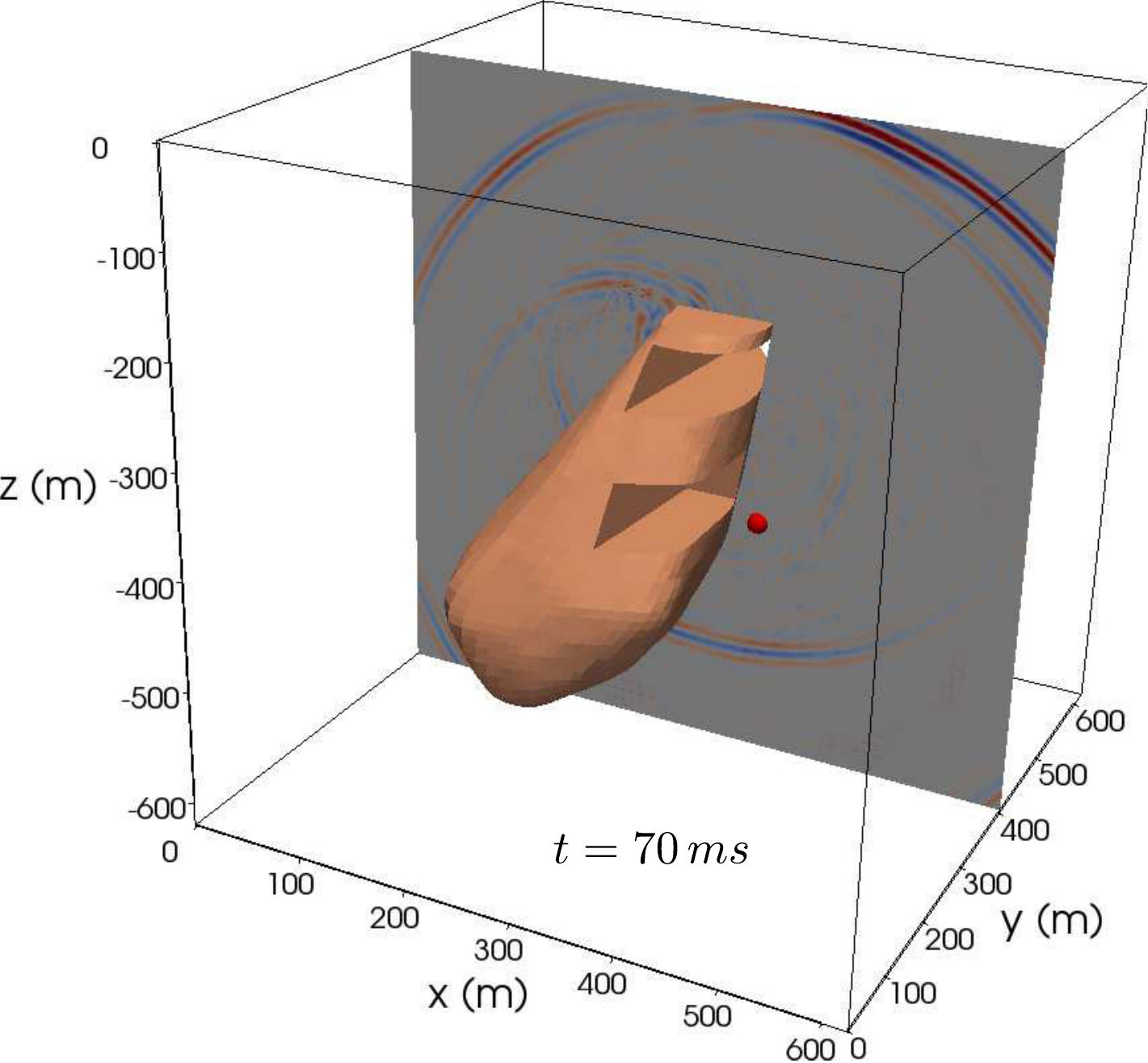}}
\subfloat{\includegraphics[scale=0.32]{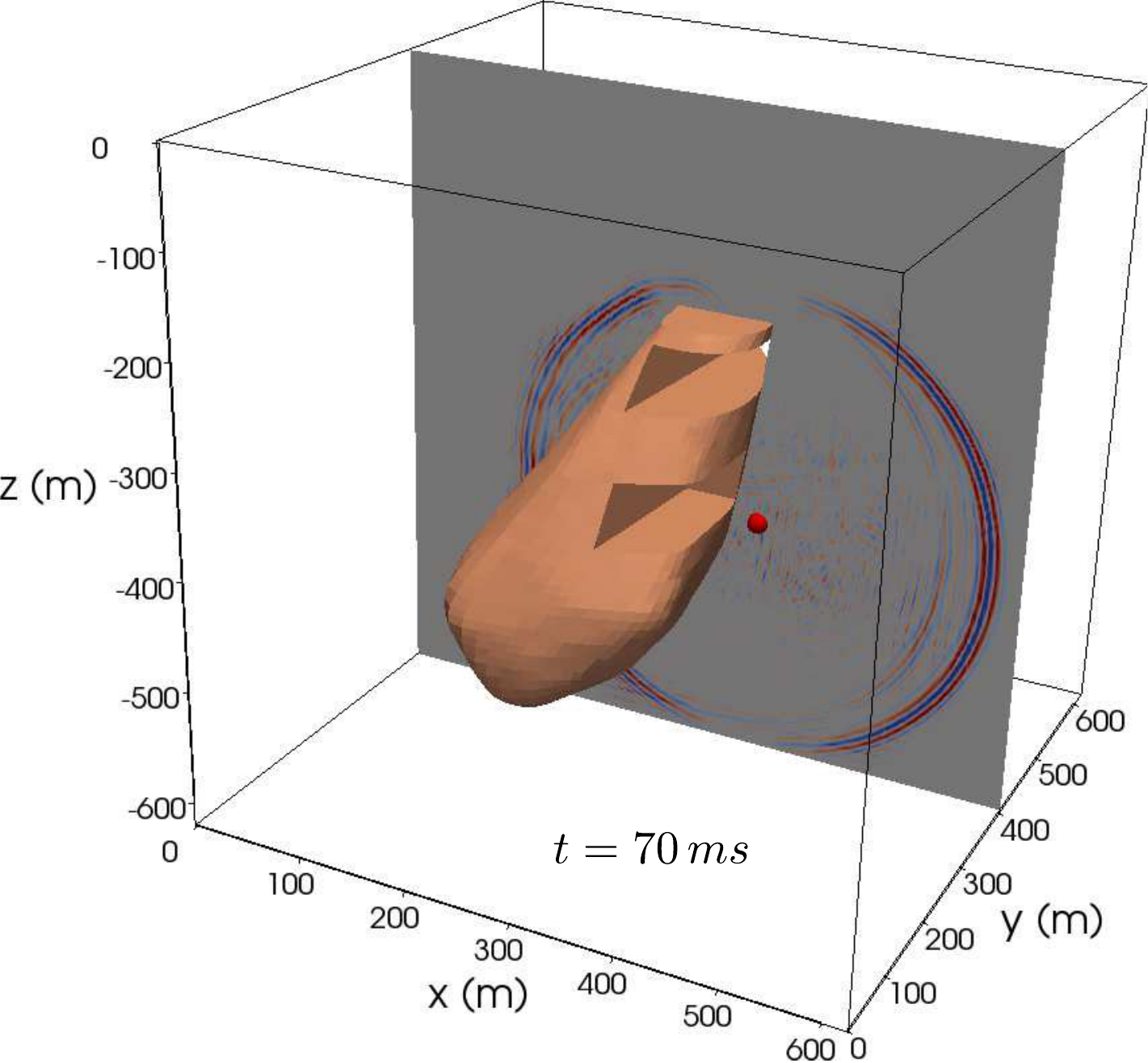}}
\caption{P-wave potential (left) and S-wave potential (right). High values are shown in red and low values are shown in blue.}
\label{fig:py3d_wave}
\end{figure}

\begin{figure}[htbp]
\centering
\subfloat{\includegraphics[scale=0.3]{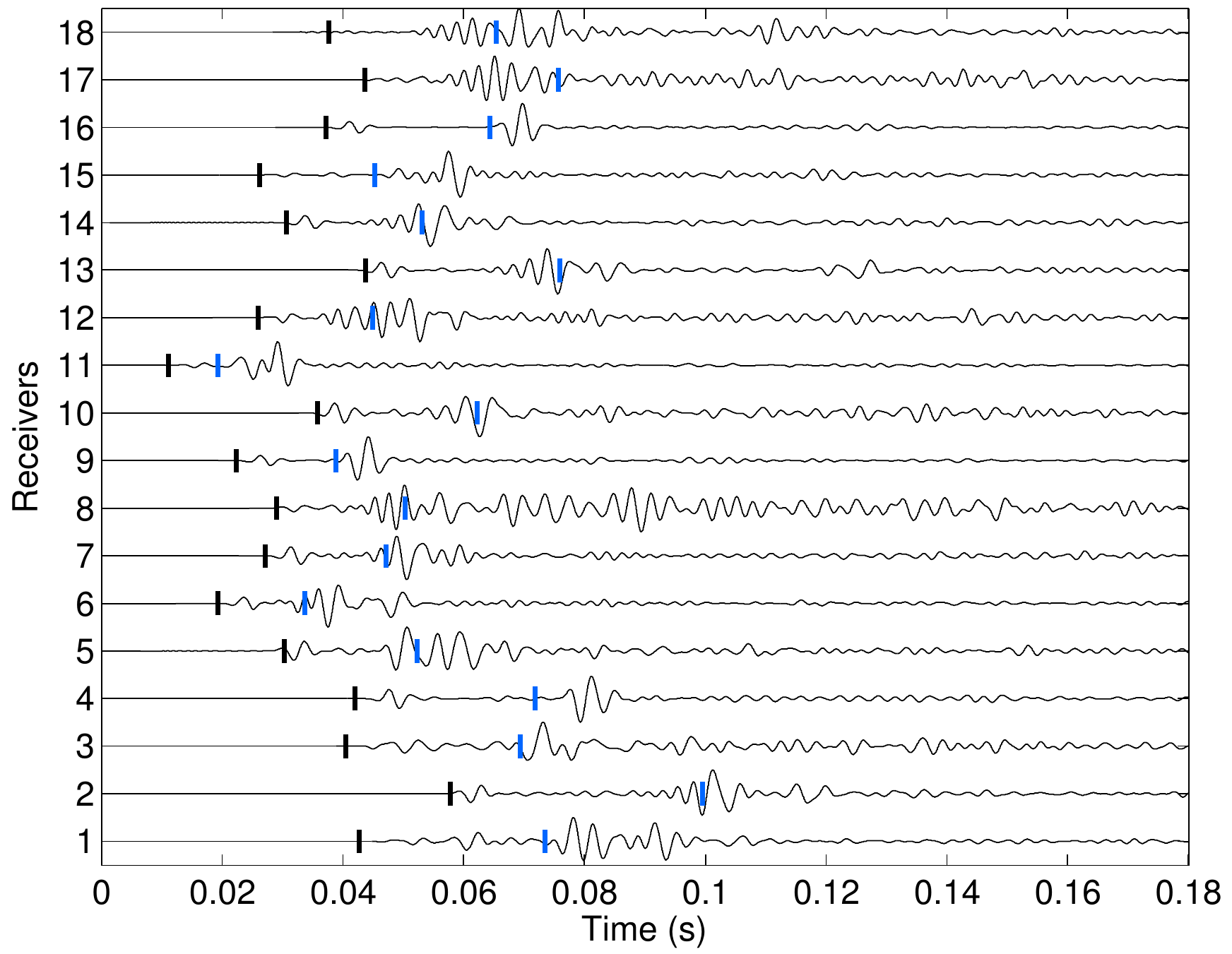}}
\subfloat{\includegraphics[scale=0.3]{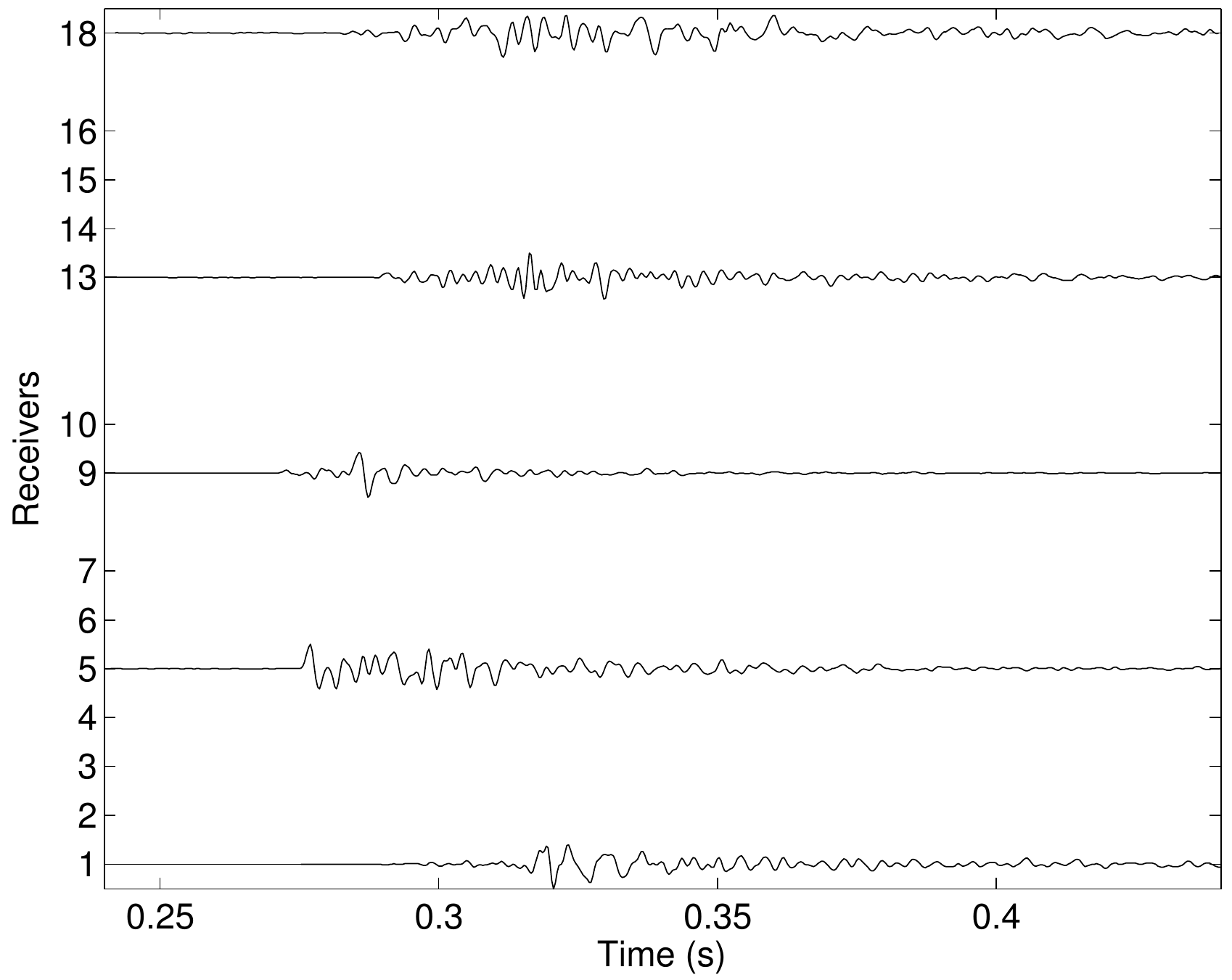}}\\
\subfloat{\includegraphics[scale=0.3]{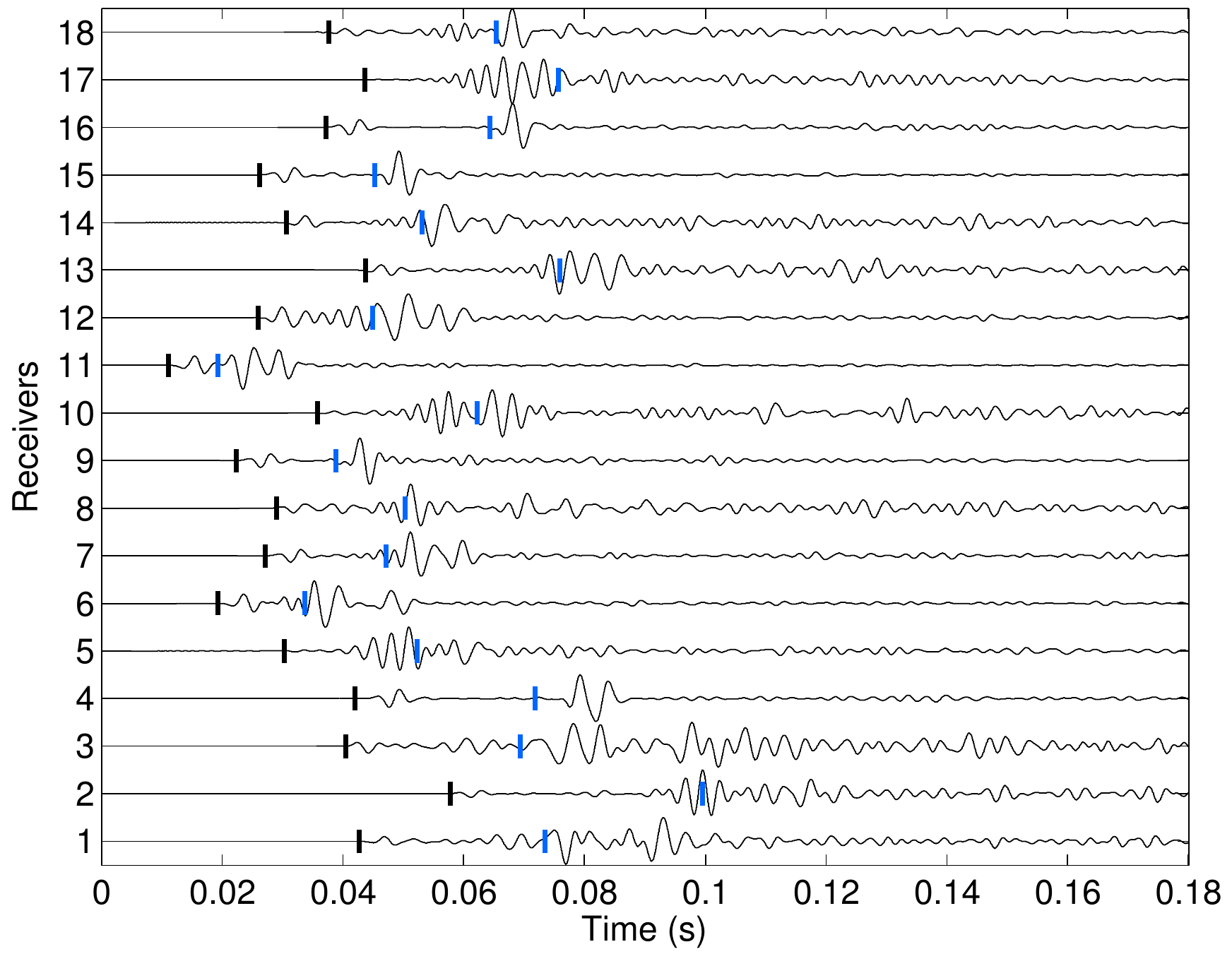}}
\subfloat{\includegraphics[scale=0.3]{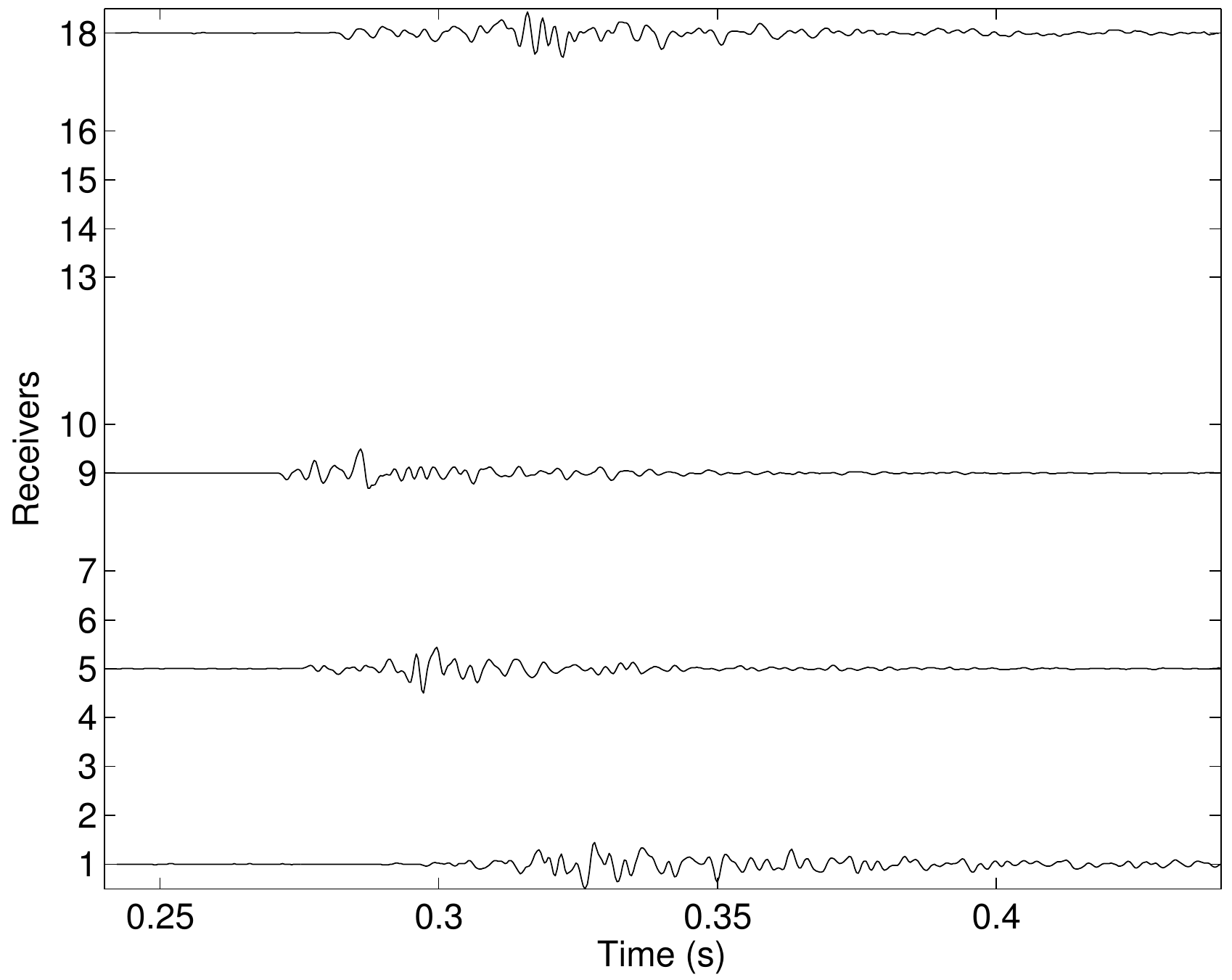}}\\
\subfloat{\includegraphics[scale=0.3]{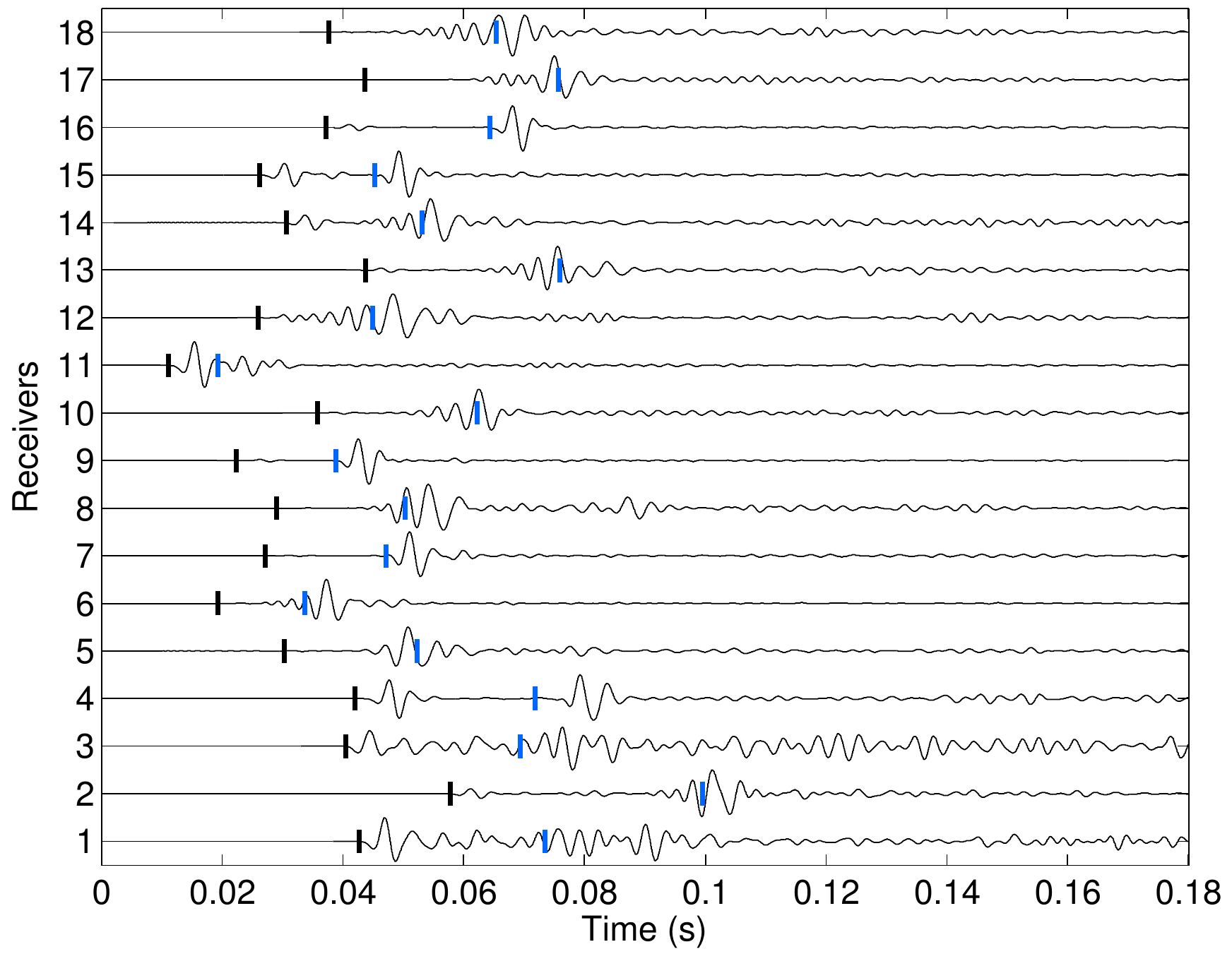}}
\subfloat{\includegraphics[scale=0.3]{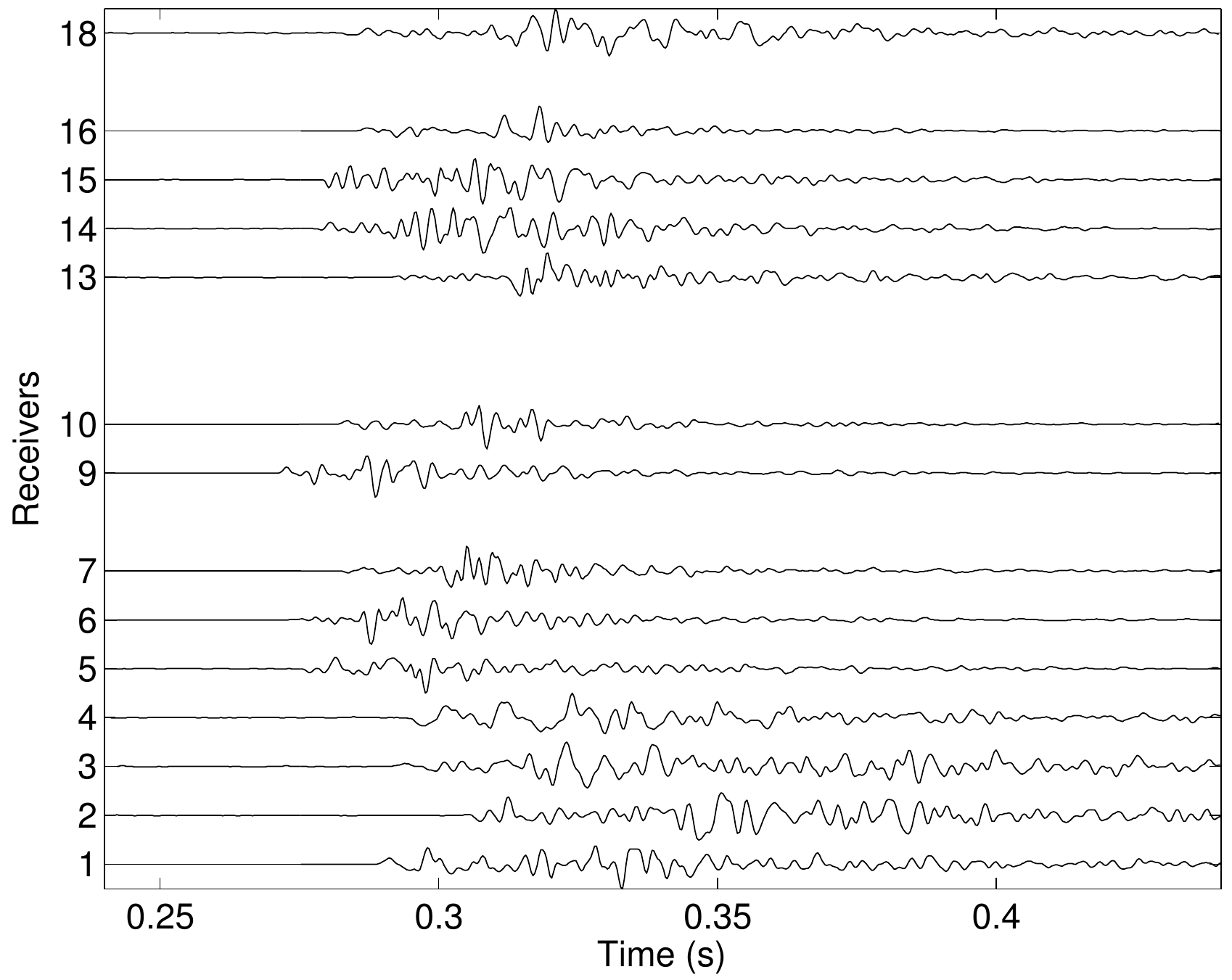}}
\caption{Synthetic (left column) and observed (right column) waveforms for the Pyh\"asalmi mine. Top to bottom are East, North, and vertical components, respectively. Superimposed are the P- (black) and S-wave (blue) first-arrival times computed with a finite-difference eikonal solver. Seismograms are normalized to trace maximum.}
\label{fig:py3d_waveform}
\end{figure}

Figure~\ref{fig:py3d_wave} shows snapshots of P-wave and S-wave potentials. We again observe reflected and converted waves. The wavefield is distorted by the stopes but gradually heals away from them. Figure~\ref{fig:py3d_waveform} illustrates the synthetic and observed seismograms. During this particular event, geophones 8, 11, 12, 17, and the North channel of geophone 13 were not functioning. The first-arrival times and coda durations for computed and observed waveforms generally match. Due to the complex velocity model and an imprecise source mechanism, we do not expect an accurate match. From the synthetic data, e.g., geophones 5-10, we observe that it would be difficult to identify and pick the P and S phases if the geophones were only the single component. Due to the shadow zones created by the voids or the source radiation pattern, first arrivals have small amplitudes at some of the receivers. Therefore, 3-component data are necessary for reliable data processing. First-arrival times computed by a finite-difference eikonal solver using structured grid are in reasonable agreement with respective arrivals in the computed waveforms. However, there is always a discrepancy in velocity model when converting unstructured mesh into structured grids.

\section{Unstable rock slope at \r Aknes}

\begin{figure}[htbp]
\centering
\includegraphics[scale=0.3]{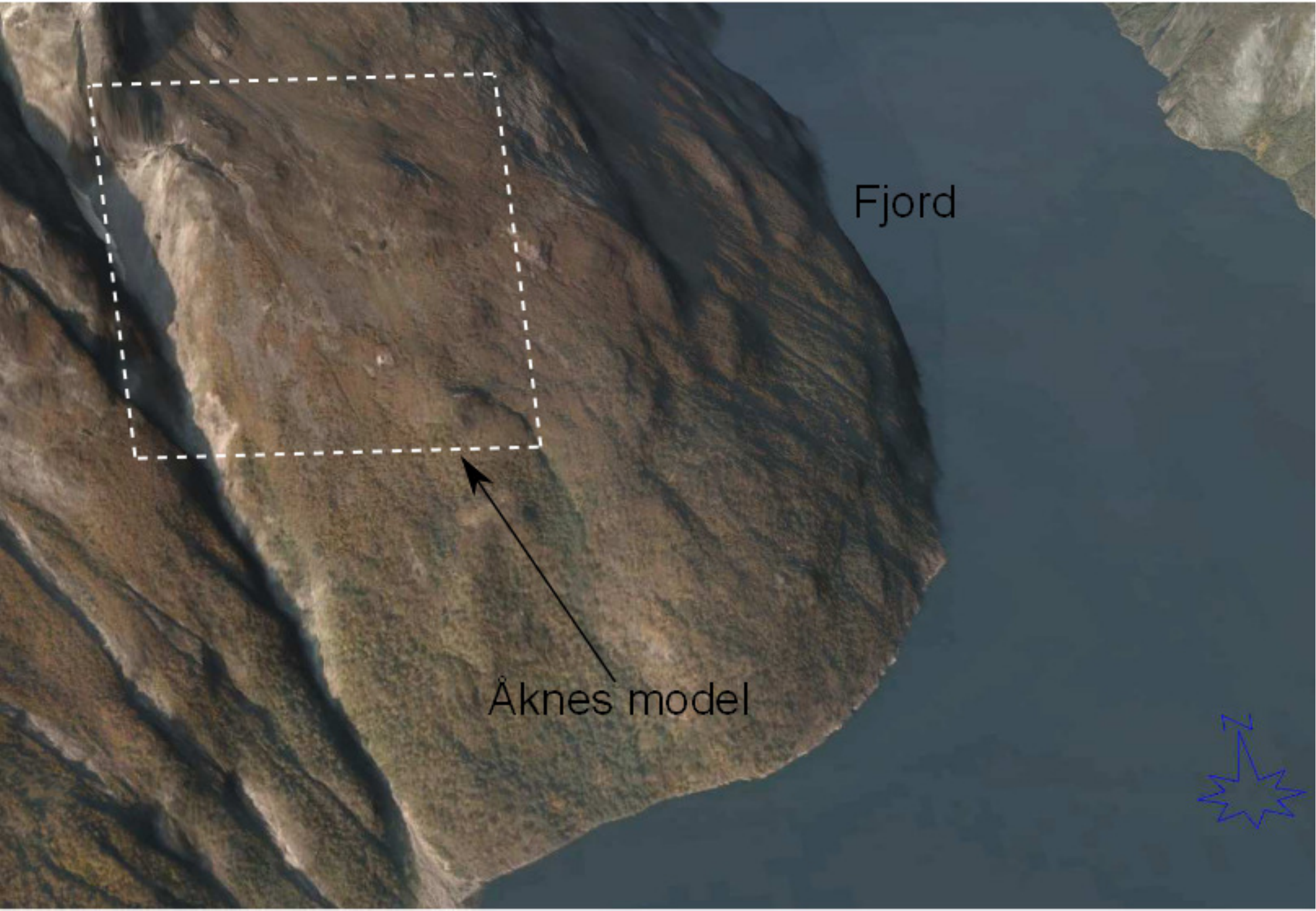}
\caption{\r Asknes rock slope. Dotted line represents the approximate outline of the \r Aknes model used for the simulation.}
\label{fig:aaknes}
\end{figure}

In this example, we simulate wave propagation in a rock slope at \r Aknes located in Western Norway,
which is monitored because of its instability (Figure~\ref{fig:aaknes}).
Its unstable flank is moving at a mean rate of about 6~cm/year and as fast as 14~cm/year~\cite{kveldsvik2008a}.
The slope mainly consists of gneiss, and it is densely jointed along the foliation. The mass of the sliding volume has been estimated to be $\sim$~35-40~million m\tsup{3} (e.g., \cite{ganerod2008,kveldsvik2008b}).
Due to its massive volume and proximity to the fjord (i.e., a narrow inlet of the sea), it poses a potential tsunami risk.
Several monitoring instruments including a seismic network (e.g., \cite{roth2009}) are in place to monitor the slope movement. The seismic network consists of 8 three-component geophones covering an area of about $250 \times 150$ m\tsup{2}. In addition, a high-sensitive broadband seismometer has been installed in the middle of the slope. Microearthquakes are frequently observed at the site (see \cite{roth2009}). 

\begin{figure}[htbp]
\centering
\subfloat[]{\includegraphics[scale=0.4]{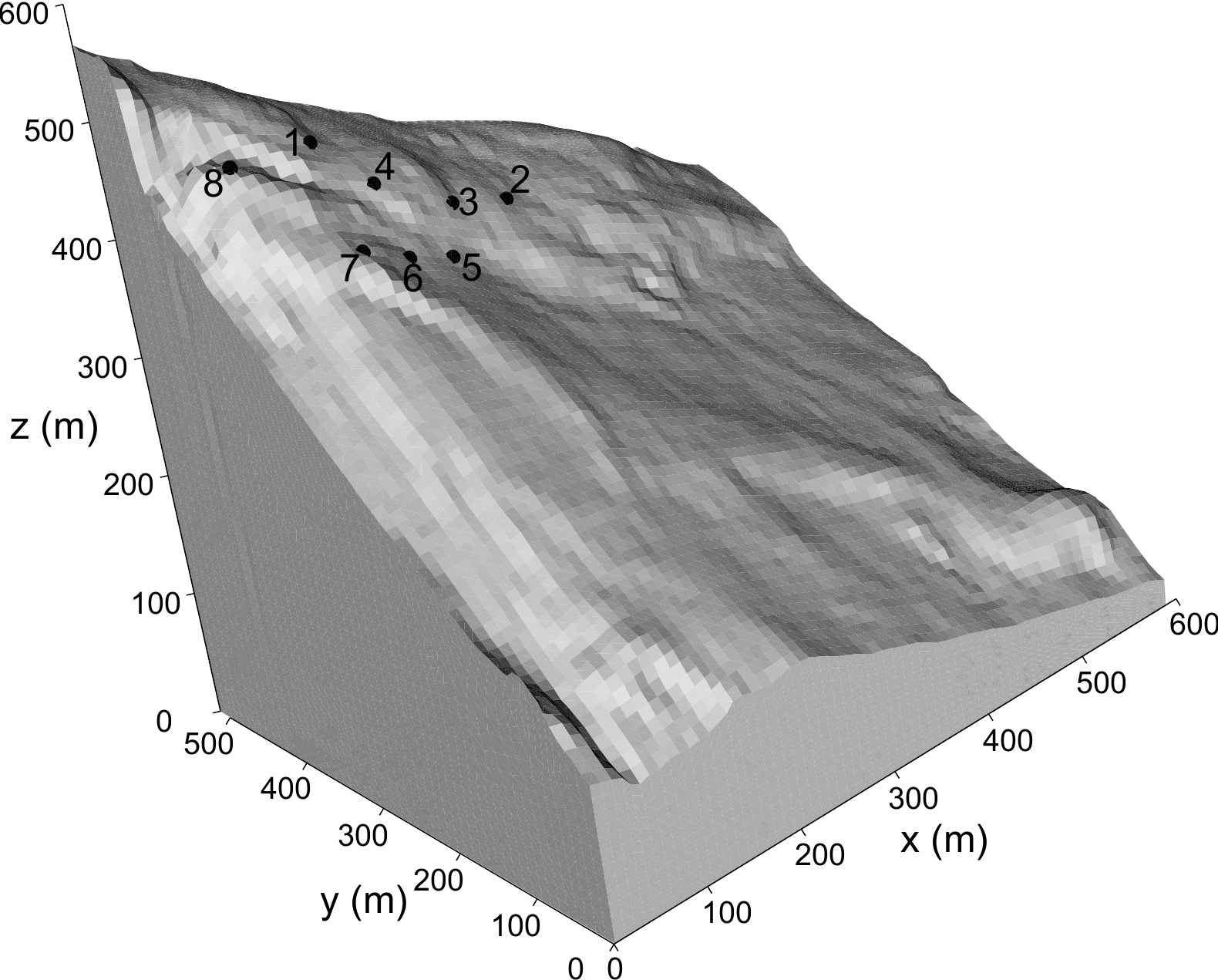}}\qquad
\subfloat[]{\includegraphics[scale=0.42]{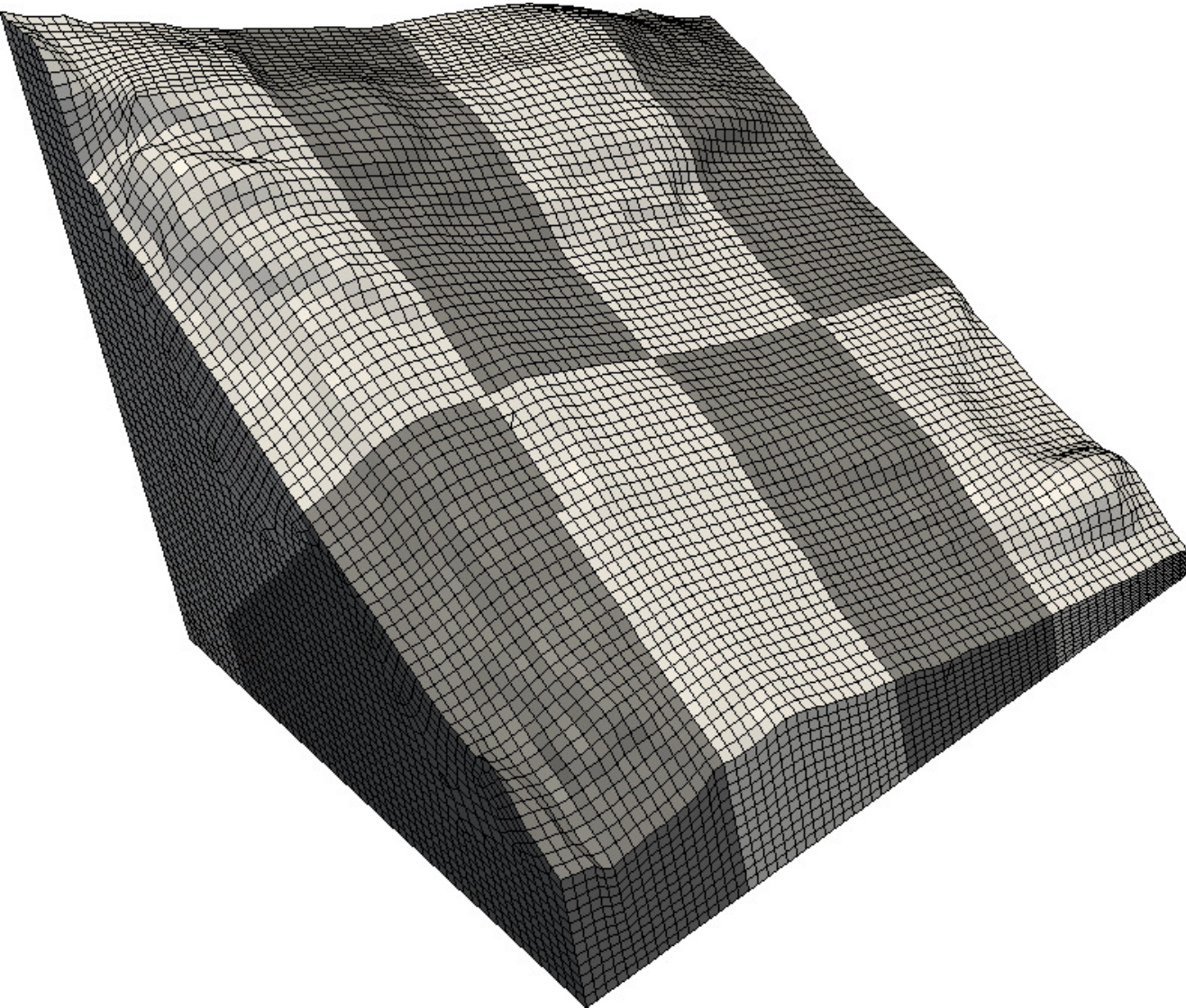}}
\caption{a) 3D model of the \r Aknes rock slope. Geophones are numbered (solid black). The $x$, $y$, and $z$ axes represent the East, North, and vertical directions respectively. b) Spectral-element mesh for the \r Aknes slope model. The mesh is partitioned into 8 domains.}
\label{fig:aaknes_model}
\end{figure}

Due to the rough topography and unavailability of a realistic velocity model, we cannot locate the microearthquakes. It is nevertheless important to identify the characteristic features of the synthetic wavefield. Once a realistic velocity model is available, we can utilise these synthetic results to locate and characterise microearthquakes. We have access to a Digital Elevation Map (DEM) of the slope. Based on the DEM of the slope, we build a 3D model including a realistic topography (Figure~\ref{fig:aaknes_model}a). Since we do not have detailed seismic properties of the slope, we use a homogeneous model with a P-wave velocity ($V_p$) of 3000~m/s, a S-wave velocity ($V_s$) of 1732~m/s, and a mass density ($\rho$) of 2000~kg/m\tsup{3}. We position the source on the free surface with $x=171$~m, $y=470$~m, and $z=461$~m so that all complexities due to the free surface topography can be captured. The source is represented by an explosion mechanism given by the moment-tensor components as $M_{xx}=M_{yy}=M_{zz}=10^{15}$~Nm, and is characterised by a Ricker wavelet having a central frequency of 120~Hz. The sampling interval for the seismogram recordings is set to 10~$\mu$s.

Using the CUBIT, we mesh the model with an average element size of 10~m resulting in a total of 107,712 spectral elements and a total of 7,161,572 spectral nodes. The mesh is partitioned into 8 domains for the parallel processing (Figure~\ref{fig:aaknes_model}b).

\begin{figure}[htbp]
\centering
\subfloat{\includegraphics[scale=0.38]{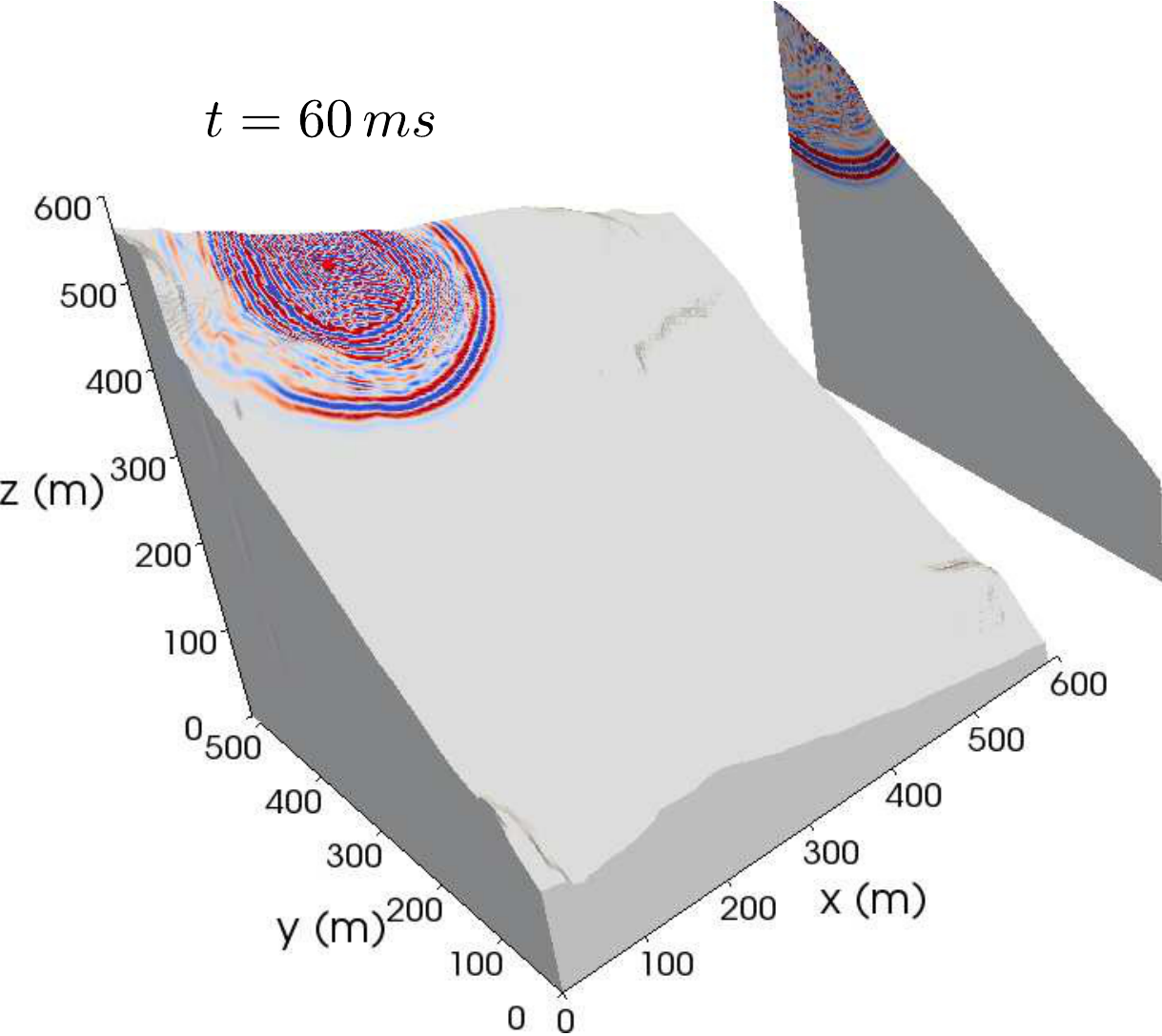}}
\subfloat{\includegraphics[scale=0.38]{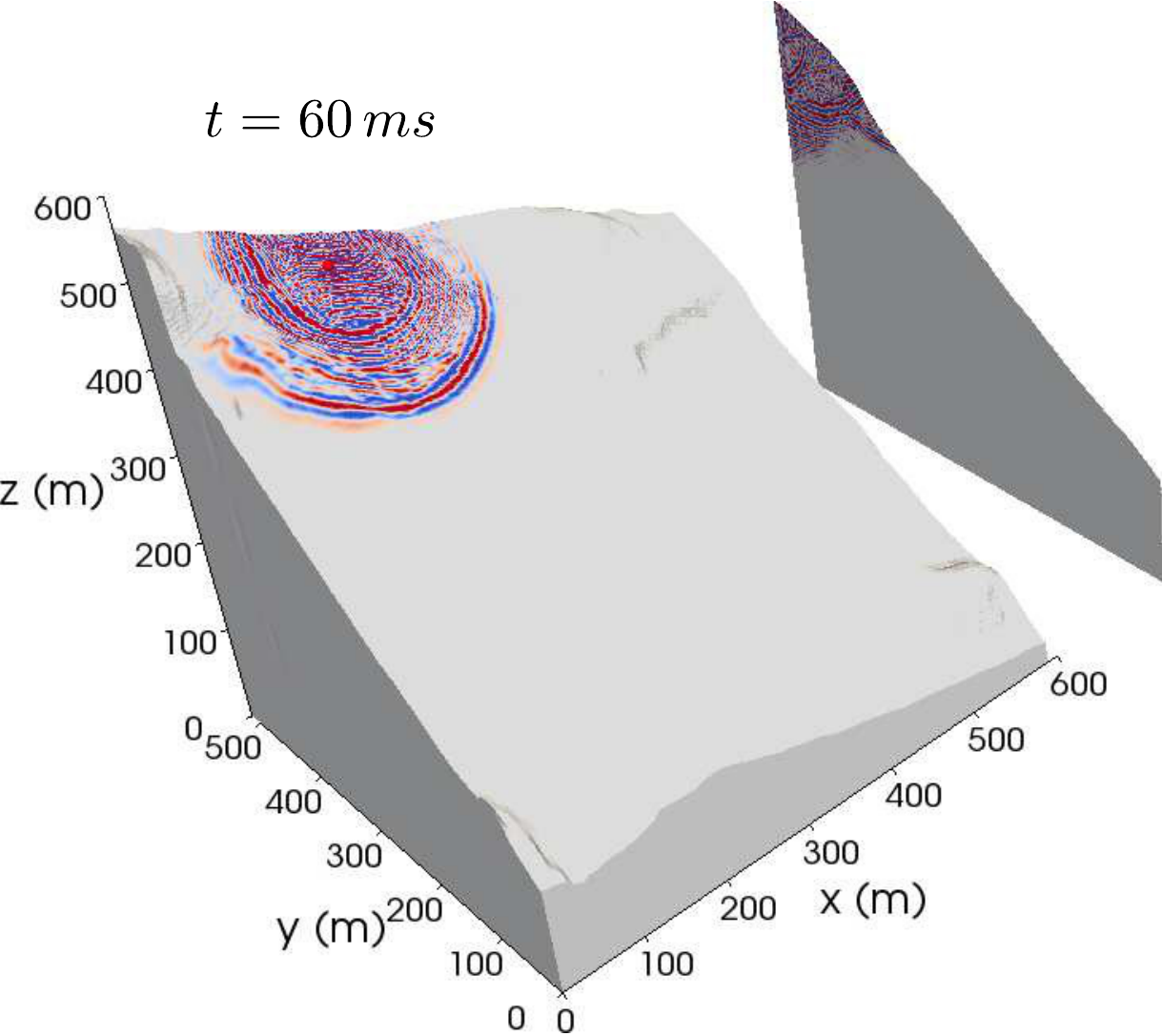}}\\
\subfloat{\includegraphics[scale=0.38]{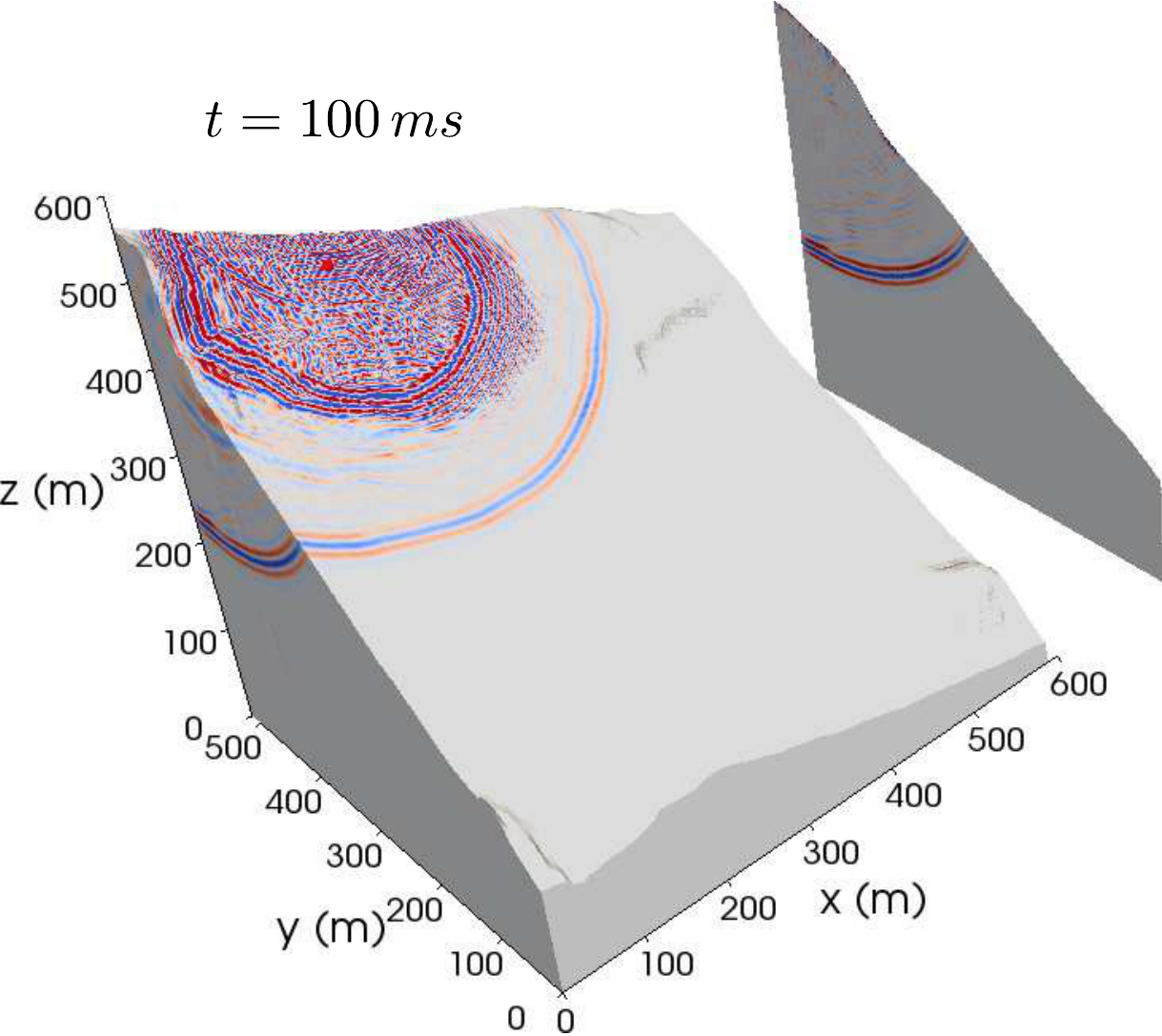}}
\subfloat{\includegraphics[scale=0.38]{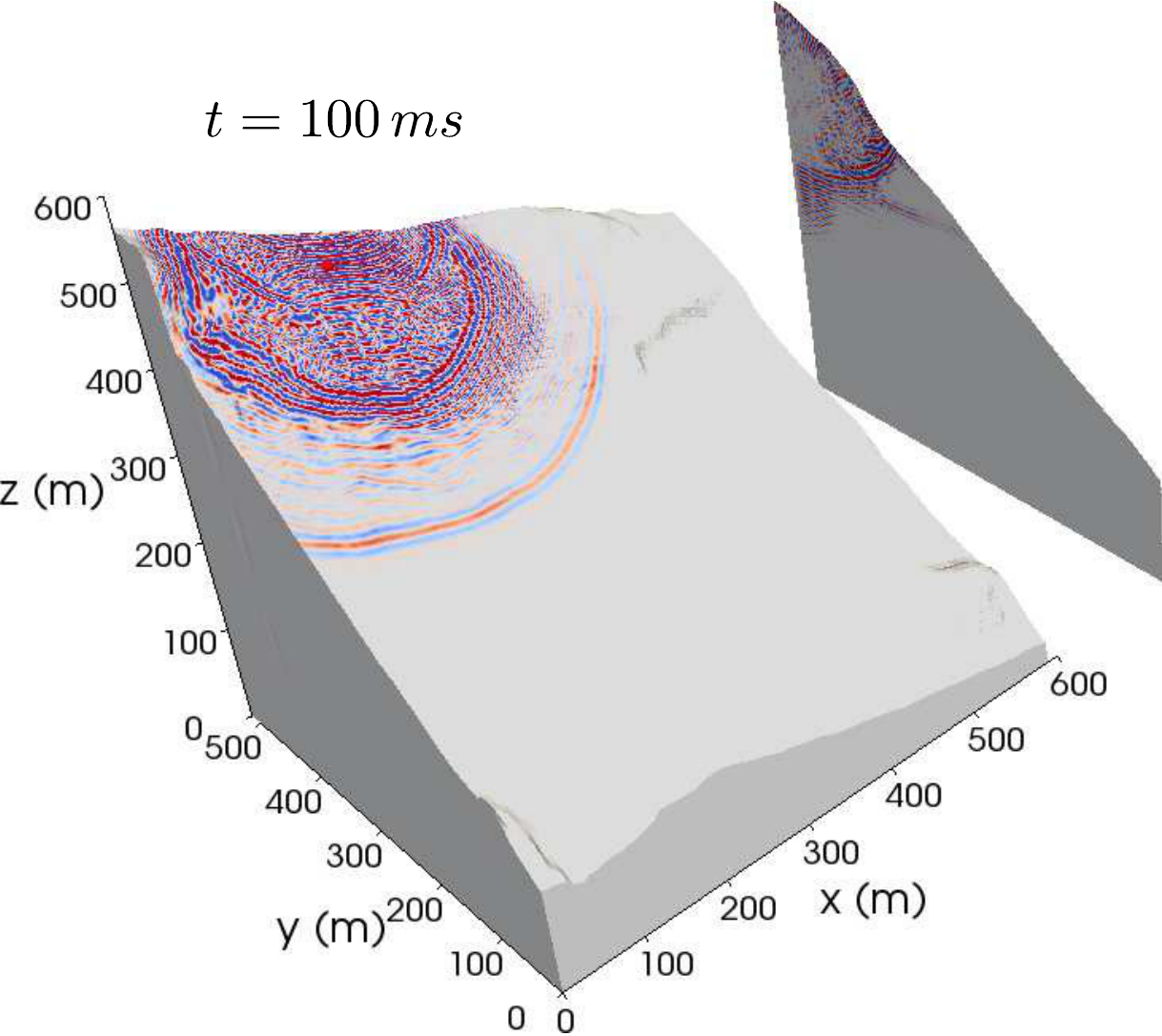}}\\
\subfloat{\includegraphics[scale=0.38]{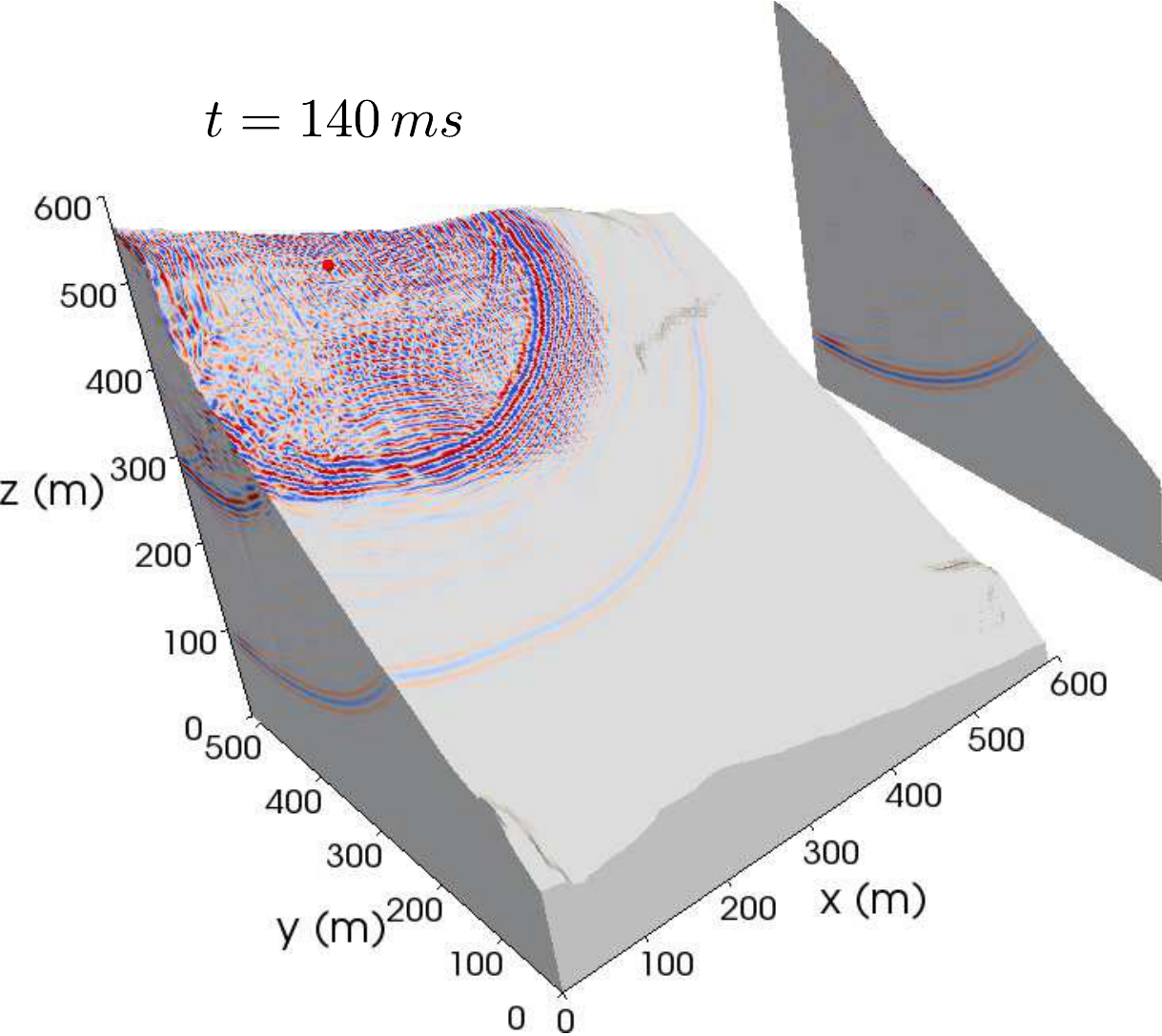}}
\subfloat{\includegraphics[scale=0.38]{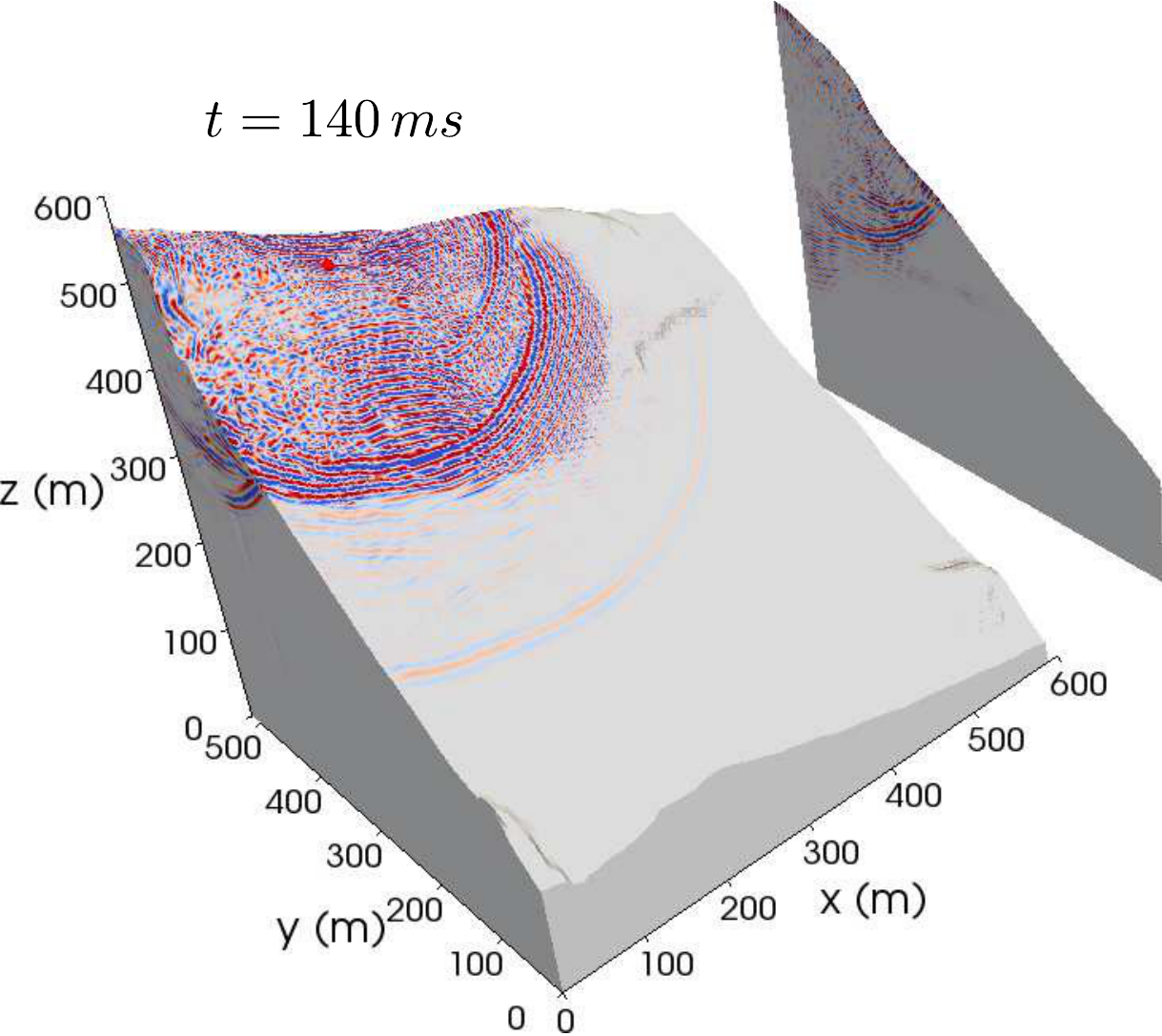}}\\
\subfloat{\includegraphics[scale=0.38]{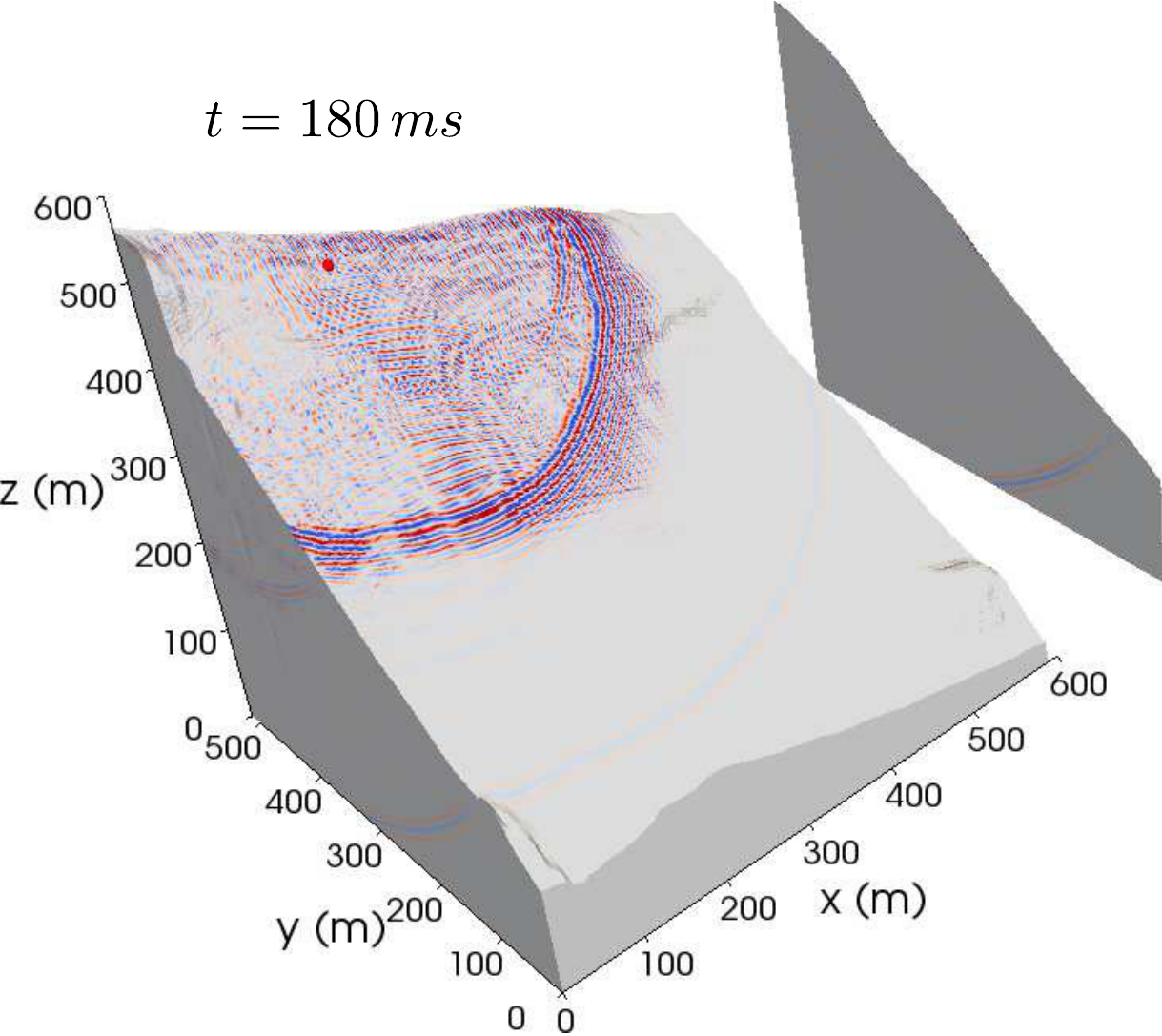}}
\subfloat{\includegraphics[scale=0.38]{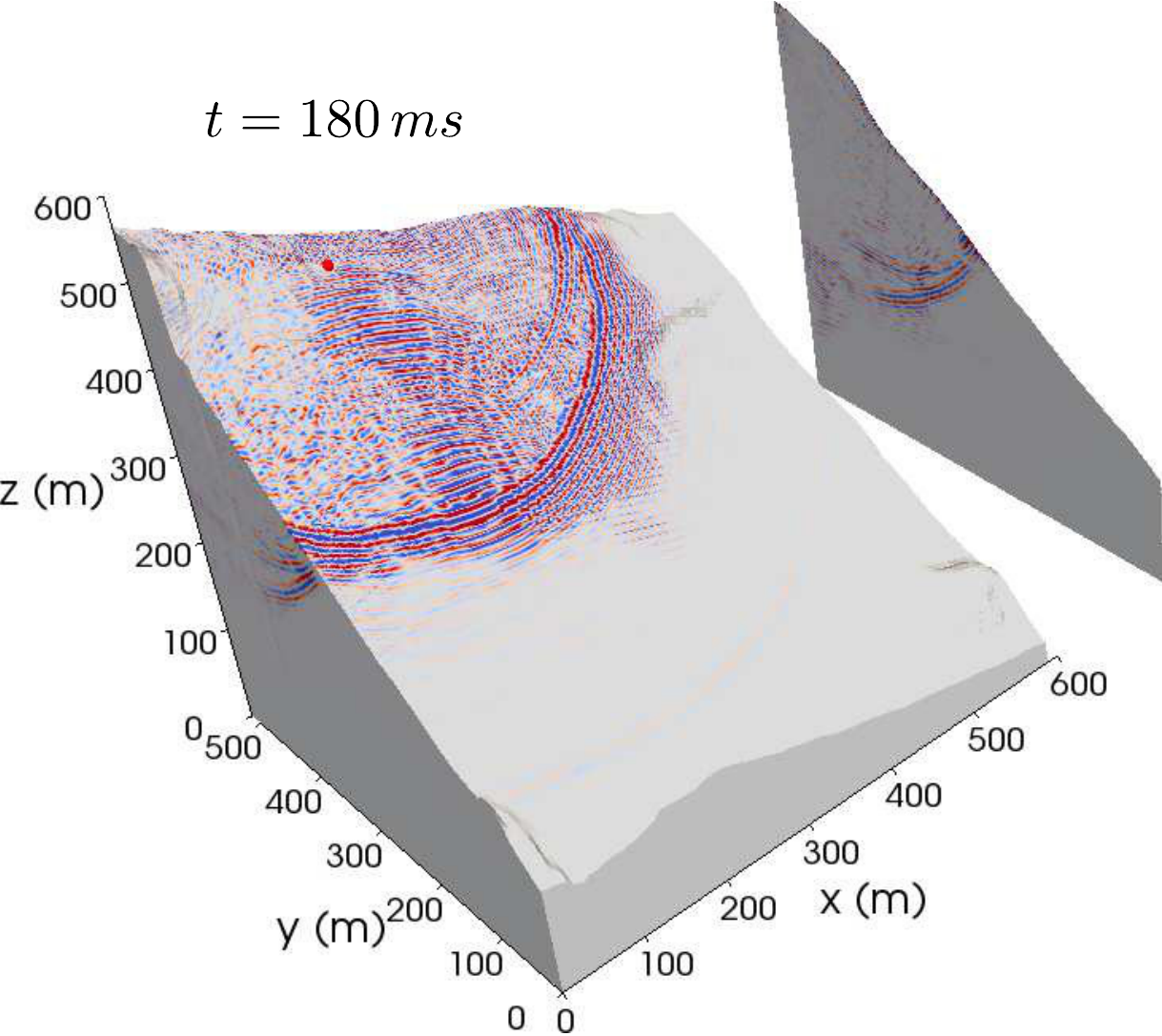}}
\caption{P-wave potential (left) and S-wave potential (right). High values are shown in red and low values are shown in blue. The side plane in each plot represents a $yz$-slice through the source location (red sphere).}
\label{fig:aaknes_wave}
\end{figure}

For visualization of the wavefield, we plot P-wave and S-wave potentials (Figure~\ref{fig:aaknes_wave}). Multiple reflections and conversions on the free surface result in a complicated wavefield, despite the simple source mechanism.

\begin{figure}[htbp]
\centering
\includegraphics[scale=0.3]{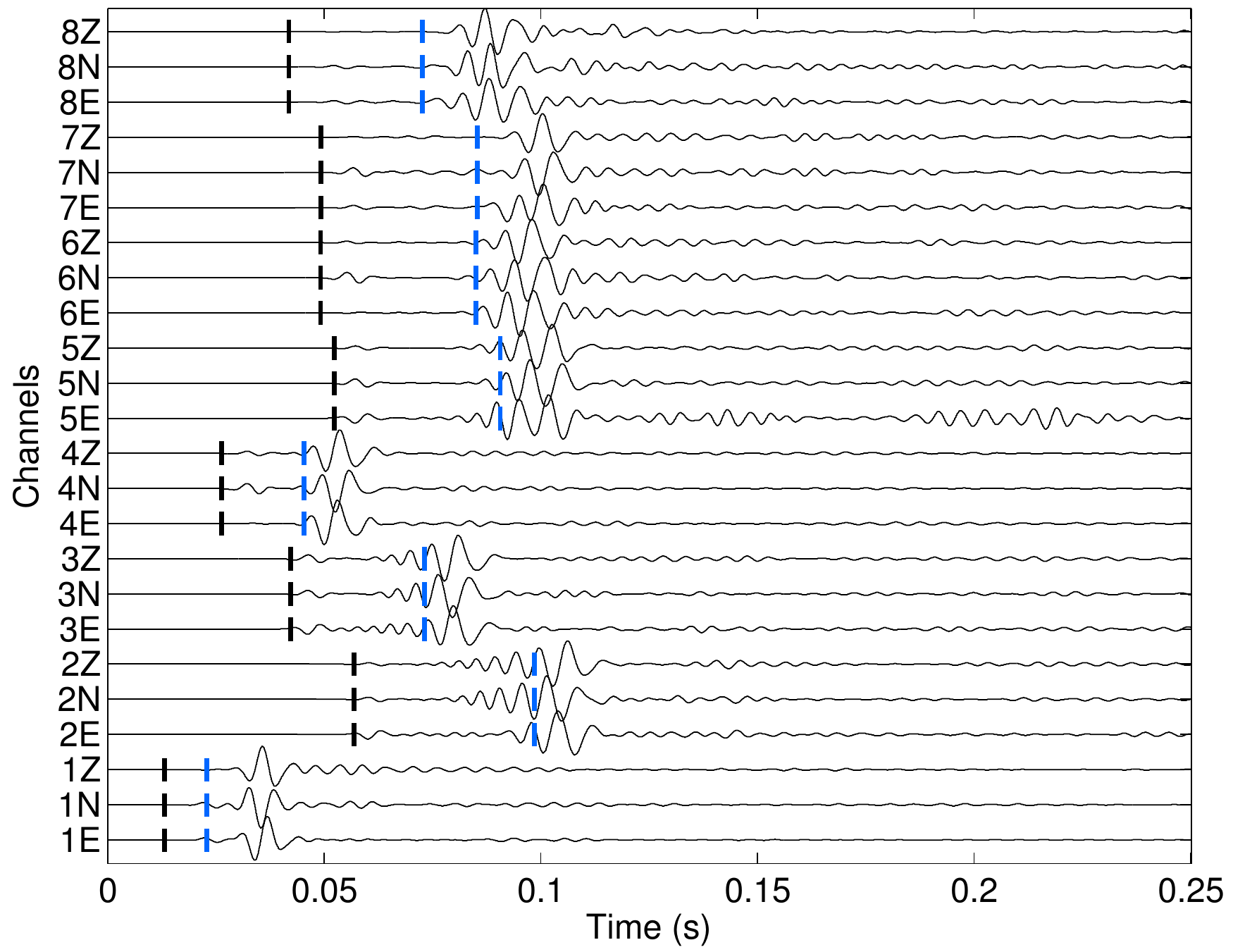}
\caption{Synthetic waveforms computed for the \r Aknes slope model. Superimposed are the P- (black) and S-wave (blue) first-arrival times computed with a finite-difference eikonal solver. Seismograms are normalized to geophone maximum.}
\label{fig:aaknes_waveform}
\end{figure}

Figure~\ref{fig:aaknes_waveform} shows the synthetic waveforms. Although we have used an explosion source, S waves are generated by free-surface reflections and conversions, which are stronger than the P waves. First-arrival times computed by a finite-difference eikonal solver are generally in good agreement with the respective arrivals in the computed waveforms, but we observe some discrepancies in a few geophones, namely 1, 7, and 8. It is difficult to accurately compute arrival times using the finite-difference eikonal solver due to the rough topography, especially when the receivers are situated on the surface.

\section{Weakly anisotropic cylinder at laboratory scale}

\begin{figure}[htbp]
\centering
\subfloat[]{\includegraphics[scale=0.25]{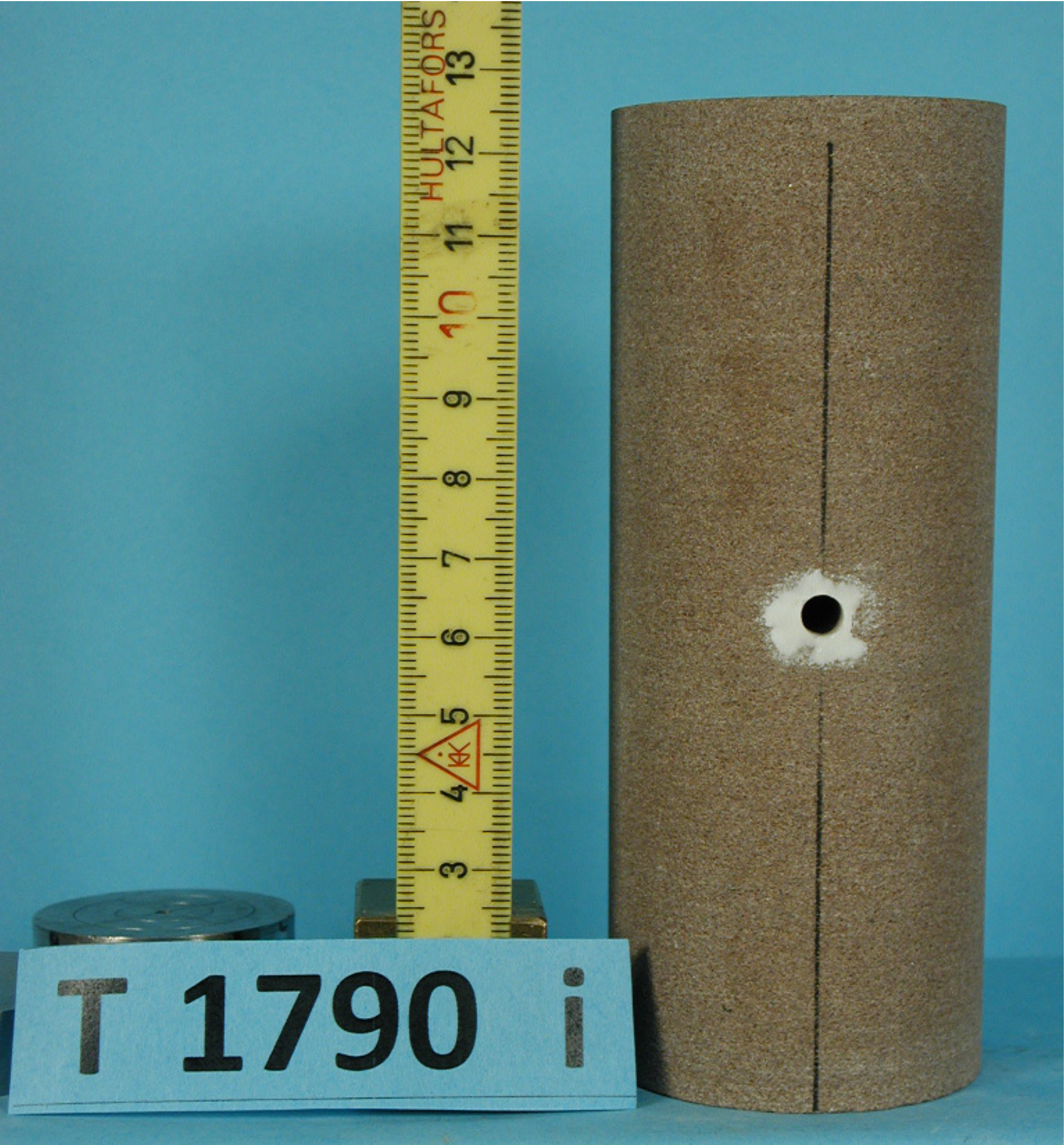}}
\subfloat[]{\includegraphics[scale=0.2]{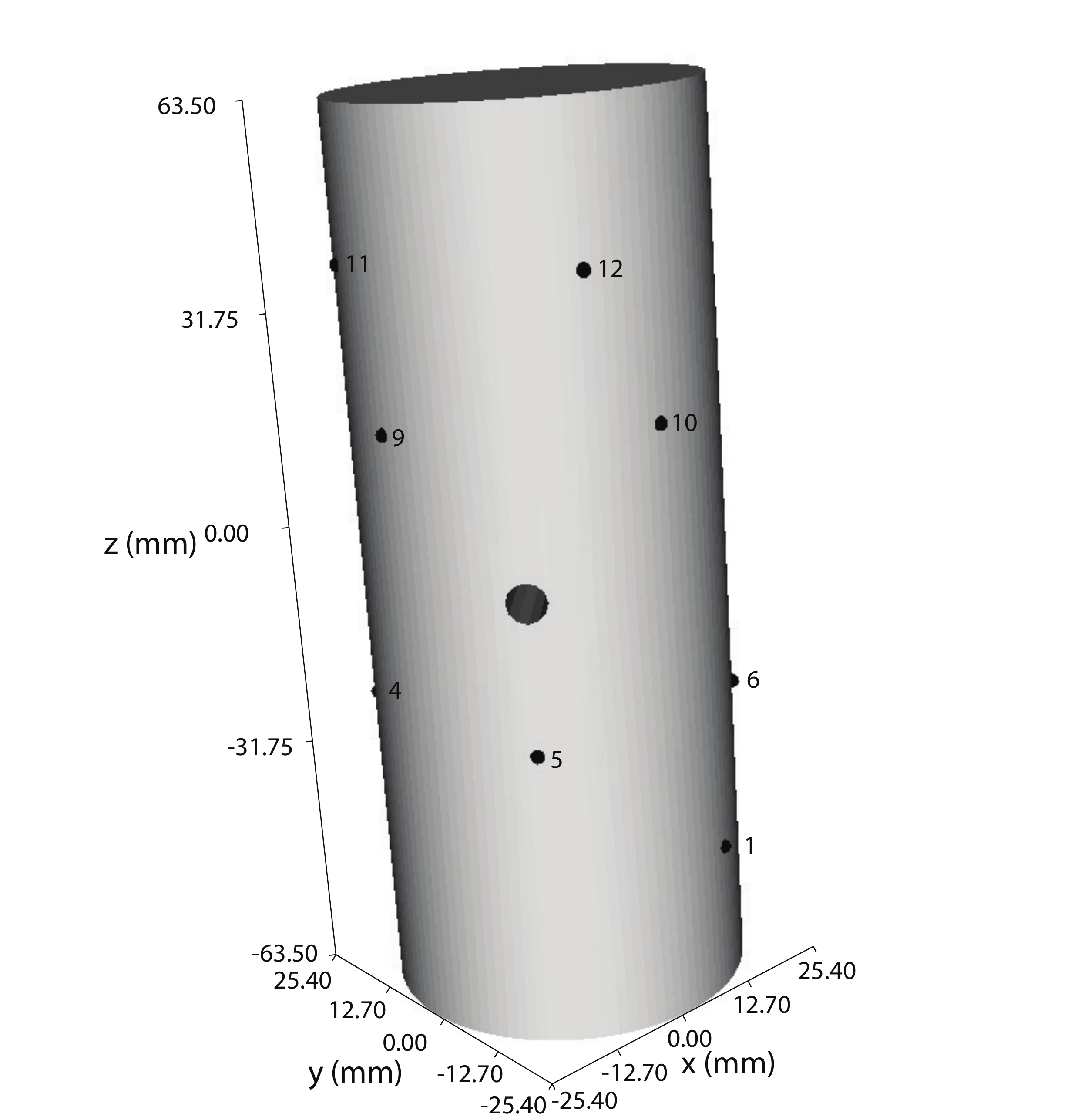}}
\caption{a) Sandstone sample for the acoustic emission experiment. b) Model of the acoustic emission experiment sample. The borehole has a diameter of 5.2~mm. The piezoelectric sensors are numbered (solid black dots).}
\label{fig:ngi_sample}
\end{figure}

Finally, we simulate wave propagation for acoustic emissions observed during a laboratory experiment (Figure~\ref{fig:ngi_sample}a). The laboratory sample consists of a Vosges sandstone cylinder with a diameter of 25.4~mm and a height of 63.5~mm. A small cylindrical hole of 5.2~mm diameter is drilled through the centre of the sample mimicking a borehole in a real field problem. The triaxial experiment was performed to fracture the sample thereby inducing the acoustic emissions. Twelve piezoelectric sensors are mounted on the surface of the specimen~\cite{aker2009} to record the radial displacement. Full waveforms are recorded with a sampling rate of 10~MHz. About 2500 events were detected during a triaxial experiment (due to limitations of the data acquisition system, this does not represent all occurred events) with automatic processing (e.g., \cite{kuhn2010,oye2010}). 

\begin{figure}[htbp]
\centering
\includegraphics[scale=0.5]{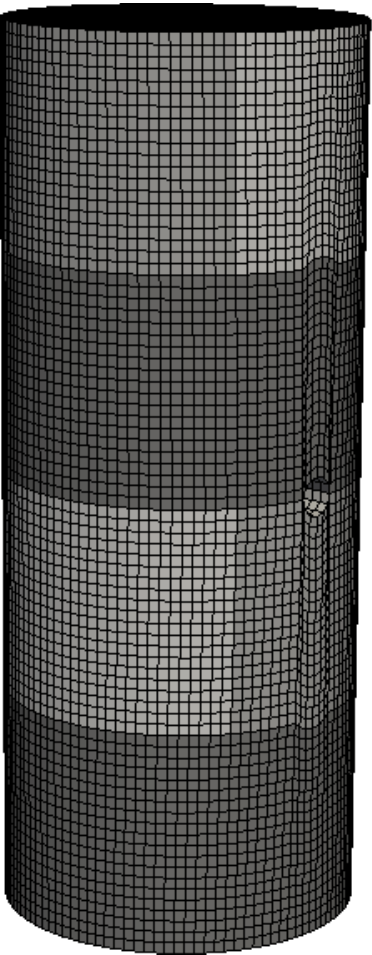}
\caption{Spectral-element mesh for the acoustic emission experiment sample. The mesh is partitioned into 8 domains.}
\label{fig:ngi_mesh}
\end{figure}

We use a homogeneous velocity model with a P-wave velocity ($V_p$) of 3660~m/s and a S-wave velocity ($V_s$) of 2286~m/s estimated from experimental data. We take a mass density ($\rho$) of 2000~kg/m\tsup{3}. For anisotropy, two Thomsen's parameters $\epsilon$ and $\gamma$ are estimated to be about -0.0869 and 0.07613, respectively, from the experimental data; and the third Thomsen's parameter $\delta$ is assumed to be -0.1 (e.g., \cite{thomsen1986}). We again use the CUBIT to mesh the model. The mesh consists of 82,240 spectral elements with an average element size of 1.5~mm resulting in a total of 5,462,768 spectral nodes. The mesh is partitioned into 8 domains using SCOTCH for parallel processing (Figure~\ref{fig:ngi_mesh}). We select a source located near the bore hole with $x=-9.0$~mm, $y=-5.1$~mm, and $z=1.8$~mm. Full moment-tensor inversion was performed for this source considering a homogeneous isotropic model using first motion polarities and amplitudes~\cite{manthei2005,kuhn2010}.
The moment tensor components for this source were estimated as $M_{xx}=-0.0673\times 10^6$~Nm, $M_{yy}=1.4297\times 10^6$~Nm, $M_{zz}=1.9070\times 10^6$~Nm, $M_{xy}=-0.8306\times 10^6$~Nm, $M_{yz}=-0.4332\times 10^6$~Nm, and $M_{zx}=-0.5052\times 10^6$~Nm. Due to the complex waveforms and the discrepancy between the homogeneous model and the real fractured model, uncertainty in the moment tensor components was large. The presence of both isotropic and deviatoric components in the moment tensor indicates a complex source mechanism. We assume a Ricker wavelet source time function with a central frequency of 500~kHz. The sampling interval for the seismogram recordings is set to 4~ns. We compute full waveforms for both isotropic and anisotropic models.

\begin{figure}[htbp]
\centering
\subfloat{\includegraphics[scale=0.4]{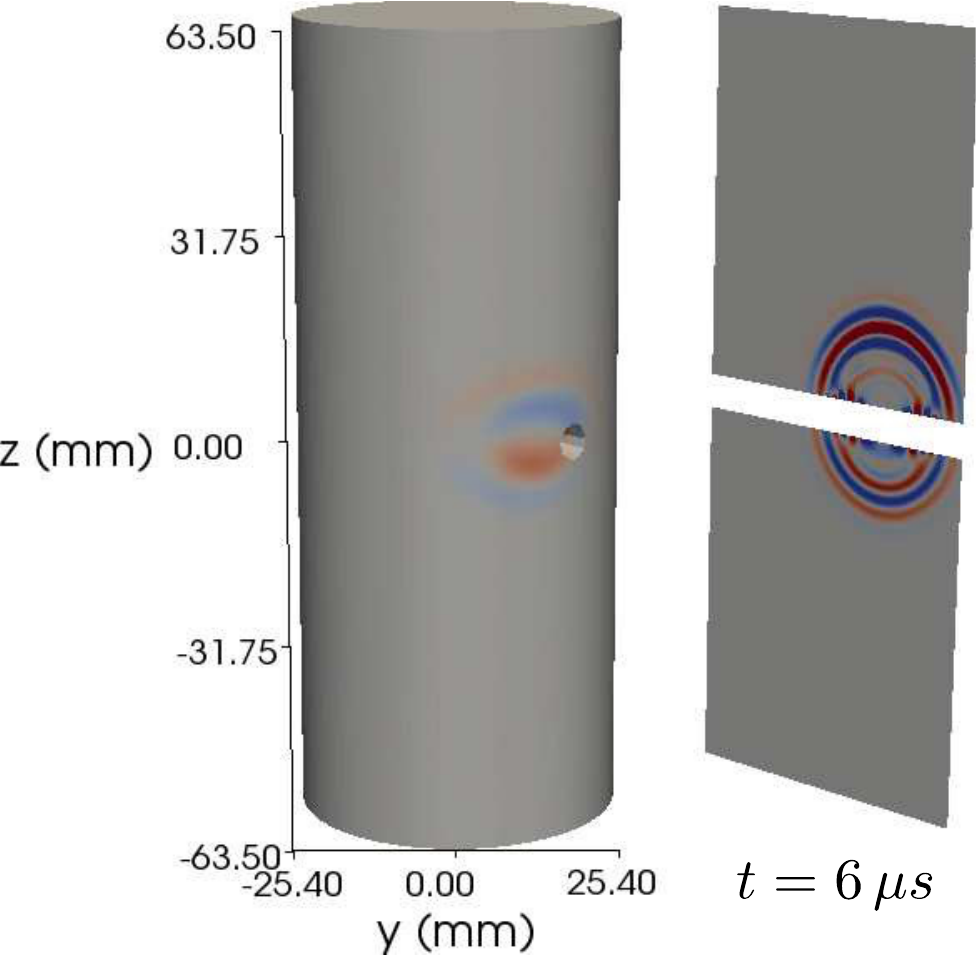}}
\subfloat{\includegraphics[scale=0.4]{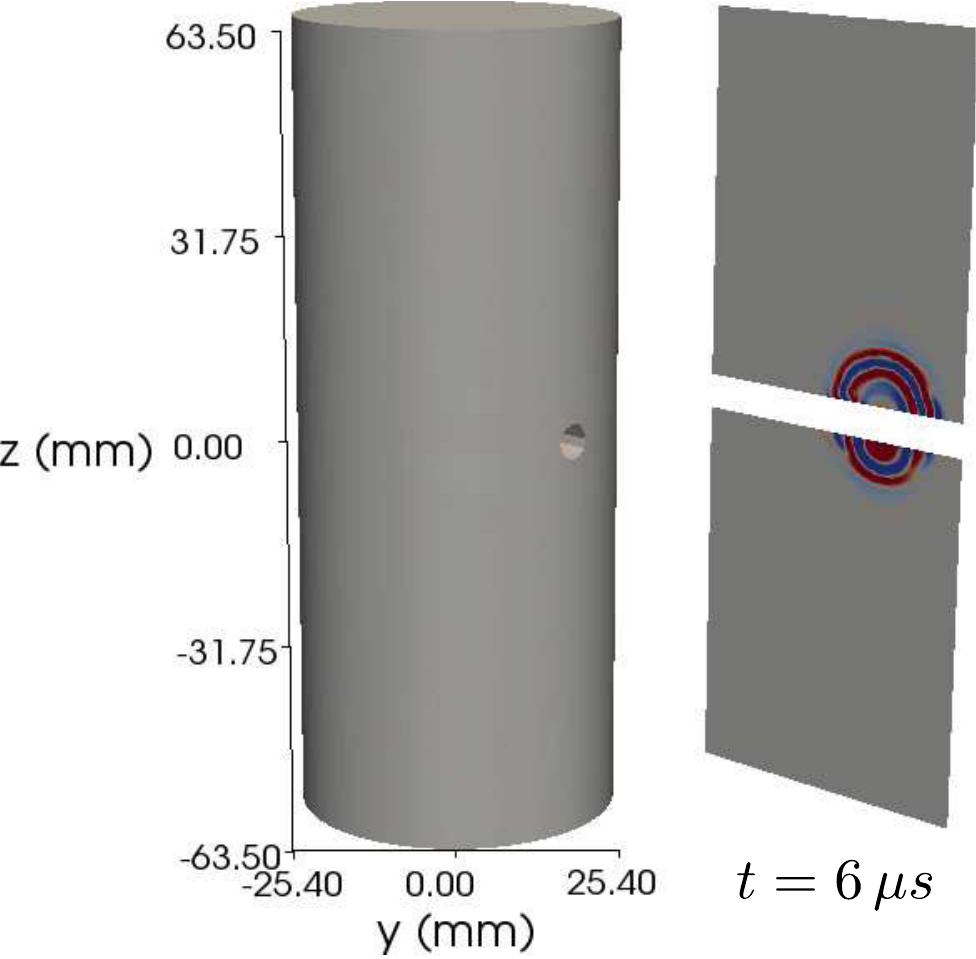}}\\
\subfloat{\includegraphics[scale=0.4]{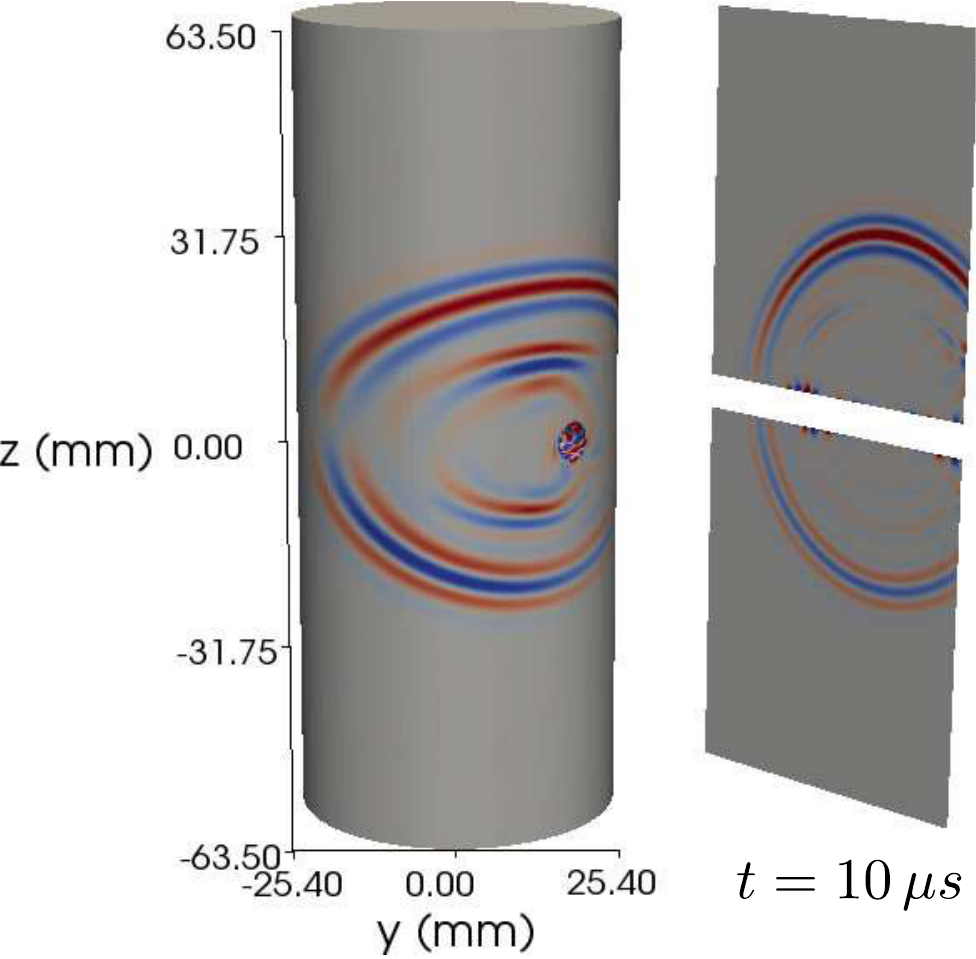}}
\subfloat{\includegraphics[scale=0.4]{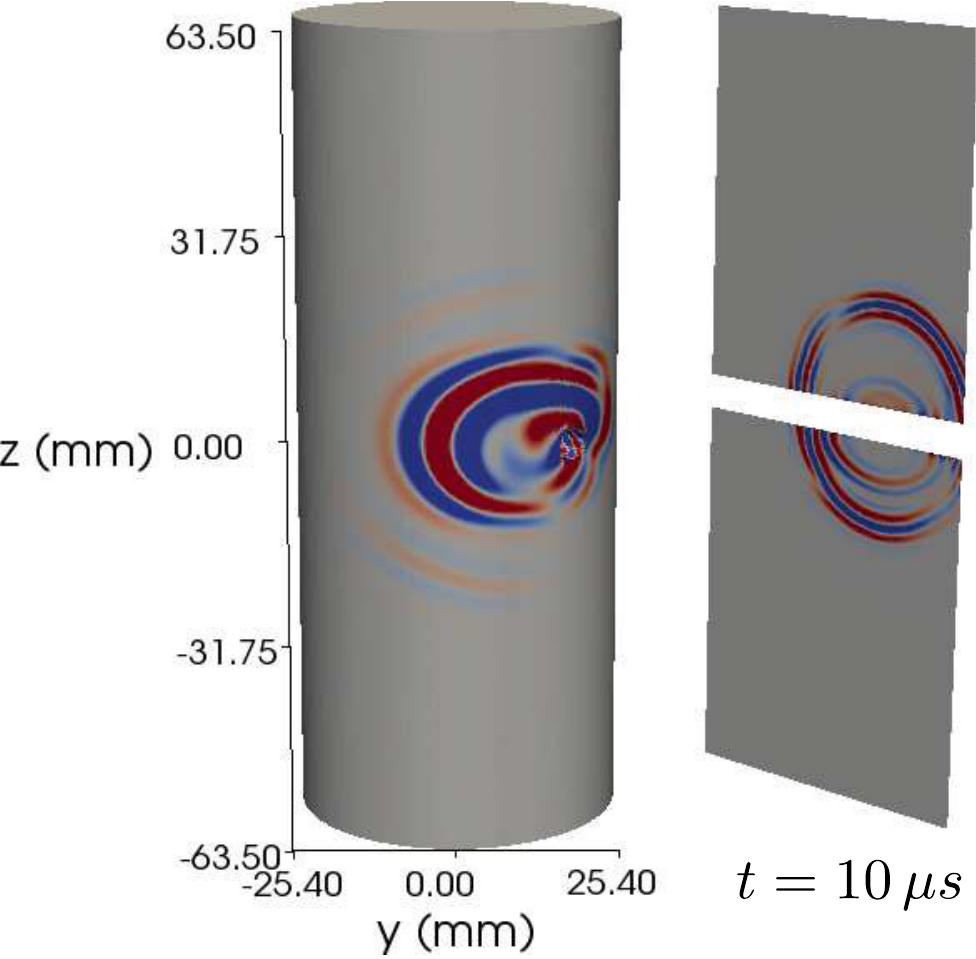}}\\
\subfloat{\includegraphics[scale=0.4]{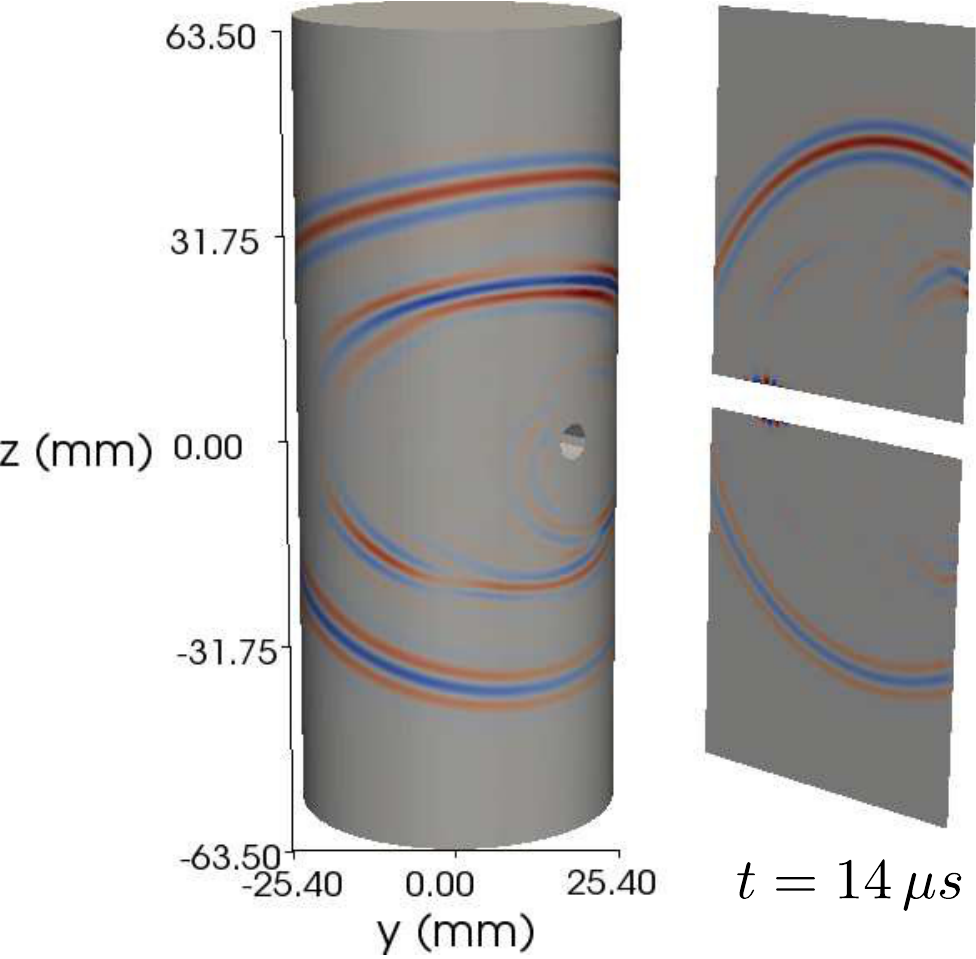}}
\subfloat{\includegraphics[scale=0.4]{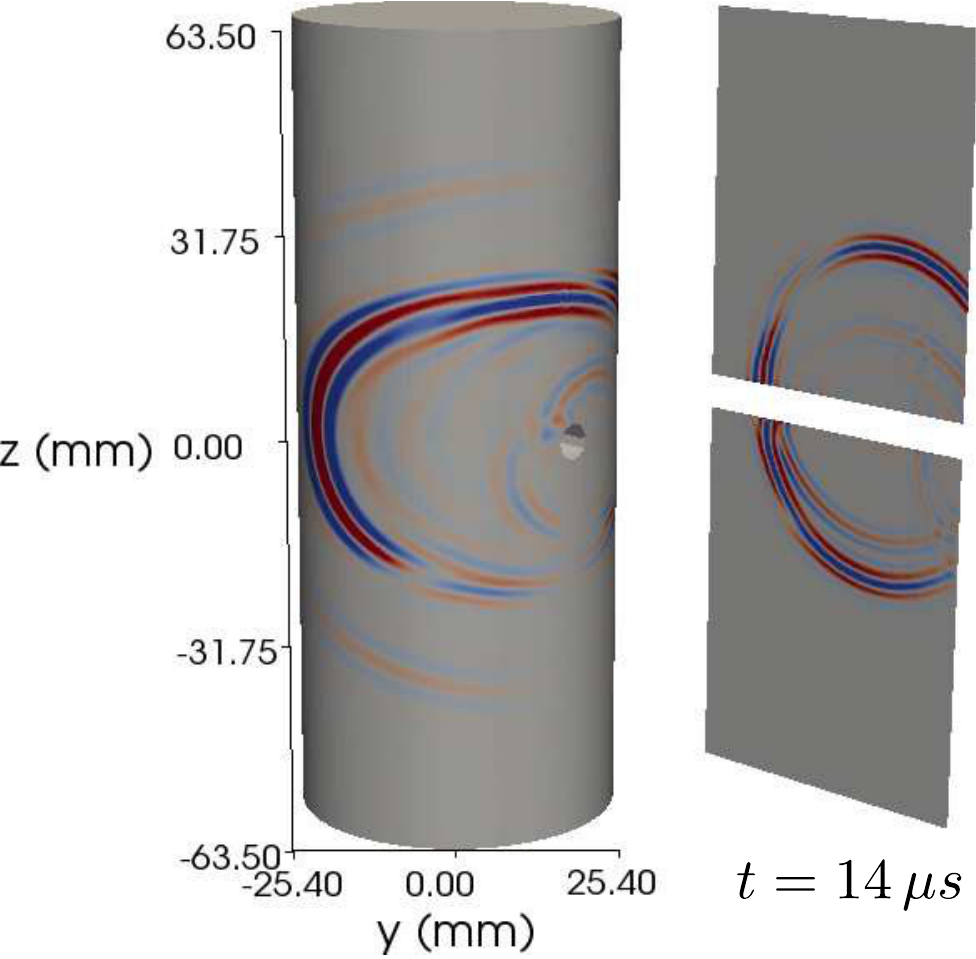}}\\
\subfloat{\includegraphics[scale=0.4]{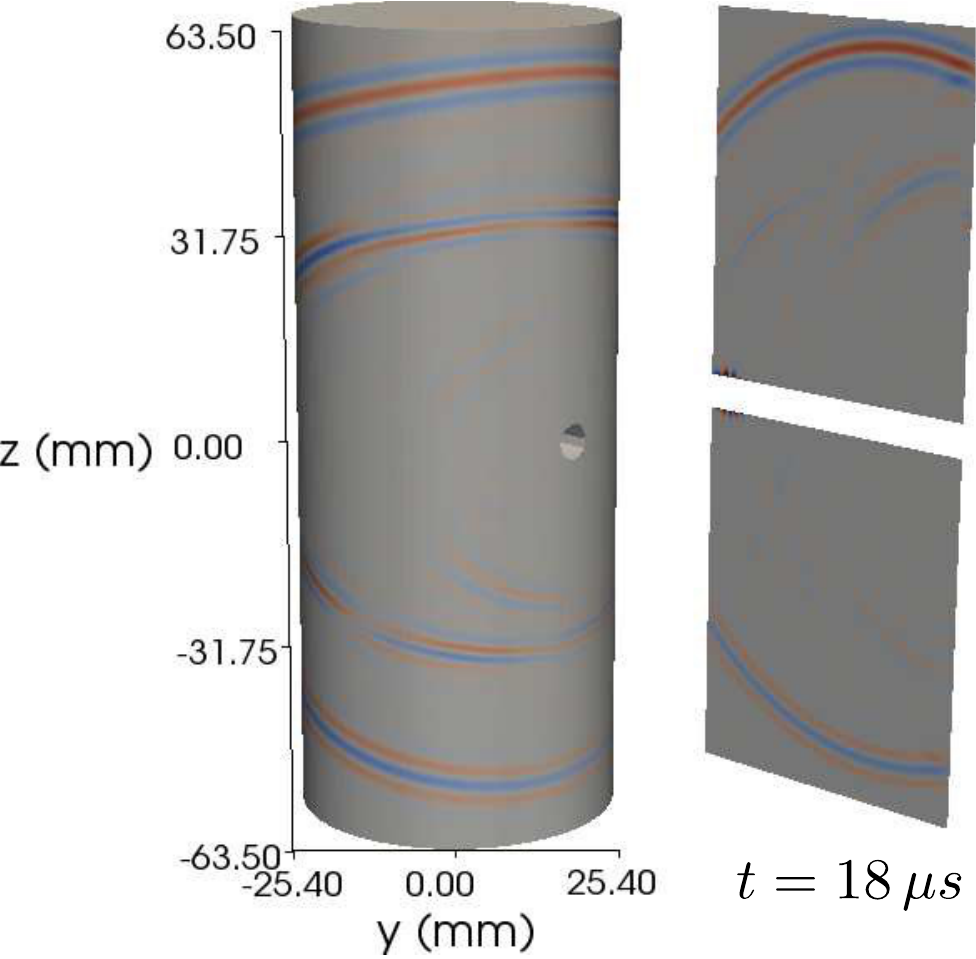}}
\subfloat{\includegraphics[scale=0.4]{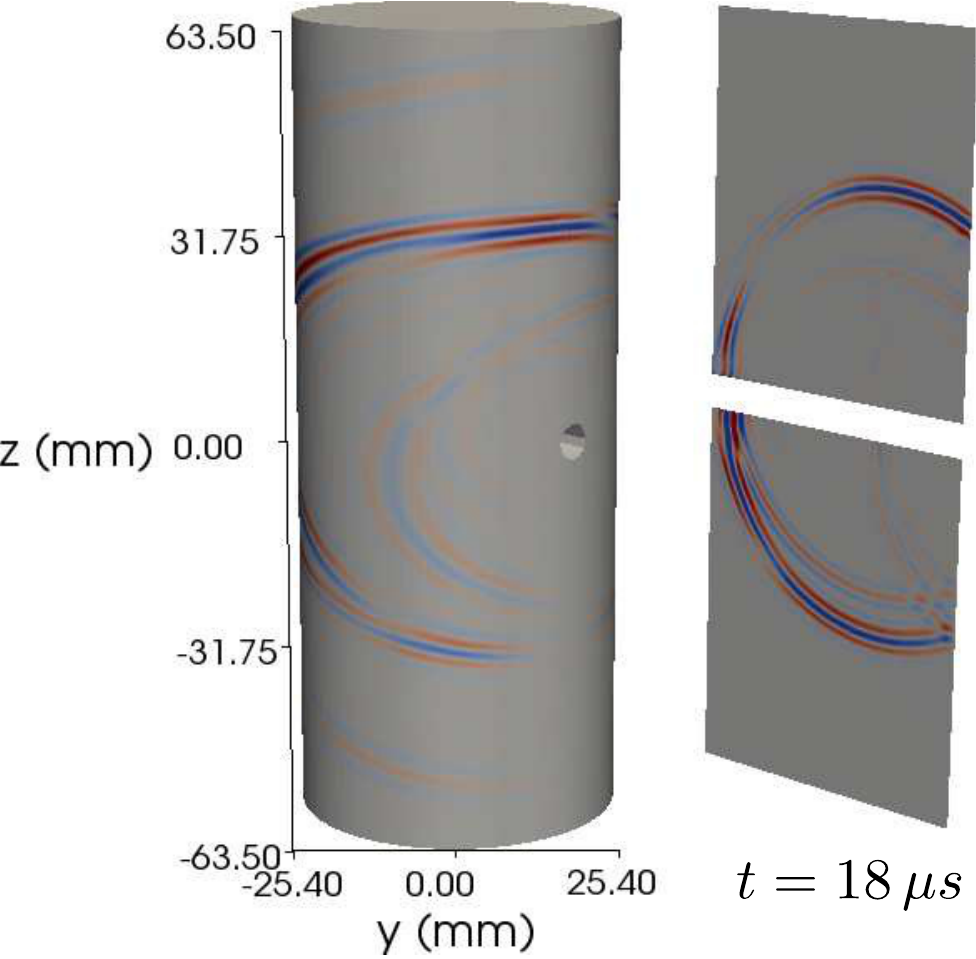}}
\caption{P-wave potential (left) and S-wave potential (right). High values are shown in red and low values are shown in blue. The side plane in each plot represents a slice through the borehole.}
\label{fig:ngi_wave}
\end{figure}

\begin{figure}[htbp]
\subfloat{\includegraphics[scale=0.3]{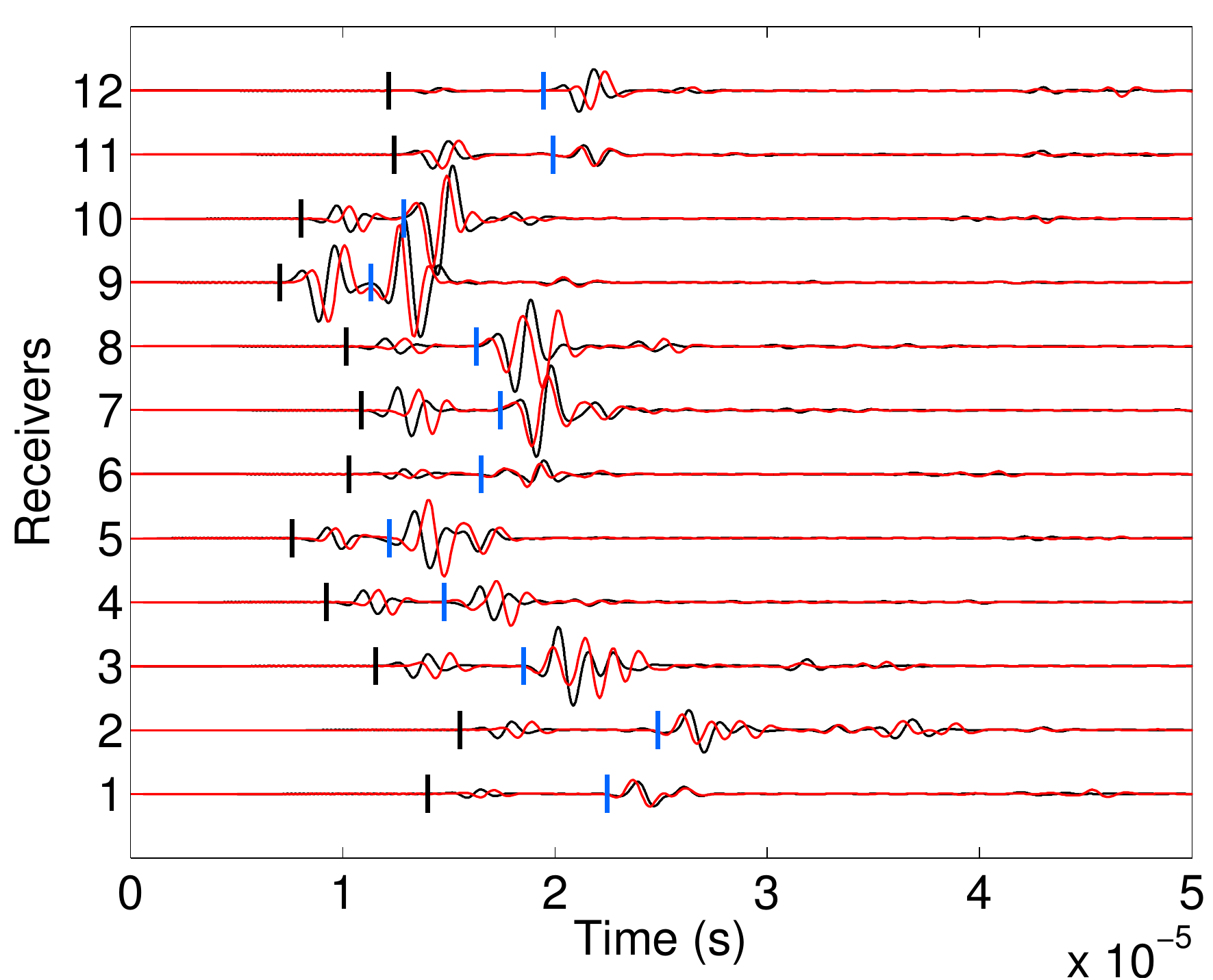}}
\subfloat{\includegraphics[scale=0.3]{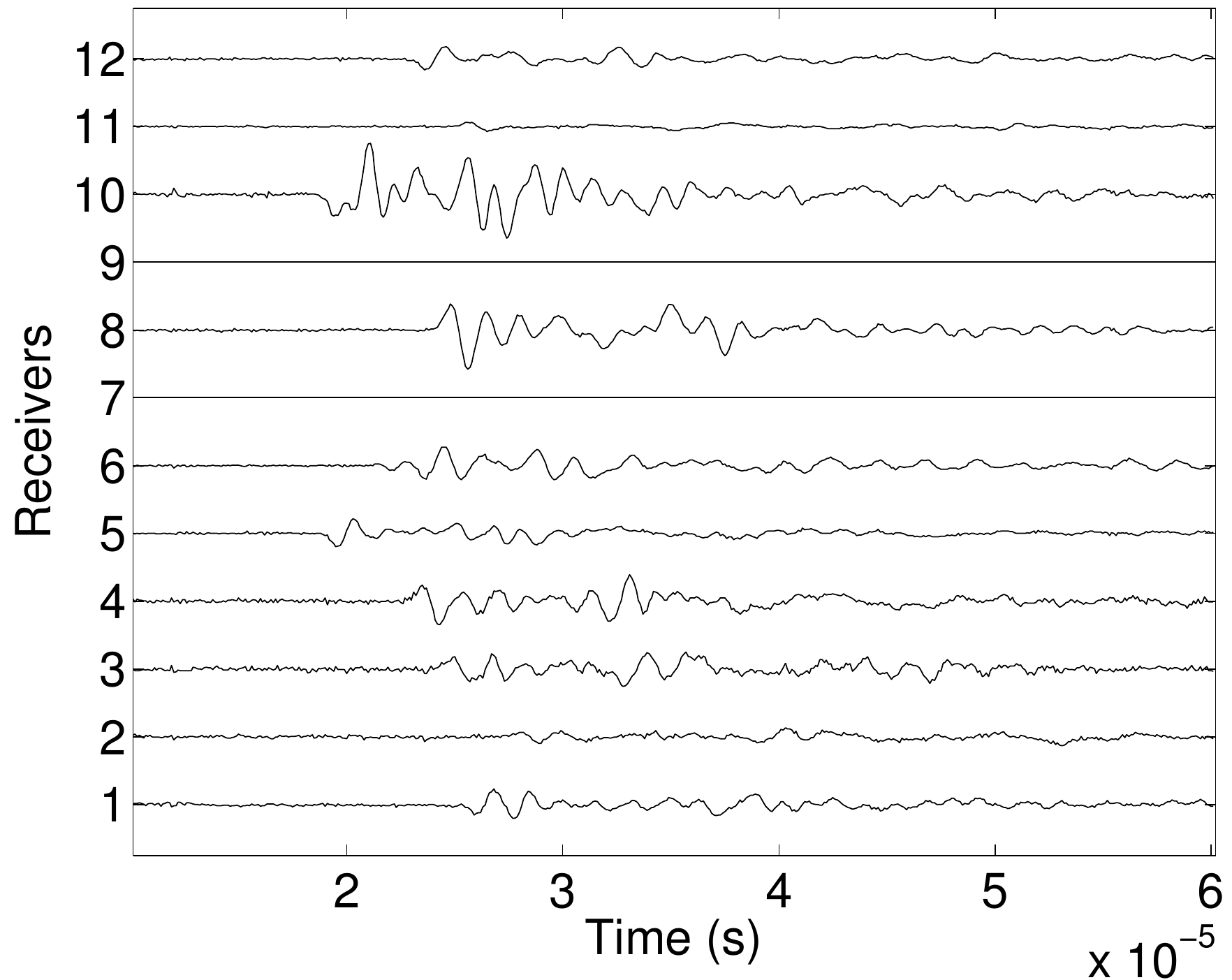}}\\
\subfloat{\includegraphics[scale=0.3]{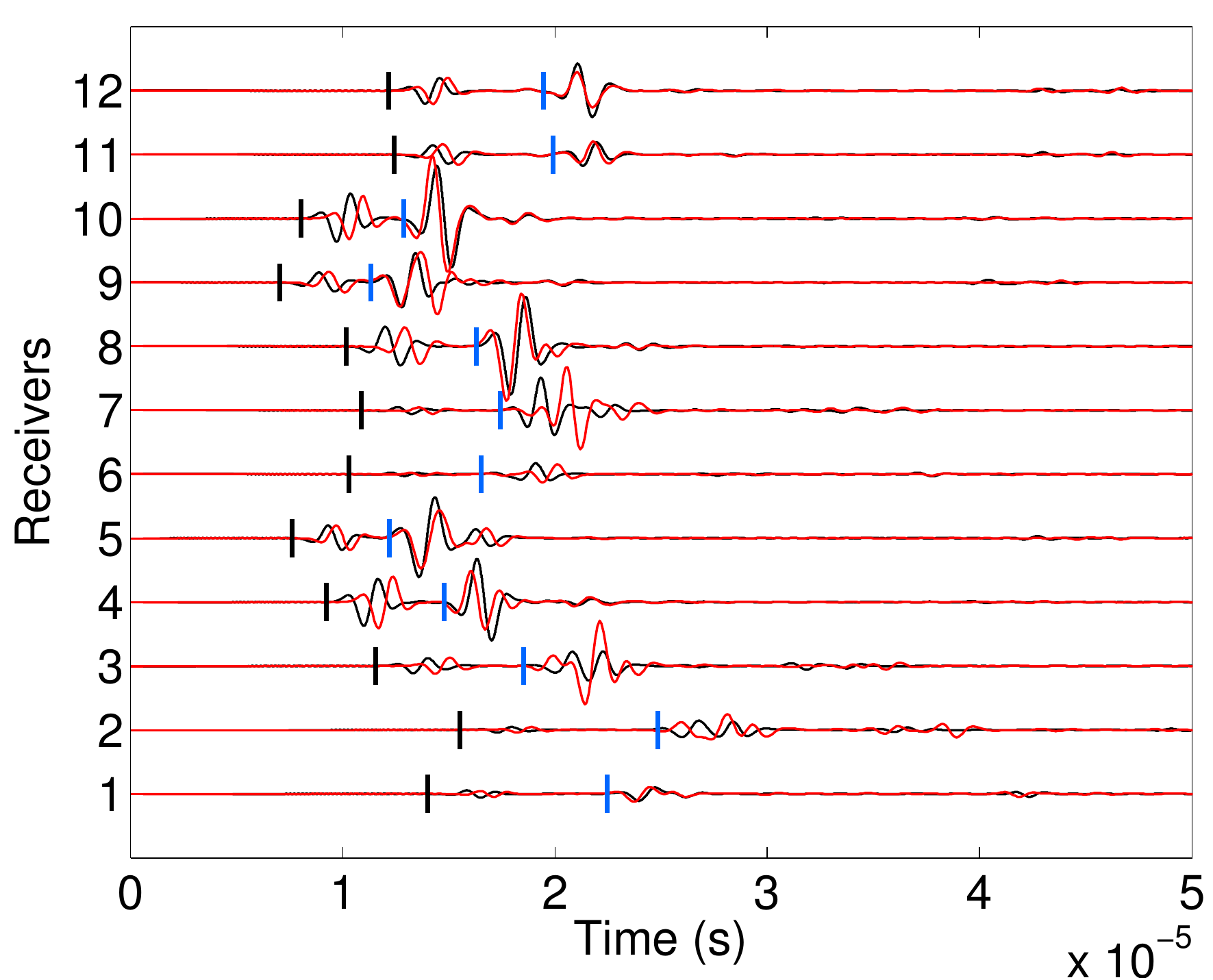}}
\subfloat{\includegraphics[scale=0.3]{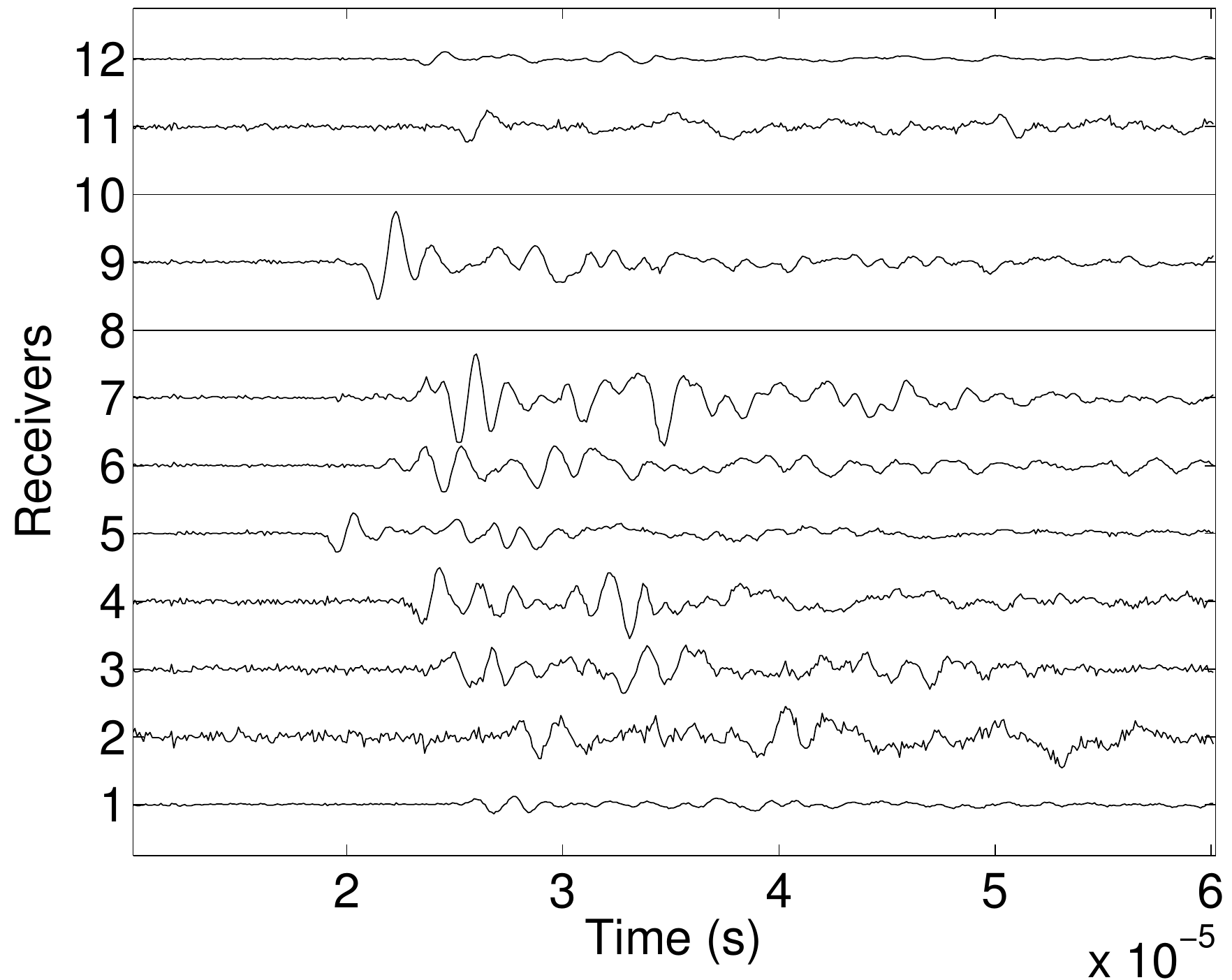}}\\
\subfloat{\includegraphics[scale=0.3]{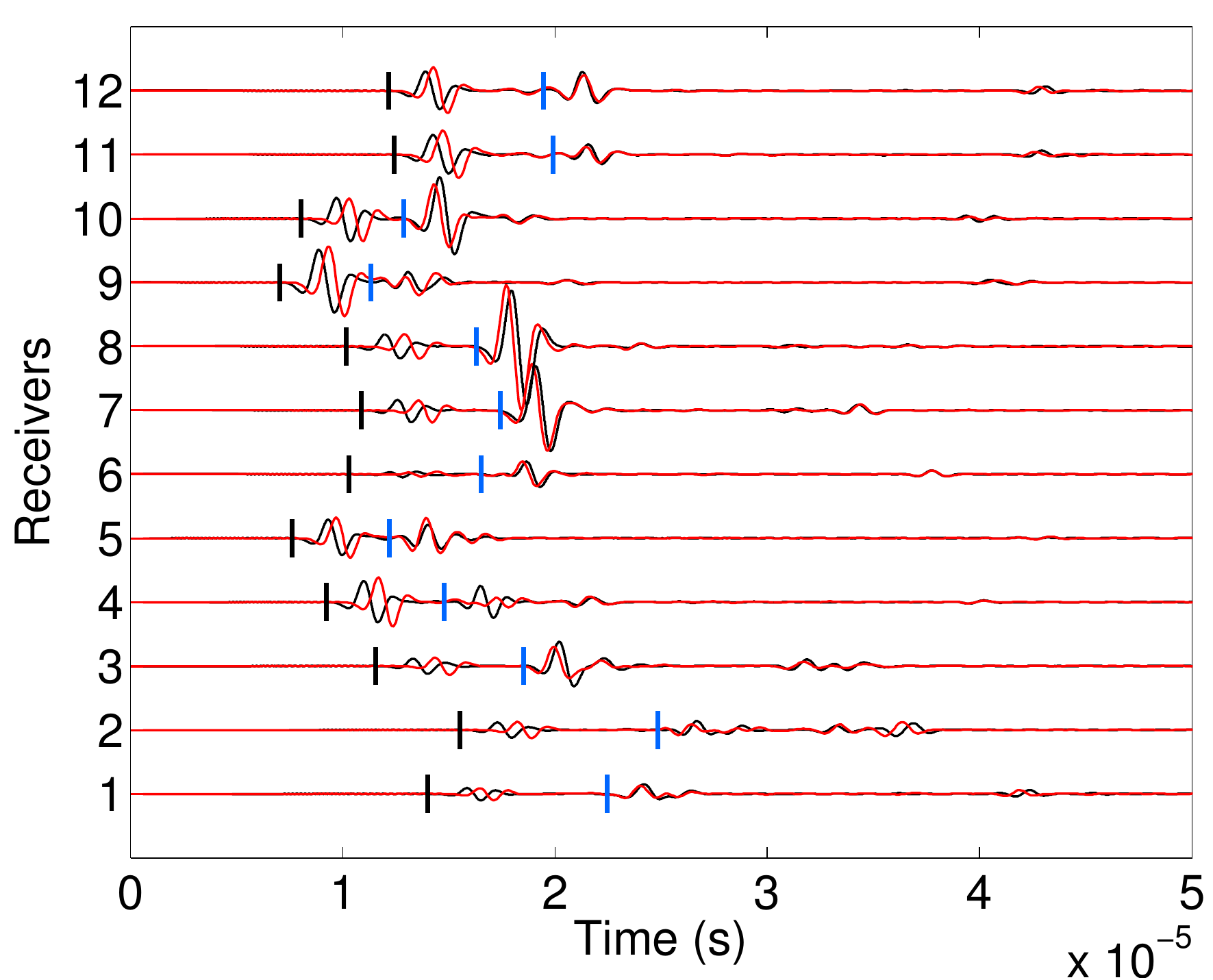}}
\caption{Synthetic waveforms (left column) computed for an isotropic (black) and an anisotropic (red) model of triaxial acoustic-emission experiment sample, and observed waveforms (right column) for the same sample. Top to bottom are $x$, $y$, and $z$ components respectively. For the observed data only the $x$ and $y$ components were available. Superimposed are the P- (black) and S-wave (blue) first-arrival times computed analytically. Seismograms are normalized to their absolute maximum.}
\label{fig:ngi_waveform}
\end{figure}

Figure~\ref{fig:ngi_wave} shows snapshots of P-wave and S-wave potentials. We observe reflected and converted wavefields from the borehole surface. Shear wave splitting is not clearly visible due to a weak anisotropy. We observe only delays in arrival times due to the lower velocity of the anisotropic model (Figure~\ref{fig:ngi_waveform}a). First-arrival times generally match with the observed data (Figure~\ref{fig:ngi_waveform}b). Since the piezoelectric sensors used in the experiment are only sensitive to radial motion, S-wave motion is not fully captured. As a result, P-wave amplitudes appear stronger than the S-wave amplitudes in the observed waveforms. Sensors 8 and 10 point towards the $x$ direction, and 7 and 9 towards the $y$ direction; therefore those sensors show signals only on the respective components. We observe  more complex waveforms in the real data. The model used to generate synthetic data does not represent all features of the original fractured sample. In the actual sample, the wavefield may interact with the borehole as well as induced fractures resulting in more complex waveforms. 

\section{Discussion}

We simulated wave propagation for microearthquakes in an underground ore mine and a rock slope, and acoustic emissions in a laboratory experiment using the spectral-element method. Although the hexahedral meshing may be a challenging task for complex velocity and structural models, the SEM is a stable and efficient tool.

For the mine model, structural complexity and high velocity contrasts caused by the mined-out cavities pose the main challenges for the wave propagation simulation. Some of the structural heterogeneities are in the order of the wavelength of the typically observed sources. As a result, a strong influence of those small-scale structures is observed in the complexity of the signals. The results of the 2D simulations show that unless we are interested in the acoustic wavefield inside the mined-out voids, we can safely discard those voids from meshing thereby reducing the computational costs significantly. Excluding the voids from meshing also significantly increases the numerical stability. We plan to investigate the possibility of full waveform tomography in the mine model using the adjoint capabilities of the spectral-element method \cite{tromp2008}. A reliable velocity model is important not only to locate microearthquakes more accurately but also to optimise mining production. 

For the \r Aknes rock slope model, we have used a homogeneous velocity model due to the unavailability of the real velocity model. Therefore, the actual topography represents the only complexity for the simulation of wave propagation. Even with a simple explosion source mechanism, the computed wavefield is complicated due to its interaction with the free surface topography. Location of microearthquakes in this model is challenging due to the rough topography and unavailability of a reliable velocity model. Joint inversion of microearthquakes and velocity structure using, for example, the adjoint method could be an interesting option.

For the cylindrical sample of the acoustic emission experiment, even a single borehole interacts with the wavefield and produces fairly complex signals. The effect of anisotropy is not clearly observed, in particular, in shear wave splitting, because the anisotropy is very weak. Synthetic Green's functions may be used to invert the source mechanism of microearthquakes. It would be interesting to see how closely the microearthquake mapping resembles the actual fracturing of the sample.

\section{Acknowledgments}

We thank Jeroen Tromp and Dimitri Komatitsch for helpful suggestion, Yang Luo and Christina Morency for their help on 2D meshing and modelling, and Ricardo M. Garcia, Jr.\, and Emanuele Casarotti for their help on hexahedral meshing with CUBIT. We thank \r Aknes/Tafjord Beredskap IKS (\texttt{www.aknes.no}) for the \r Aknes digital elevation map. We thank
Katja Sahala and ISS for access to the mine model and the in-mine data, and Eyvind Aker and Fabrice Cuisiat for access to the model and data of acoustic emission experiment. Parallel programs were run on the Titan cluster owned by the University of Oslo and the Norwegian metacenter for High Performance Computing (NOTUR), and operated by the Research Computing Services group at USIT, the University of Oslo IT-department. The 3D data were visualized using the open-source parallel visualization software ParaView/VTK (\texttt{www.paraview.org}). This work was funded by the Norwegian Research Council,
and supported by industry partners BP, Statoil and Total.

\bibliographystyle{plain}

\end{document}